\documentclass[english,usenatbib]{mn2e}
\pdfoutput=1
\usepackage[]{fontenc}
\usepackage[latin1]{inputenc}
\usepackage{amsmath}
\usepackage{graphicx}
\usepackage{amssymb}
\usepackage[authoryear]{natbib}
\usepackage{longtable}

\newcommand{\msun}{\mbox{${\rm M}_\odot$}}
\newcommand{\rsun}{\mbox{${\rm R}_\odot$}}

\newcommand{\apj}{\mbox{ApJ}}
\newcommand{\apjl}{\mbox{ApJL}}
\newcommand{\apjs}{\mbox{ApJS}}
\newcommand{\araa}{\mbox{ARA\&A}}
\newcommand{\aj}{\mbox{AJ}}
\newcommand{\mnras}{\mbox{MNRAS}}
\newcommand{\aap}{\mbox{A\&A}}
\newcommand{\nat}{\mbox{Nature}}

\title[Hydrodynamics of three-body interactions]{On the onset of runaway stellar collisions in dense star
  clusters - II. Hydrodynamics of three-body interactions}

\author[E. Gaburov, J. Lombardi and S. Portegies Zwart]
       {Evghenii Gaburov$^{1,2,3}$, James C.\ Lombardi, Jr.$^{4}$ and Simon Portegies Zwart$^{1,2,3}$\\
$^1$ Leiden Observatory, Leiden Observatory, the Netherlands\\
$^2$ Astronomical Institute ``Anton Pannekoek'', University of Amsterdam, the Netherlands \\ 
$^3$ Section Computational Science, University of Amsterdam, the Netherlands \\ 
$^4$ Department of Physics, Allegheny College,  USA
}

\begin{document}

\maketitle

\begin{abstract}
The onset of runaway stellar collisions in young star clusters is more
likely to initiate with an encounter between a binary and a third star
than between two single stars. Using the initial conditions of such
three-star encounters from direct $N$-body simulations, we model the
resulting interaction by means of Smoothed Particle Hydrodynamics
(SPH).  Our code implements new equations of motion that allow for
efficient use of non-equal mass particles and is capable of evolving
contact binaries for thousands of orbits.  We find that, in the
majority of the cases considered, all three stars merge together.  In
addition, we compare our SPH calculations against those of the
sticky-sphere approximation.  If one is not concerned with mass loss,
then the sticky sphere approach gives the correct qualitative outcome
in approximately 75\% of the cases considered.  Among those cases in
which the sticky-sphere algorithm identifies only two particular stars
to collide, the hydrodynamic calculations find the same qualitative
outcome in about half of the instances.  If the sticky-sphere approach
determines that all three stars merge, then the hydrodynamic
simulations invariable agree.  However, in such three star mergers,
the hydrodynamic simulations reveal that: (1) mass lost as ejecta can
be a considerable fraction of the total mass in the system (up to
$\sim25$\%); (2) due to asymmetric mass loss, the collision product
can sometimes recieve a kick velocity that exceeds 10 km/s, large
enough to allow the collision product to escape the core of the
cluster; and (3) the energy of the ejected matter can be large enough
(up to $\sim 3\times 10^{50}$ erg) to remove or disturb the inter
cluster gas appreciably.
\end{abstract}

\section{Introduction}\label{sect:introduction}

Stars are born in clusters, which upon formation are generally dense
and massive. In recent years it has become clear that clusters remain
bound even after losing a considerable fraction of their mass due to
primordial out-gassing \citep{2007MNRAS.380.1589B}.  The subsequent
dynamical evolution of these clusters leads to a state of core
collapse \citep{2007MNRAS.378L..29P}, almost irrespective of the
number of primordial binaries \citep{2004Natur.428..724P}; primordial
binaries do, however, appear to delay the collapse of the core
\citep{2003ApJ...593..772F,2006MNRAS.368..677H}.  In addition,
clusters with appropriate initial conditions may form a very massive
star by means of runaway stellar collisions
\citep{2004Natur.428..724P}. Such an object has been hypothised to be
a progenitor of an intermediate mass black hole (however, see
\cite{2009arXiv0902.1753G}).

Even if binaries are not present at the birth of a star cluster, they
can form via three-body encounters during the process of core
collapse.  Indeed, the expansion of the cluster core after deep
gravothermal collapse \citep{1983MNRAS.204P..19S} is mediated by
binaries, regardless of the presence or absence of a primordial
population.  During post-core collapse evolution, a cluster may enter
a phase of gravothermal oscillations \citep{1989ApJ...342..814C},
allowing periods of high interaction rate and providing further
opportunity for binaries and single stars to interact closely.

Analytic expressions describing encounters between a binary and a
third star, all treated as point masses, have been derived for various
portions of parameter space
\citep{1975MNRAS.173..729H,1983ApJ...268..342H,1993ApJS...85..347H}.
In addition, complementary numerical surveys have been performed in
the point mass approximation by a number of authors
\citep{1970AJ.....75.1140H,1983ApJ...268..319H, 1992AJ....103.1955H}.
During triple encounters, however, individual stars may approach close
enough to each other that the approximation of point-particle dynamics
breaks down: the size and internal structure of the stars then play a
major role in determining the outcome of the encounter. Consequently,
some numerical studies have augmented the point-mass treatment with
simplified models that incorporate several hydrodynamic effects
\citep{1986ApJ...306..552M,2004MNRAS.352....1F}. Large-scale $N$-body
simulations of clusters have demonstrated the ubiquity of resonance
interactions in dynamically unstable triples
\citep{1999A&A...348..117P}--the scenarios that ultimately may lead to
the coalescence of all three stars \citep{2004MNRAS.352....1F}. The
accurate modelling of the details under which triples merge, and
whether or not two or all three stars in an encounter participate in
the merger, has a profound consequence for the occurrence of collision
runaways \citep{2002ApJ...576..899P,2006MNRAS.368..141F} and whether
or not such runaways can lead to the formation of binaries among
intermediate mass black holes \citep{2006ApJ...640L..39G}.

The first three-dimensional hydrodynamic calculations of encounters
between a binary and a single star were performed by
\citet{1990ApJ...349..150C} with the smoothed particle hydrodynamics
(SPH) method. However, computational constraints at that time limited
their work to a very small number of SPH particles (usually 136 per
star) and to $n=1.5$ polytropes, appropriate only for white dwarfs or
extremely low mass main sequence stars.  Subsequent hydrodynamic
treatments of three-body interactions typically confined themselves to
scenarios in which at least one of the stars was a compact object and
therefore could be treated as a point mass
\citep[e.g.,][]{1993ApJ...411..285D,1994ApJ...424..870D}. \citet{1998MNRAS.301..745D}
and \citet{2004MNRAS.348..469A} consider three-body encounters between
a binary and a red giant star as a mechanism for destroying red giants
near the centres of dense stellar systems.  Their hydrodynamic
simulations follow the fluid of the red giant envelope during the
encounter, with the red giant core and both components of the binary
being treated as point masses.  Because only the red giant envelope is
treated hydrodynamically, the only mergers that can result are those
which form a binary of the two point masses surrounded by a common
envelope donated from the red giant envelope.

Numerous hydrodynamic simulations of colliding stars have studied the
structure of the merger product \citep{1987ApJ...323..614B,
  1994ApJ...424..870D, 1995ApJ...445L.117L, 1998MNRAS.301..745D,
  2002ApJ...568..939L, 2005MNRAS.358.1133F, 2008MNRAS.383L...5G}.  In
some cases the evolution of these collision products is studied
further \citep{2007ApJ...668..435S, 2009arXiv0902.1753G}, especially
within the context of the formation and evolution of blue stragglers
\citep{2001ApJ...548..323S,2005MNRAS.358..716S,2008arXiv0811.2974S,
  2008A&A...488.1007G, 2008A&A...488.1017G}.  Such collision studies,
however, have been focused on encounters between two single stars,
ignoring for the time being that collisional cross sections and rates
can be large for systems consisting of three or more stars.

The scenario of triple-star mergers among low-mass main-sequence stars
has been previously considered by \citet{2003MNRAS.345..762L} using
SPH.  Their calculations indicate that the collision product always
has a significantly enhanced cross-section and that the distribution
of most chemical elements within the final product is not sensitive to
many details of the initial conditions.  They, however, concentrated
solely on low mass stars and treated the triple star merger as two
separate, consecutive parabolic collisions.

Recently, \cite{2008MNRAS.384..376G} performed an extensive and
detailed study to investigate the circumstances under which a first
collision between stars occurs. Using direct $N$-body integration with
sticky spheres of realistic stellar sizes, they argued that binaries
tend to catalyse collisions. In their simulations the binaries that
are formed during core collapse tend to interact with an incoming
star, which subsequently merges with one of the binary components.
The results of \cite{2003MNRAS.345..762L} suggest that the
hydrodynamics of such interactions are unlikely to keep the binary
itself undamaged. Instead, it is quite likely that the stellar
material that is expelled during a collision engulfs the system in a
common envelope, leading to the merger of all three stars.

In this paper, we introduce a new implementation of SPH and apply it
to follow accurately the hydrodynamics of encounters between hard
binaries and intruders.  We concentrate on cases involving massive
main-sequence stars, such as those found in young star clusters,
treating all three stars simultaneously and with realistic orbital
parameters determined from a dynamical cluster calculation. In
particular, the initial conditions are selected from the set of
$N$-body simulations carried out by \cite{2008MNRAS.384..376G}, but
with the internal structure of the stars now being determined by a
stellar evolution code. A comprehensive survey of triple-star
collisions would need to explore an enormous amount of parameter
space, but here we focus on a number of representative cases. In
total, we selected 40 encounters from the simulations of
\cite{2008MNRAS.384..376G}. Among these are random selections, as well
as some that are specifically chosen because of their relevance for
the subsequent $N$-body evolution or because of their uncertain
outcome given the relatively simple treatment of mergers in the
$N$-body simulations.

This paper is structured as follows. In \S \ref{sect:methods} we
introduce our new formulation of SPH, which allows efficient use of
non-equal mass particles, as well as our approach for relaxing single
stars. In \S \ref{sect:relax_binary} we describe how we model close
and contact binary star systems, and we demonstrate the stability of
these systems for at least the time interval of interest. The set of
initial conditions for the three-body collisions are presented in \S
\ref{sect:initial_conditions}. Finally, \S \ref{sect:results}
presents, while \S \ref{sect:discussion} discusses, the results of our
calculations.

\section{Methods and Conventions}\label{sect:methods}

\subsection{SPH code}\label{sect:sph_code}

Smoothed Particle Hydrodynamics is the most widely used hydrodynamics
scheme in the astrophysics community. It is a Lagrangian particle
method, meaning that the fluid is represented by a finite number of
fluid elements or ``particles.''  Associated with each particle $i$
are, for example, its position ${\bf r}_i$, velocity ${\bf v}_i$, and
mass $m_i$.  Each particle also carries a purely numerical smoothing
length $h_i$ that determines the local spatial resolution and is used
in the calculation of fluid properties such as acceleration and
density.  See \citet{1992ARA&A..30..543M} and
\citet{1999JCoAM.109..213R} for reviews on SPH. The code which we used
in this work was presented in \citet{2006ApJ...640..441L}. However, we
modify the dynamical equations to allow the efficient use a of range
in particle masses. This modification is presented in Appendix
\ref{appendix:derivation}

\subsection{Choice of units}

Throughout this paper, numerical results are given in units where
$G=M_\odot=R_\odot=1$, where $G$ is the Newtonian gravitational
constant and $M_\odot$ and $R_\odot$ are the mass and radius of the
Sun.  The units of time, velocity, and energy are then
\begin{eqnarray}
  t_u & = & \left({R_\odot^3\over G M_\odot}\right)^{1/2} = 1594\,{\rm
    s},\\ v_u & = & \left({G M_\odot\over R_\odot}\right)^{1/2} =
  437\,{\rm km}\,{\rm s}^{-1},\label{vu} \\ E_u & = & {G M_\odot^2\over R_\odot} =
  3.79 \times 10^{48}\,{\rm erg}.\label{Eu}
\end{eqnarray}

\subsection{Relaxing a single star}\label{sect:relax_single}

\begin{figure}
  \begin{center}
    \includegraphics[width=\columnwidth]{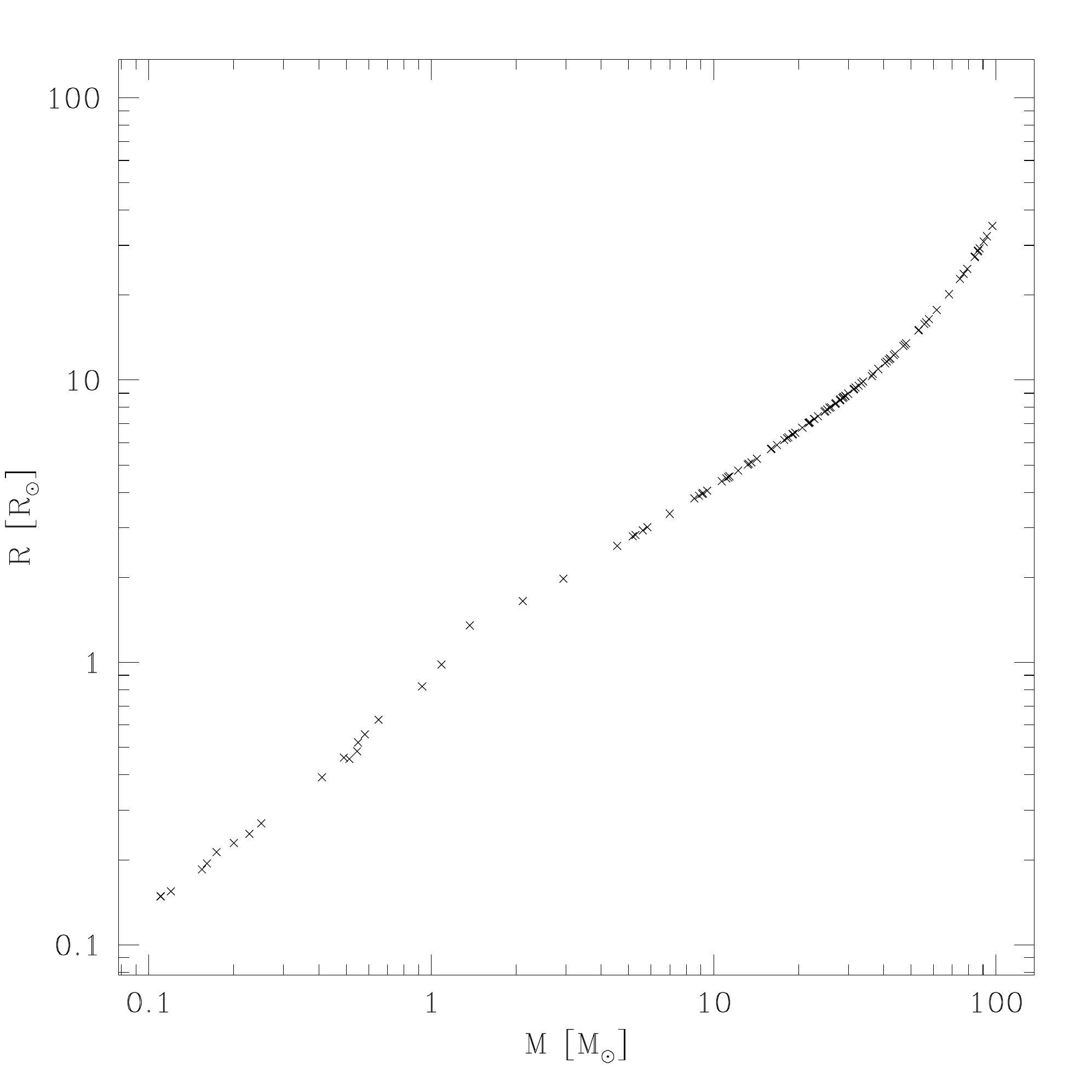}
  \end{center}
  \caption{Stellar radius versus mass at 2 Myr, as given by the TWIN stellar evolution code, for the stars
    considered in this paper.}
  \label{fig:ms}
\end{figure}

Before initiating a triple star collision, we must first prepare an
SPH model for each star in isolation.  To compute the structure and
composition profiles of our parent stars, we use the TWIN stellar
evolution code \citep{1971MNRAS.151..351E, 2008A&A...488.1017G,
  2008PHD_GLEBBEEK} from the MUSE software environment
\citep{2009NewA...14..369P}\footnote{{\tt http://muse.li}}. We evolve
main-sequence stars with initial helium abundance $Y=0.28$ and
metallicity $Z=0.02$ for a time $t=2\,$Myr, a small enough age that
even the most massive stars in a star cluster are still on the main
sequence. The mass-radius relation which results from these
calculations is shown in Figure \ref{fig:ms}.

Initially, we place the SPH particles on a hexagonal close packed
lattice, with particles extending out to a distance only a few
smoothing lengths less than the full stellar radius. After the initial
particle parameters have been assigned according to the desired
profiles from TWIN, we allow the SPH fluid to evolve into hydrostatic
equilibrium.  During this calculation, we include the artificial
viscosity contribution to the SPH acceleration equation so that energy
is conserved, and we do not include a drag force on the particles. For
the relaxation calculations of massive stars, we do, however,
implement a method to keep low mass particles from being pushed to
large radii: namely, during the initial stages of the relaxation, we
implement a variation on the XSPH method \citep{1992ARA&A..30..543M,
  2002MNRAS.335..843M}, in which the velocity used to update positions
is the average of the actual particle velocity and the desired
particle velocity (zero). All our relaxed models remain static and
stable when left to dynamically evolve in isolation.

\begin{figure}
  \begin{center}
    \includegraphics[width=84mm]{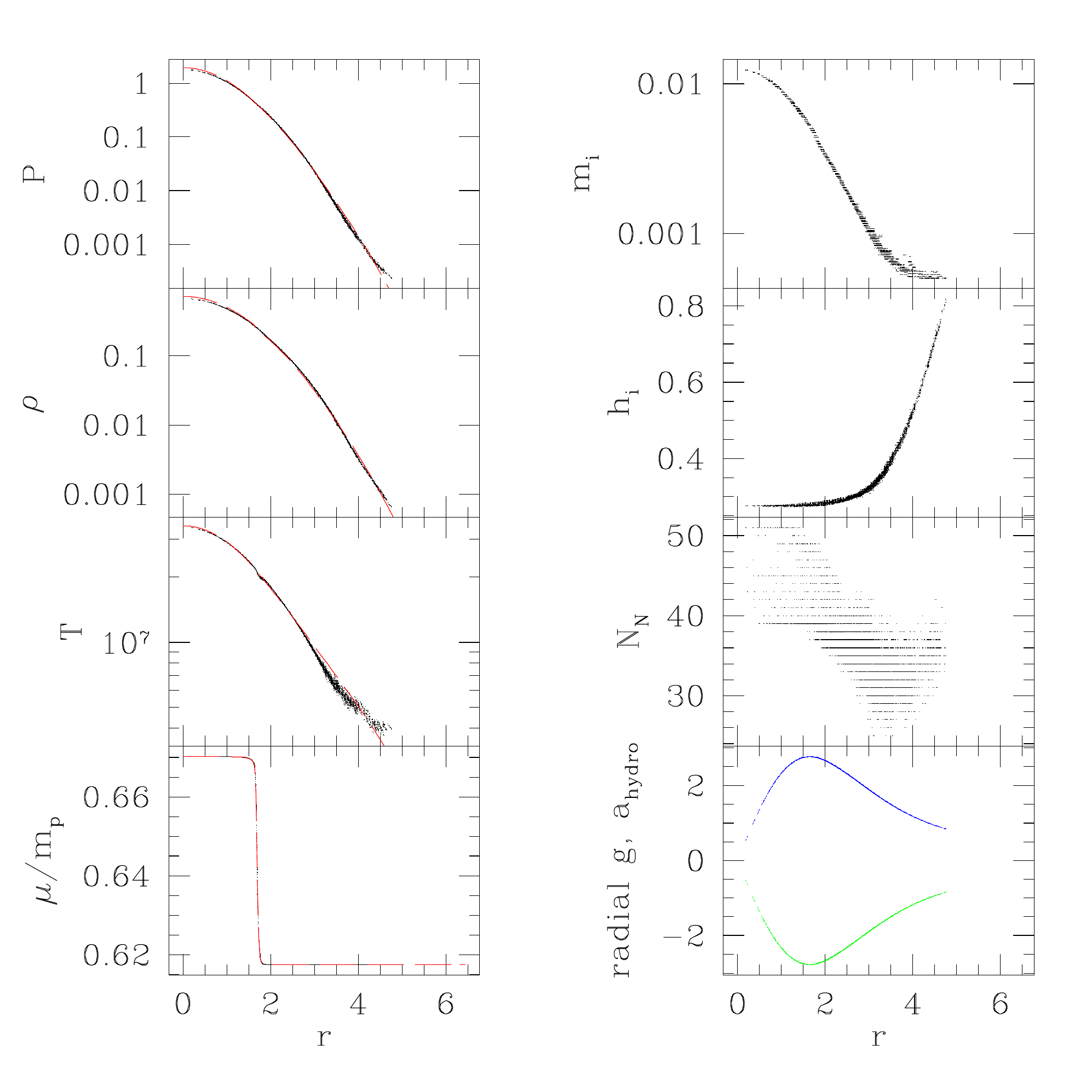}
  \end{center}
  \caption{Properties of the SPH model for a $19.1 M_\odot$ star.
    Profiles are shown as a function of radius, after relaxation for
    730 time units.  The frames in the left column show profiles of
    pressure $P$, density $\rho$, temperature $T$ (in Kelvin), and
    mean molecular weight $\mu$ in units of the proton mass $m_p$,
    with the dashed curve representing results the TWIN evolution code
    and dots representing particle data from our SPH model.  The right
    column provides additional SPH particle data: individual SPH
    particle mass $m_i$, smoothing length $h_i$, number of neighbours
    $N_N$, and radial component of the hydrodynamic acceleration
    $a_{\rm hydro}$ (upper data) and gravitational acceleration $g$
    (lower data).}
  \label{fig:plnb073_m19}
\end{figure}

This approach allows us to model the desired profiles very accurately,
and we present an example in Figure \ref{fig:plnb073_m19}, where we
plot desired profiles and SPH particle data for a $19.1 M_\odot$ star.
The structure and composition profiles of the SPH model closely follow
those from TWIN profiles, and the model remains stable when
evolved dynamically.

\section{Preparing a binary}\label{sect:relax_binary}

In this section, we present our algorithm to model the close binary
systems that are used in most of the triple star collisions (see
\S\,\ref{sect:results}). The first step in creating a binary is to
relax each of the two stellar components in isolation, as described in
the previous section. In the case of detached binaries, we place these
relaxed stellar models along the $x$ axis with their centres of mass
separated by the desired separation $r$.  For contact binaries,
however, we begin with the stars well separated and gradually decrease
the semi-major axis until the desired separation is achieved, in order
to minimise oscillations initiated by tidal forces. In all cases, the
centre of mass of the system remains fixed in space, which we choose
to be the origin.

During the binary relaxation process, the positions of the particles
within each star are adjusted at each timestep by simple uniform
translations along the binary axis, such that the separation between
the centres of mass equals the desired separation $r$.
Simultaneously, the angular velocity $\Omega_{orb}$ defining the
co-rotating frame is continuously updated, such that the net
centrifugal and gravitational accelerations of the two stars cancel
exactly:
\begin{equation}
\Omega_{\rm orb}^2=-{1\over 2} \left({\sum_{\star1} m_i {\dot
    v}_{x,i}\over \sum_{\star1} m_i x_i}+{\sum_{\star2} m_i {\dot
    v}_{x,i}\over \sum_{\star2} m_i x_i}\right),
\end{equation}
where $\sum_{\star j}$ symbolizes a sum over all particles in star $j$.
Here, the Cartesian coordinate $x$ is measured along to the binary
semi-major axis; ${\dot v}_{x,i}$ is the acceleration of particle $i$
parallel to the axis of the binary in an inertial frame. A centrifugal
acceleration is given to all particles such that the system approaches
a steady state corresponding to a synchronised binary.  As in the
relaxation process of a single star, we also include the artificial
viscosity contribution to the SPH acceleration equation.

\begin{figure}
  \begin{center}
    \includegraphics[width=\columnwidth]{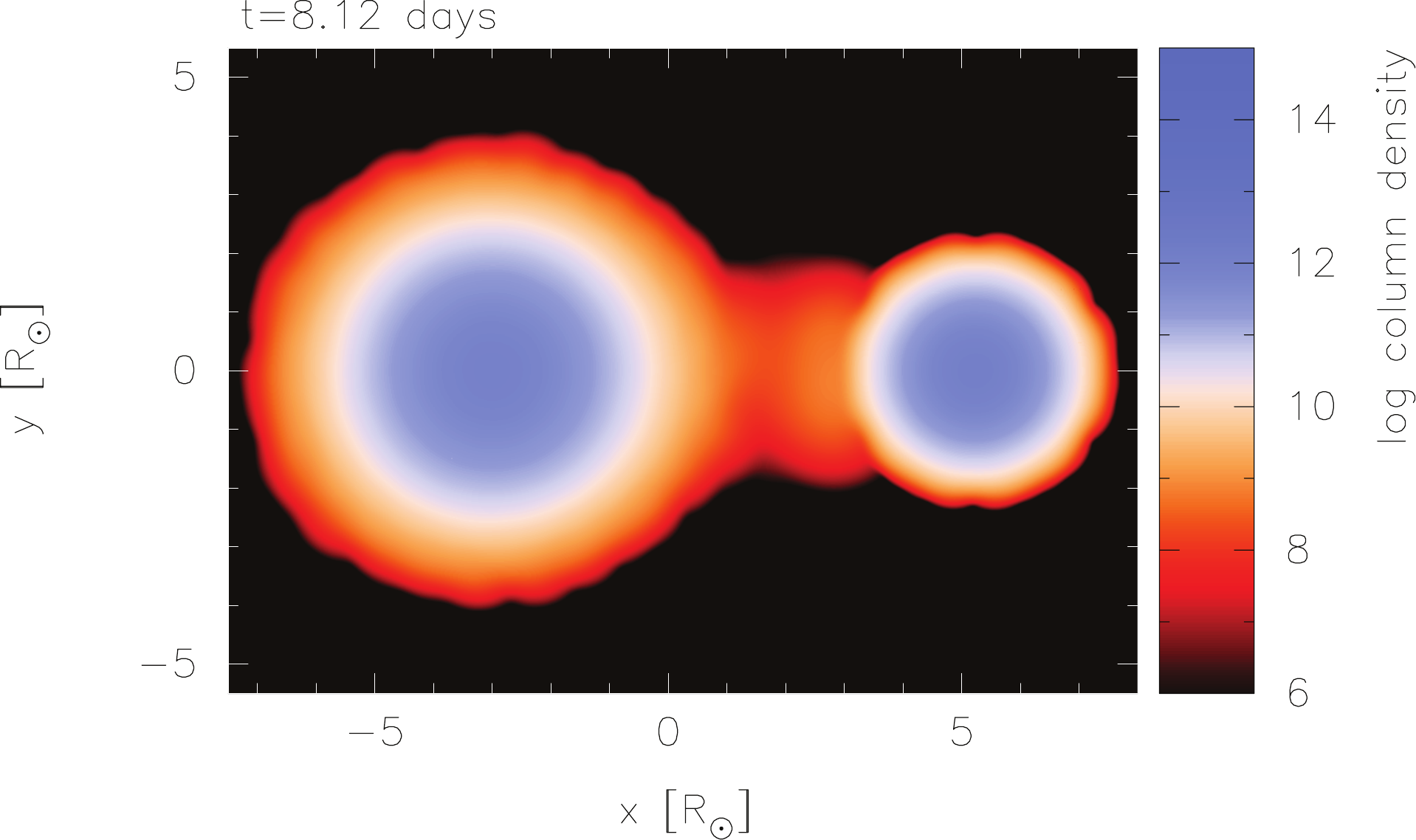}\\
    \includegraphics[width=\columnwidth]{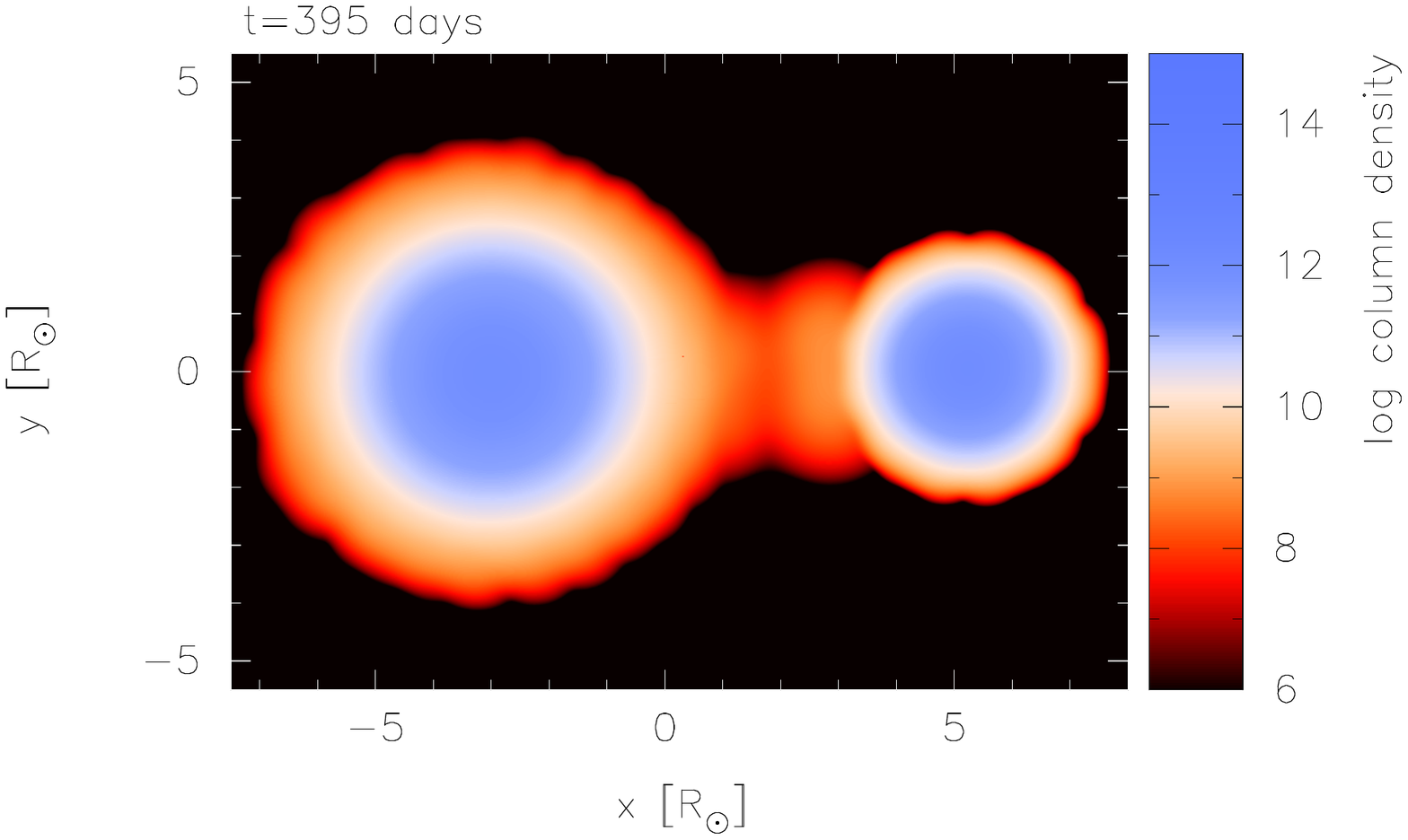}
  \end{center}
  \caption{A contact binary consisting of a $12.2 M_\odot$ primary and
    a $6.99M_\odot$ secondary both at the end of the relaxation (upper
    frame) and after dynamical evolution through more than 600 orbits
    (lower frame).  Colours represent column density, measured in g
    cm$^{-3}$ on a log scale, along lines of sight perpendicular to
    the orbital plane.}
  \label{fig:contact_binary}
\end{figure}

\begin{figure}
  \begin{center}
    \includegraphics[width=\columnwidth]{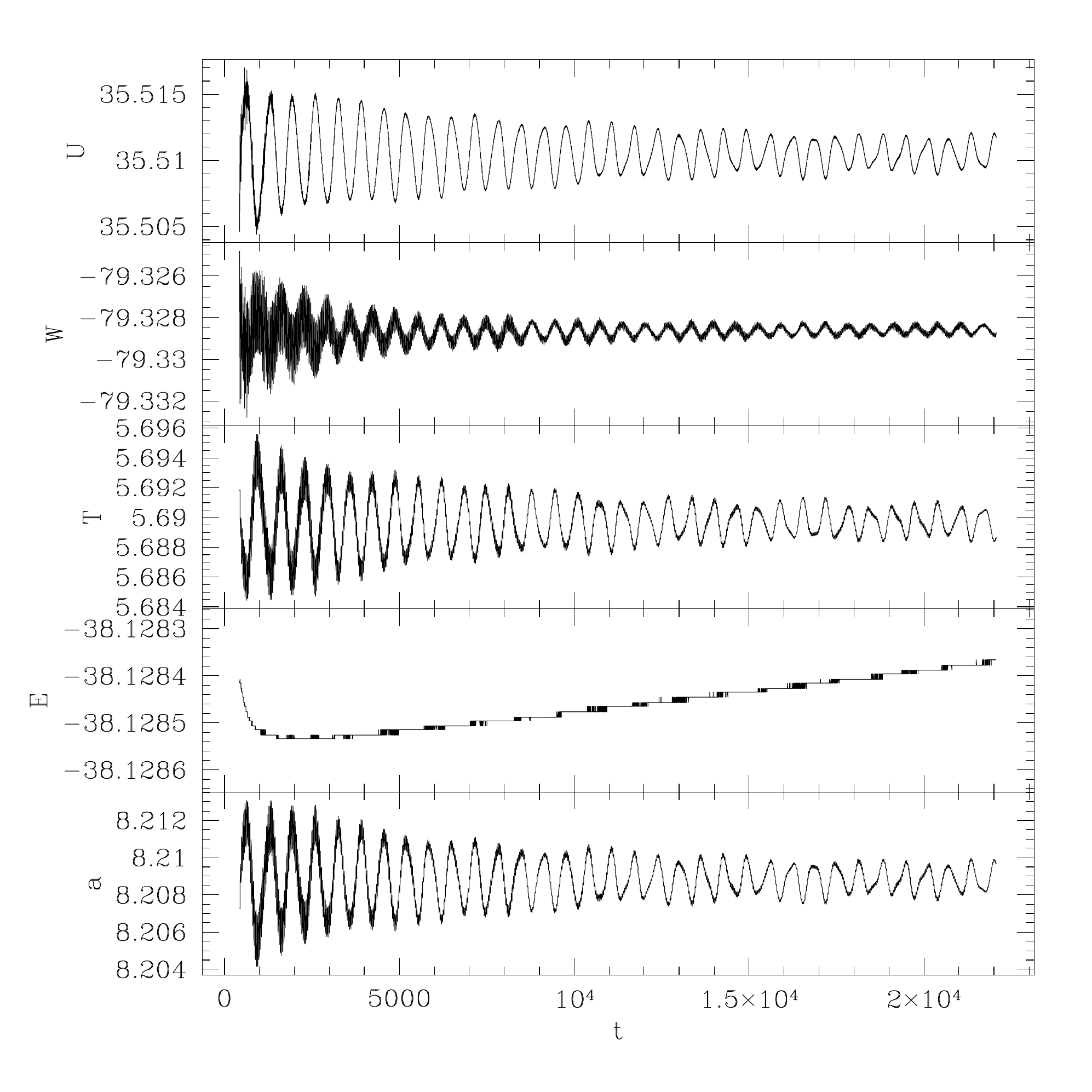}
  \end{center}
  \caption{Internal energy $U$, gravitational potential energy $W$,
    kinetic energy $T$ and total energy $E$ versus time $t$ for the
    dynamical evolution of the contact binary shown in Fig.\ \ref{fig:contact_binary}.
    The orbital
    period is 35 time units, while the epicyclic period is 650 time
    units.  }
  \label{fig:contact_0_440_99999}
\end{figure}

This approach allows us to create close binaries that remain in
dynamically stable orbits for many hundreds of orbits, if not indefinitely.
An example is presented in Fig. \ref{fig:contact_binary}
and Fig. \ref{fig:contact_0_440_99999}.  In
Fig. \ref{fig:contact_binary} we plot column densities of a contact
binary both before and after dynamical evolution through over 600
orbits.  In Fig. \ref{fig:contact_0_440_99999} we show time evolution
of various energies for the same binary.  The epicyclic oscillations,
with a period of 650 time units, are clearly visible.  The fact that
the epicyclic period is more than an order of magnitude larger than
the orbital period of 35 time units underscores how close this binary
is to the dynamical stability limit. As a binary approaches this
limit, the epicyclic period would formally approach infinity
\citep{1994ApJ...432..242R}.  The innermost dynamically stable orbit
then marks the transition when the squared frequency of the epicyclic
oscillations passes from a positive to a negative value, so that the
qualitative behaviour of perturbations changes from oscillatory to
exponential. Throughout the calculation, the perturbations remain
small and actually damp with time: the internal energy $U$ remains
constant to within about 0.03\%, the gravitational energy $W$ to
within about 0.008\%, and the kinetic energy to within about 0.2\%.
Meanwhile, the total energy is conserved 
within about 0.0004\%.

\begin{figure*}
  \begin{center}
     \includegraphics[scale=0.55]{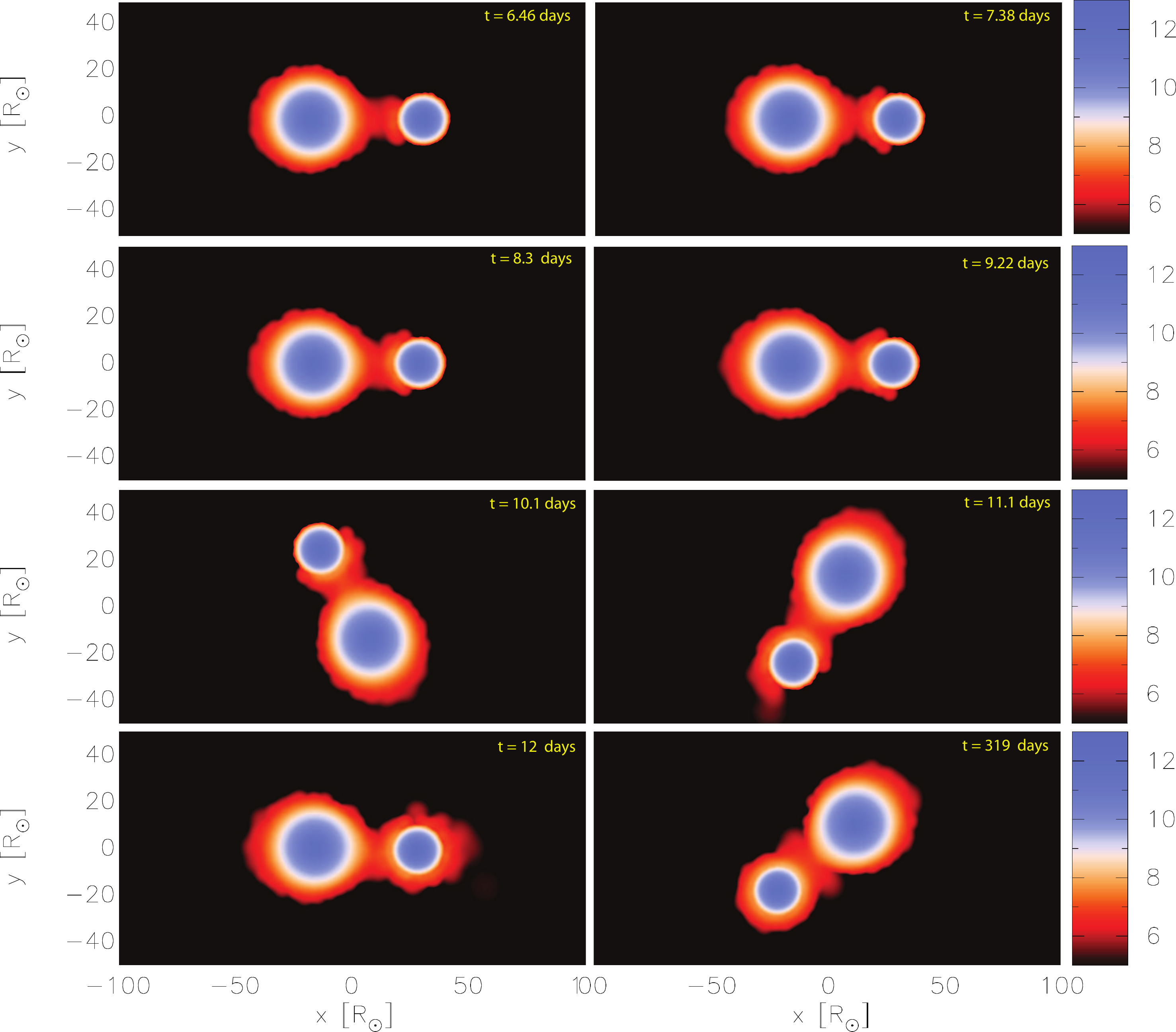}
  \end{center}
  \caption{The relaxation and dynamical evolution of a close binary
    between 92.9\msun\, and 53.3\msun\, with the semi-major axis equal
    to 43.8\rsun. The orbital period of the binary is 2.78 days (150.6 time
    units).  The calculation switches from a corotating frame to
    an inertial frame at a time of 9.22 days.}
  \label{fig:bin92+53}
\end{figure*}

\begin{figure}
  \begin{center}
    \includegraphics[width=\columnwidth]{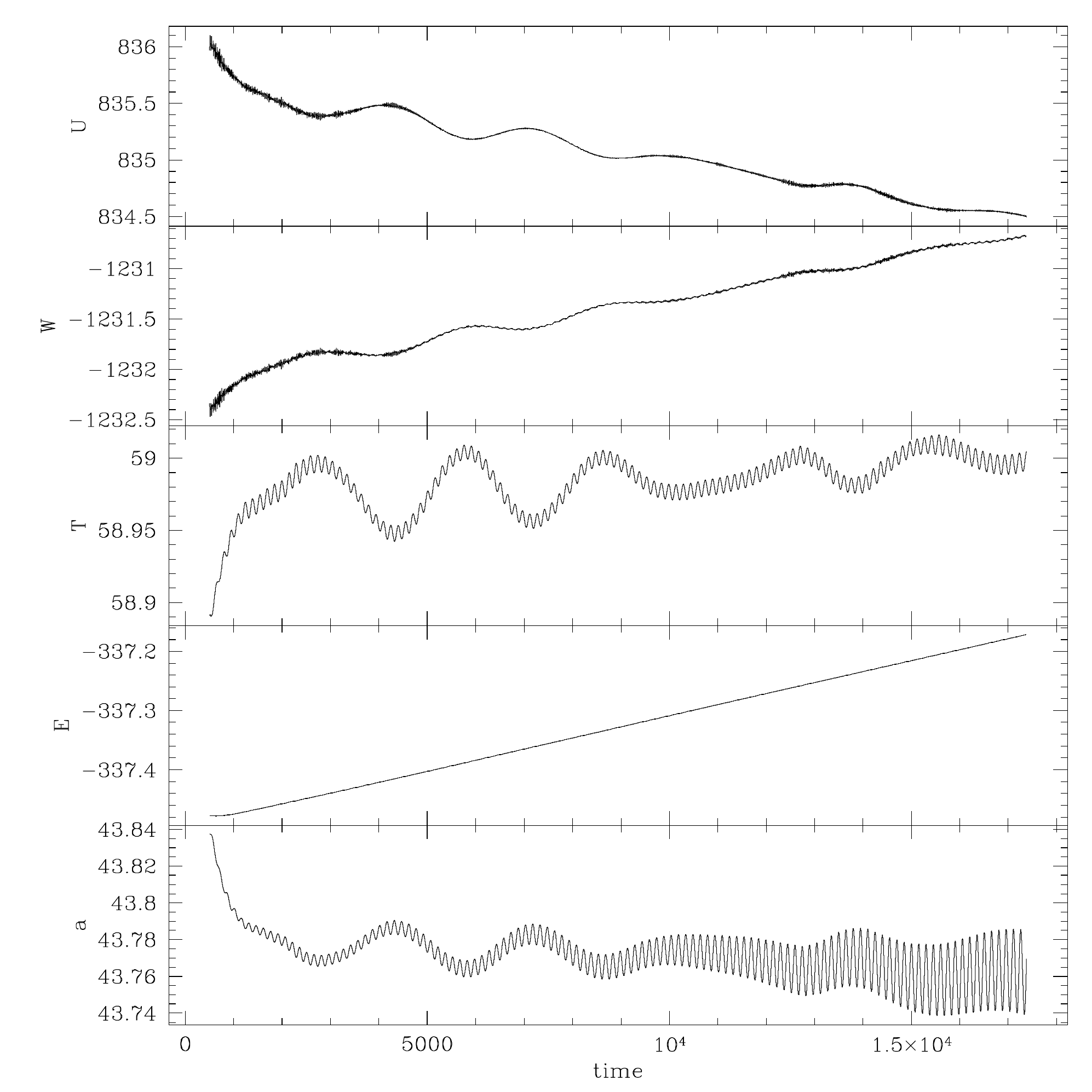}
  \end{center}
  \caption{Internal energy $U$, gravitational potential energy $W$,
    kinetic energy $T$, total energy $E$ and semi-major axis $a$
    versus time $t$ for the dynamical evolution of an isolated close
    binary consisting of a 92.9\msun\, primary and a 53.3\msun\,
    secondary star.
    All quantities remain within 0.2\% of their initial value throughout the simulation
    of more
    than 100 orbits, highlighting the abilty of our code to evolve stably even those
    binaries in deep contact. The small increase in the total energy occurs due
    to a few low mass particles that are escaping to infinity.}
  \label{fig:92+53energy}
\end{figure}

In another example, we relaxed a contact binary with a 92.9\msun\, and
a 53.3\msun\, star with semi-major axis of 43.8\rsun. In
Fig.\ref{fig:bin92+53} we show snapshots for every 50 time units (0.92
days) of the binary during and after relaxation process.
We began the relaxation process of a binary with an initial semi-major
axis of 55.8\rsun, and we decrease it to 43.8\rsun in 500 time units
(9.2 days). The top-most left panel shows a binary during relaxation
at a time of 350 units (6.5 days), and the semi-major axis at this
time is equal to 48.0\rsun. It is possible to notice commencement of
the mass transfer form the primary onto the secondary. At the time of
500 units (9.22 days), when the semi-major axis becomes 43.8\rsun, we
stop the relaxation and dynamically evolve the system in the inertial
frame. At this time half of the secondary star is already submerged in
the fluid of the primary star. By the time of 650 units (12 days), the
secondary star is completely engulfed in the fluid of the primary
star. The bottom-most right panel shows a binary at the time of 17300
units (319 days), and the semi-major axis maintains its value of
43.8\rsun. In Fig. \ref{fig:92+53energy} we show energy and semi-major
axis of the binary as a function of time.

\section{Initial conditions}\label{sect:initial_conditions}

The parameter space of three-body encounters is immense, leaving no
hope to be completely covered with SPH simulations. The approach we
take here is to study part of it by using the initial conditions
obtained from direct $N$-body simulation. In particular, we take
initial condition for three-body collisions from
\cite{2008MNRAS.384..376G} who carried out an extensive set of
$N$-body simulations of young star clusters. In these simulations the
stars were modelled as hard spheres with a given mass and
corresponding radius. A collision occurs when two spheres experience
physical contact, or in other words, when the separation between the
centre of these spheres is equal to the sum of their radii.  This
treatment of collisions, known as the sticky sphere approximation,
conserves total mass and momentum.

In this paper, however, we resolve the stellar structure and focus on
isolated close three-body interactions. This can be justified since
usually such interactions last less than a year, and therefore local
conditions hardly change on such a short timescale. All three-body
interactions we split in two groups: the interaction between a binary
and a single star, and the interaction between three single stars
which are in the middle of a resonant interaction. The latter case is
straightforward to model, as we need to prepare only relaxed single
star models, as described in \S \ref{sect:methods}, and then assign
the appropriate initial positions and velocities to each of the
stars. The actual dynamical interaction process is then modelled using
the SPH code.

In the case of an interaction between a binary and a single star, we
initially relax the binary as described in \S
\ref{sect:relax_binary}. The binary separation is taken from the
$N$-body simulations. Because most of the binaries have separations of
a few stellar radii, tidal circularisation plays an important role,
and therefore eccentricity of these binaries is nearly equal to
zero. In some of the cases, the synthetic stellar evolution part of
$N$-body calculations predict a binary separation too small to be
dynamically stable, and in such cases we relax an SPH model of the
binary near the smallest possible semi-major axis such that the binary
remains stable or quasi-stable, such that the merger time-scale is at
least a few thousand time units.

\begin{figure}
  \begin{center}
    \includegraphics[width=\columnwidth]{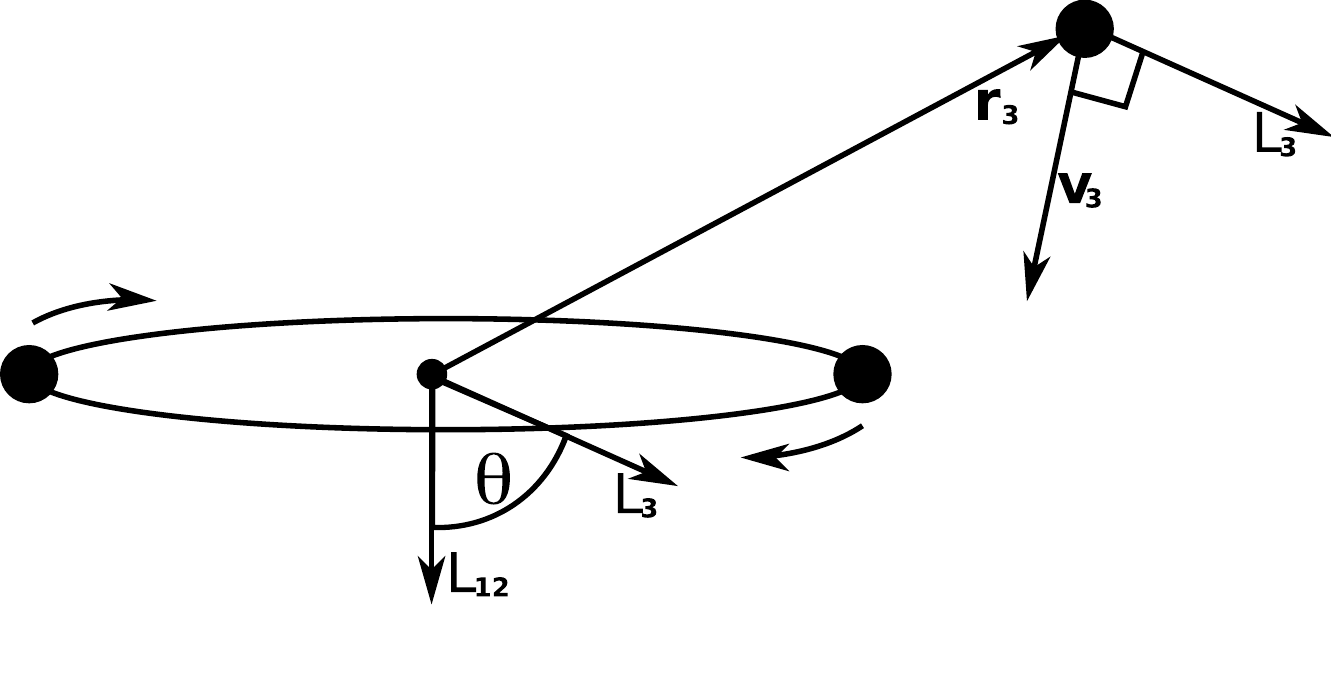}$\,$
  \end{center}
  \caption{The orientation of a binary and an intruder star}
  \label{fig:init_cond}
\end{figure}

Table \ref{tab:initial_conditions2} lists the initial positions and
trajectories in a way that is meant to aid the mental visualisation of
each case: for example, comparing the periastron separation $r_{\rm
  p,ib}$ to the binary semimajor axis $a_{12}$ provides an indication
of where within the binary the intruder strikes.  The ratio $E_{\rm
  ib}/|E_{12}|$ gives a measure of how much energy is being brought to
the system by the intruder, relative to the binding energy of the
binary. A negative value of $E_{\rm ib}/|E_{12}|$ implies that the
intruder star is bound to the binary, otherwise it is initially
unbound. We note, however, that the magnitude of this ratio is much
less than unity, which corresponds to a nearly parabolic encounter
between the intruder and the binary star.  Indeed, in almost all of
these cases the trajectory of the intruder about the binary is nearly
parabolic ($0.9 < e < 1.1$).  The last column indicates the angle of
approach of the intruder toward the binary, with
$0\le\theta\le180^\circ$.  More precisely, the angle $\theta$ is the
angle between the angular momentum vector ${\bf L_{12}}$ of the binary and the
angular momentum vector (${\bf L_3}$) of the intruder calculated about the center of
mass of the binary (Fig.\  \ref{fig:init_cond}). For example,
$\theta=0$ corresponds to coplanar trajectories with the intruder
orbiting the binary in the same direction (clockwise or
counter-clockwise) as the binary is orbiting; $\theta=90^\circ$
corresponds to the third star incident on the binary from a direction
perpendicular to the orbital plane of the binary; and
$\theta=180^\circ$ again corresponds to coplanar trajectories,
although now the intruder approaches with an angular momentum that is
antiparallel to that of the binary. All of our initial binaries are on
nearly circular orbits ($e_{12}<0.02$), with the exception of case 249
($e_{12}=0.41$).

\begin{table}
 \begin{center}
   \begin{tabular}{cccccccc}
     \hline
     \hline
     id & $m_{\rm b, 1}$ & $m_{\rm b,2}$ & $m_{\rm i}$ & $a_{12}$ & $r_{\rm p,ib}$ & $E_{\rm ib}/|E_{12}|$ & $\theta$ \\
        & \multicolumn{3}{c}{[\msun]} & \multicolumn{2}{c}{[\rsun]} &  & [$^\circ$] \\
     \hline
     203 & 47.1 & 36.3 & 1.09 & 33.8 & 33.3 &-7.6e-3 & 80 \\ 
     206 & 24.6 & 21.9 & 20.6 & 13.3 & 7.46 &+1.7e-1 & 68 \\
     207 & 42.2 & 18.2 & 0.65 & 30.6 & 16.3 &+3.0e-1 & 124 \\
     208 & 86.7 & 0.16 & 0.51 & 54.0 & 6.27 &+3.8e-2 & 41  \\
     211 & 61.7 & 8.89 & 18.4 & 23.3 & 52.8 &+1.1e-1 & 42  \\
     212 & 87.6 & 27.1 & 22.7 & 35.1 & 10.9 &-7.0e-2 & 102 \\
     213 & 76.8 & 13.6 & 0.23 & 32.3 & 2.77 &+2.2e-3 & 160 \\
     217&  86.4    &  28.9    & 0.11    &  51.1    &  9.58     &+1.2e-3 &       123 \\%carrv58
     222 & 22.8 & 11.1 & 5.28 & 34.8 & 10.5 &+3.4e-1 & 73 \\
     223&  28.6    &  4.57    &  19.4    &  13.0    &  12.1    &-4.3e-3 &       107 \\%carrv58
     224 & 48.1 & 22.0 & 0.2  & 22.5 & 5.85 &+2.6e-3 & 147 \\
     227&  16.0    & 0.17    &  5.62    &  25.6    &  4.05     &-1.2e-1 &        59 \\%carrv60
     231&  25.8    & 0.411    &  26.1    &  26.0    &  12.9    &-4.4e-1 &        43 \\%carrv59
     232&  12.2    &  6.99    &  19.1    &  8.26    &  2.10    &+1.7e-4 &        20 \\%carrv60
     233 & 28.9 & 2.94 & 47.6 & 26.7 & 12.2 &+1.0e-2 & 91 \\
     236&  40.5    &  31.4    &  29.3    &  23.5    &  19.2    &+1.4e-2 &        84 \\%carrv58
     241 & 28.1 & 11.3 & 41.7 & 13.6 & 12.8 &-1.6e-2 & 51 \\
     242&  41.1    &  23.5    & 0.490    &  21.3    &  1.54    &-2.5e-3 &       142 \\%carrv60
     245 & 43.5 & 16.0 & 79.1 & 33.2 & 32.7 &-2.0e-2 & 97 \\
     246&  42.2    &  38.3    &  1.37    &  28.4    &  19.1    &+8.7e-3 &       161 \\%carrv60
     249&  74.7    &  0.11    &  0.15    &  101     &  5.88    &-1.6e-1 &        51 \\%carrv59
     250&  44.0    &  31.9    & 0.550    &  36.4    &  3.48    &-2.1e-1 &        63 \\%carrv59
     253&  53.4    &  8.55    & 0.583    &  22.4    &  12.5    &-2.5e-2 &        35 \\%carrv59
     256&  33.4    &  2.11    &  5.84    &  18.6    &  7.50    &-5.2e-2 &        58 \\%carrv59
     257&  97.3    &  24.9    &  5.18    &  51.5    &  2.69    &-2.9e-1 &       132 \\
     258&  90.4    &  0.55    & 0.93     &  28.6    &  10.2    &-3.5e-3&        86 \\
     259&  55.9    &  21.7    &  11.4    &  26.3    &  6.72    &+2.9e-2&        46 \\%carrv59
     260 & 92.9 & 53.3 & 13.3 & 43.9 & 54.6 &+5.7e-1 & 14 \\
     267 & 28.6 & 14.2 & 19.1 & 26.3 & 0.28 &-5.3e-1 & 126 \\
     298 & 56.7    &  25.3    &  28.1    &  26.2    &  18.9    &-1.8e-1&       143 \\%carrv58
     299 & 52.3    &  16.9     & 52.3    &  26.2    & 0.0  &    -0.952e-4 &   94  \\
     \hline
     \hline
   \end{tabular}
 \end{center}
 \caption{In the first column, we present the case identification
   number. The second and third columns show the masses of the binary
   components, while the fourth column gives the mass of the intruder.
   The fifth column gives the semimajor axis $a_{12}$ of the binary.
   Columns 6 and 7 gives the periastron separation $r_{\rm p,ib}$ and
   eccentricity $e_{\rm ib}$, respectively, of the equivalent two-body
   Kepler orbit between the intruder and the center of mass of the
   binary.  Column 8 gives the ratio of the energy $E_{\rm ib}$ in
   this orbit of the intruder and binary to the binding energy
   $|E_{12}|$ of the binary itself.  Column 9 gives the angle
   $\theta$, in degrees, between the angular momentum of the binary
   and the angular momentum of the intruder about the binary (see
   Fig.\,\ref{fig:init_cond}). }
 \label{tab:initial_conditions2}
\end{table}

We initiate two types of hydrodynamic calculations.  The first type, which comprises the majority of our
calculations, consists of a co-rotating binary intruded upon by a
third star. In these situations, a circular binary is relaxed as we
described in \S \ref{sect:relax_binary}. If it is a contact binary,
then the circular orbit is maintained, with the orbital plane and phase
being shifted to match those of the desired initial conditions. If the
binary is detached, then the velocity of each star is adjusted to give
not only the desired orbital orientation and phase, but the
eccentricity as well. In this way, we account for tidal bulging in the
binary components. The third star, relaxed as described in \S
\ref{sect:methods}, is initially not rotating and separated from the
binary by many times the radius of the larger star, which allows us to
neglect its tidal effects in the initial configuration.

The second type of hydrodynamic calculation involves the collision of
three individual stars (Table \ref{tbl_eg_jcl_spz08:3s_runs}). These
also represent cases in which a binary is disrupted by an
intruder. In these cases the three stars are caught in a long-lived
resonant interaction that would be too computational expensive to
follow entirely with the hydrodynamic code. Each of the three stars is
first relaxed by the means described in \S \ref{sect:methods}. Their
initial positions and velocities in the collision calculation
represent a snapshot from the point mass dynamical calculation in
which the stars were widely separated but nearing the end of their
resonant interaction.

\begin{table}
  \begin{center}
    \begin{tabular}{cccc}
     \hline
     \hline
     id &  $m_1$ & $m_2$ & $m_3$ \\
        &  \multicolumn{3}{c}{[\msun]} \\
     \hline
     201&  84.1    & 0.25     &  27.1  \\
     202&  57.9    & 0.12     &  29.9  \\
     204&  42.4    & 11.5    & 16.5  \\
     214&  9.49    &  16.8    &  17.8  \\
     219&  36.6    &  9.10    &  10.7  \\
     220&  84.3    &  68.3    &  32.7   \\
     257&  5.18    &  24.9    &  97.3  \\
     261&  33.9    &  13.2    &  9.17  \\
     262&  29.3    &  31.5    & 18.4 \\
     \hline
     \hline
    \end{tabular}
  \end{center}
  \caption{The masses, in solar masses, of single stars which
    participate in the resonance interaction. The first column show
    the case number, while the following columns give the masses of
    participating stars}
  \label{tbl_eg_jcl_spz08:3s_runs}
\end{table}

\section{Results}\label{sect:results}

In this section we report on the results of 40 simulations of
different encounters between three stars. In terms of computational
time, most of the runs are performed with $N\sim 10^4$ and lasted
somewhere between one and two weeks on a modern PC equipped with an
MD-GRAPE3 \citep{1996ApJ...468...51F} card or an NVIDIA GPU
\citep{2007astro.ph..3100H, 2007NewA...12..641P, 2008NewA...13..103B,
  2009arXiv0902.4463G} for both self-gravity calculation. Our higher
resolution calculations ($N\sim 10^5$) typically require a few months to
complete; the total number of integration steps are usually between
$10^5$ and $10^6$.

\onecolumn
%\begin{landscape}
\begin{longtable}[c]{lrlrll}
  % \begin{minipage}{180mm}
  % \begin{tabular}{lrl}
  \hline
  \hline
  id& method & outcome & speed & $f_L$ & $E_{ej}$\\ 
  & & & [km/s] &  & [$10^{48}$erg]\\ 
  \hline 
  \hline
  201 & pm    & (1,2,3) $\rightarrow$ (1,2),3     & 0.8, 352\\
      & ss    & (1,2,3) $\rightarrow$ (\{1,3\},2) & 0\\
      & 14118 & (1,2,3) $\rightarrow$ (\{1,3\},2) & $ < 0.1$& $ < 0.001$ & $ < 0.1$ \\  % t=  449.0%carrv59:/runs/triples/three_singles/in_201_n-62 - checked traj3d by eye
      & 28296 & (1,2,3) $\rightarrow$ (\{1,3\},2) & $ < 0.1$& $ < 0.001$ & $ < 0.1$ \\  % t=  451.0%carrv59:/runs/triples/three_singles/in_201
      & 113046& (1,2,3) $\rightarrow$ (\{1,3\},2) & $ < 0.1$& $ < 0.001$ & $ < 0.1$ \\  % t=  464.0%carrv60:/runs/triples/three_singles/in_201/run201
  \hline
  202 &  pm   & (1,2,3) $\rightarrow$ (1,2),3     & 0.7, 529\\
      &  ss   & (1,2,3) $\rightarrow$ (\{1,3\},2) & 0\\
      &  6138 & (1,2,3) $\rightarrow$ (\{1,3\},2) &  $< 0.1$&  $< 0.001$&  $ <0.1$\\  % t=  311.0%carrv59:/runs/triples/three_singles/in_202_n-62
      & 11466 & (1,2,3) $\rightarrow$ (\{1,3\},2) &  $< 0.1$&  $< 0.001$ & $ <0.1$\\  % t=  311.0 %carrv59:/runs/triples/three_singles/in_202_n-125
      & 22380 & (1,2,3) $\rightarrow$ (\{1,3\},2) &  $< 0.1$&  $<0.001$ &  $ <0.1$\\  % t=  312.0 %carrv59:/runs/triples/three_singles/in_202 - checked by eye
      & 91956 & (1,2,3) $\rightarrow$ (\{1,3\},2) &  $< 0.1$&  $<0.001$ &  $< 0.1$\\  % t=  312.0 %carrv60:/runs/triples/three_singles/in_202/run202
  \hline
  203 & pm    & (1,2),3 $\rightarrow$ (1,2,3) $\rightarrow$ (1,2),3 & 6.24, 480   \\
      & ss    & (1,2),3 $\rightarrow$ (1,2,3) $\rightarrow$ (\{1,3\},2) & 0\\
      & 10398 & (1,2),3 $\rightarrow$ (1,2,3) $\rightarrow$ (\{1,3\},2) & $< 0.1$ & $< 0.001$ & $0.13$ \\
  \hline
  204 & pm    & (1,2,3)  $\rightarrow$ (1,3), 2 & 40.6, 143     \\
      & ss    & (1,2,3)  $\rightarrow$ (\{1,3\},2)  $\rightarrow$ \{2,\{1,3\}\} & 0\\
      & 11946 & (1,2,3)  $\rightarrow$ (\{1,2\},3\} $\rightarrow$ \{\{1,2\},3\}  & 7.2 & 0.13 & 12.\\
      & 60024 &  (1,2,3) $\rightarrow$ (\{1,2\},3\} $\rightarrow$ \{\{1,2\},3\}  & 3.8 & 0.082 & 12. \\
  \hline
  206 & pm    & (1,2),3 $\rightarrow$ (1,2,3)     $\rightarrow$ (1,2),3 & 60.7, 136  \\
      & ss    & (1,2),3 $\rightarrow$ (1,\{2,3\}) $\rightarrow$ \{\{2,3\},1\}& 0 \\
      & 14475 & (1,2),3 $\rightarrow$ (1,2,3)     $\rightarrow$ (\{1,2\},3) $\rightarrow$ \{\{1,2\},3\} & $1.2$ & $0.048$ & $13.$ \\
  \hline
  207 & pm    & (1,2),3  & 3.31, 308\\
      & ss    & (1,2),3 $\rightarrow$ (1,\{2,3\}) & 0\\
      & 9492  & (1,2),3  & 3.7, 345 & 0 & 0 \\
  \hline
  208 & pm    & (1,3),2 & 1.02, 555\\
      & ss    & (1,3),2 $\rightarrow$ (\{1,2\},3) & 0 \\
      & 15018 & (1,3),2 $\rightarrow$ (\{1,2\},3) & $< 0.1$ & $ < 0.001$ & $0.22$ \\
  \hline
  211 & pm    & (1,3),2  & 91.4, 824\\
      & ss    & (1,3),2 $\rightarrow$ \{1,3\},2 & 53.1, 203\\ 
      & 11028 & (1,3),2 $\rightarrow$ \{1,3\},2 & $39$, $146$ & $0.026$ & $6.7$\\
  \hline
  212 & pm    & (1,2),3 $\rightarrow$ (1,2,3)     $\rightarrow$ (1,3),2 & 34.3, 139\\
      & ss    & (1,2),3 $\rightarrow$ (\{1,3\},2) $\rightarrow$ \{\{1,3\},2\}& 0 \\
      & 22080 & (1,2),3 $\rightarrow$ (1,2,3)     $\rightarrow$ (\{1,3\},2) $\rightarrow$ \{\{1,3\},2\} & $3.6$ & $0.17$ & $66.$ \\
  \hline
  213 & pm     & (1,2),3  & 1.82, 724\\
      & ss     & (1,2),3 $\rightarrow$ (\{1,3\},2) & 0\\
      & 13314  & (1,2),3 $\rightarrow$ (\{1,3\},2) & $< 0.1$ &  $< 0.001$ & $< 0.1$ \\
      & 125130 & (1,2),3 $\rightarrow$ (\{1,3\},2) & $< 0.1$ &  $< 0.001$ & 1.5 \\
  \hline
  214 &    pm & (1,2,3) $\rightarrow$ (1,2),3  & 95, 348\\
      &   ss  & (1,2,3) $\rightarrow$ (\{1,2\},3)& 0 \\
      & 11016 & (1,2,3) $\rightarrow$ (\{2,1\},3) $\rightarrow$ \{\{2,1\},3\}  &  1.4    &    0.14 &  3.9  \\      % t= 1921.0 %carrv59: /runs3/triples/three_singles/new_integrator/in_214  -checked by eye
  \hline
  217 &  pm    & (1,2),3 $\rightarrow$ (1,2,3) $\rightarrow$ (1,2),3 & 0.7, 713 \\
      &  ss    & (1,2),3 $\rightarrow$ (\{1,3\},2) & 0\\
      &  17442 & (1,2),3 $\rightarrow$ (\{1,3\},2) & $< 0.1$ &  $< 0.001$ & $< 0.1$ \\ % t=  470.0 %carrv58
      &  139656& (1,2),3 $\rightarrow$ (\{1,3\},2) & $< 0.1$ &  $< 0.001$ & $< 0.1$ \\ % t=  470.0 %carrv58:/runs/triples/bin/in_217_hr/run217_hr
  \hline
  219 & pm    & (1,2,3) $\rightarrow$ (1,2),3 & 49, 255\\
      & ss    & (1,2,3) $\rightarrow$ (\{1,3\},2) $\rightarrow$ \{\{1,3\},2\}  &  0  \\
      & 14160 & (1,2,3) $\rightarrow$ (\{1,2\},3) $\rightarrow$ \{\{1,2\},3\}  &  8.1    &    0.038&  33.   \\      % t=  344.0 %carrv59:/runs3/triples/three_singles/new_integrator/in_219_redge4 - checked by eye
  \hline
  220 & pm    & (1,2,3) $\rightarrow$ (1,2),3 & 166, 778\\
      & ss    & (1,2,3) $\rightarrow$ (\{1,2\},3)$\rightarrow$ \{\{1,2\},3\} & 0\\
      & 20178 & (1,2,3) $\rightarrow$ (1,\{2,3\}) $\rightarrow$ \{\{2,3\},1\} &  14.    &    0.062 & 130 \\  % t= 1422.0 %carrv59:/runs3/triples/three_singles/new_integrator/n-62/in_220 - gives essentially the same result as N=46296 simulation
      & 46296 & (1,2,3) $\rightarrow$ (1,\{2,3\}) $\rightarrow$ \{\{2,3\},1\} &  11.    &    0.062 & 130 \\  % t= 1426.0 %carrv59:/runs3/triples/three_singles/new_integrator/in_220  - checked by eye
  \hline
  222 & pm    & (1,2),3 & 38.3, 246\\ 
      & ss    & (1,2),3 $\rightarrow$ (\{1,3\},2) $\rightarrow$ \{\{1,3\},2\} & 0\\
      & 17076 & (1,2),3  & 48, 310 & 0 & 0  \\
  \hline
  223 &  pm    & (1,3),2 $\rightarrow$ (1,2,3) $\rightarrow$ (1,2),3 & 43, 454\\
      &  ss    & (1,3),2 $\rightarrow$ (\{1,2\},3) $\rightarrow$  \{\{1,2\},3\}& 0\\
      &  12456 & (1,3),2 $\rightarrow$ (1,2,3) $\rightarrow$ (\{2,1\},3) $\rightarrow$  \{\{2,1\},3\} &  12. & 0.25 & 14. \\     % t=18483. %carrv58
  \hline
  224 & pm    & (1,2),3 & 0.858, 323\\
      & ss    & (1,2),3 $\rightarrow$ (\{1,3\},2)& 0\\
      & 14472 & (1,2),3 $\rightarrow$ (1,2,3) $\rightarrow$ (\{1,3\},2) $\rightarrow$ \{\{1,3\},2\} & $0.94$ & $0.023$ & $3.4$\\
  \hline
  227&  pm & (1,3),2 $\rightarrow$ (1,2,3) $\rightarrow$ (1,3),2 & 19, 55\\
     &  ss & (1,3),2 $\rightarrow$ (\{1,2\},3) $\rightarrow$ \{\{1,2\},3\}& 0\\
     &  10008 & (1,3),2 $\rightarrow$ (1,2,3) $\rightarrow$ (\{1,2\},3) $\rightarrow$ \{\{1,2\},3\} &  3.6    &    0.027&  2.3\\      % t=  706.0 %carrv60 - checked by eye.  It seems fine to say there is a short resonance phase because intruder and primary do not merge immediately.
  \hline
  231&  pm & (2,3),1 $\rightarrow$ (1,2,3) $\rightarrow$ (1,2),3 & 71, 72\\
     &  ss    & (2,3),1 $\rightarrow$ \{1,2\},3& 1.3, 163\\
     &  13554 & (2,3),1 $\rightarrow$ (1,2,3) $\rightarrow$ (\{1,3\},2) $\rightarrow$ \{2,\{1,3\}\}  & 0.25    &    0.036&  1.9\\      % t=23070.0 %carrv59:/runs3/triples/bin/in_231/run231b110
  \hline
  232 & pm  & (2,3),1 $\rightarrow$ (1,2,3) $\rightarrow$ (1,2),3 & 106, 107\\
      &  ss & (2,3),1 $\rightarrow$ (1,\{2,3\}) $\rightarrow$ \{\{2,3\},1\}& 0\\
      &  13110 & (2,3),1 $\rightarrow$ (\{1,3\},2) $\rightarrow$ \{\{1,3\},2\}  &  4.0    &    0.067&  20.\\      % t=  538.0 %carrv60
  \hline
  233 & pm    & (2,3),1 $\rightarrow$ (1,3),2 & 18.7, 487\\
      & ss    & (2,3),1 $\rightarrow$ (\{1,2\},3) $\rightarrow$ \{\{1,2\},3\} & 0\\
      & 13020 & (2,3),1 $\rightarrow$ (1,2,3) $\rightarrow$ (\{1,2\},3) $\rightarrow$ \{\{1,2\},3\} & 3.7 & 0.17 & 20 \\
  \hline
  236&  pm & (1,2),3 $\rightarrow$ (1,2,3) $\rightarrow$ (2,3),1 & 39, 58\\
     &  ss & (1,2),3 $\rightarrow$ (1,\{2,3\}) $\rightarrow$ \{\{2,3\},1\}& 0\\
     &  12672 & (1,2),3 $\rightarrow$ (1,2,3) $\rightarrow$ (\{1,2\},3) $\rightarrow$ \{\{1,2\},3\} &  5.8    &    0.14&  25. \\     % t= 1596.0 %carrv58
  \hline
  241 & pm    & (2,3),1 $\rightarrow$ (1,2,3)     $\rightarrow$ (1,3),2 & 68.1, 128\\
      & ss    & (2,3),1 $\rightarrow$ (\{1,3\},2) $\rightarrow$ \{\{1,3\},2\} & 0  \\
      & 19956 & (2,3),1 $\rightarrow$ (1,2,3)     $\rightarrow$ (\{1,3\},2) $\rightarrow$ \{\{1,3\},2\} & $8.0$ & $0.086$ & $32.$ \\
  \hline
  242&  pm & (1,2),3 & 6.5, 856\\
     &  ss & (1,2),3 $\rightarrow$ (\{1,3\},2) $\rightarrow$ \{\{1,3\},2\}& 0\\
     &  10224 & (1,2),3 $\rightarrow$ (1,2,3) $\rightarrow$ (\{1,3\},2) $\rightarrow$ \{2,\{1,3\}\} & 0.26    &    0.016&  2.0\\      % t= 1685.0 %carrv60 - checked by eye
  \hline
  245 & pm    & (2,3),1 $\rightarrow$ (1,2,3)     $\rightarrow$ (1,2),3 & 31.2, 23.9\\
      & ss    & (2,3),1 $\rightarrow$ (\{1,3\},2) $\rightarrow$ \{\{1,3\},2\}& 0  \\
      & 16884 & (2,3),1 $\rightarrow$ (1,2,3)     $\rightarrow$ (\{1,2\},3) $\rightarrow$ \{\{1,2\},3\} & $5.3$ & $0.027$ & $26.$  \\
  \hline
  246& pm     & (1,2),3 & 3, 194\\
     & ss     & (1,2),3 $\rightarrow$ (1,\{2,3\}) $\rightarrow$ \{1,\{2,3\}\}& 0\\
     &   5232 & (1,2),3 $\rightarrow$ (1,2,3) $\rightarrow$ (\{1,3\},2) $\rightarrow$ \{2,\{1,3\}\}&  2.5    &    0.017&  7.6\\      % t= 1717.0 %carrv60:/runs/triples/bin/in_246_lr/run246_lr - checked by eye
     &  10554 & (1,2),3 $\rightarrow$ (1,2,3) $\rightarrow$ (1,2),3 $\rightarrow$ \{1,2\},3  &  11.8, 691 &    0.010&  3.7\\      % t= 4707.0 %carrv60 (mr1)
     &  21204 & (1,2),3 $\rightarrow$ (1,2,3) $\rightarrow$ (1,\{2,3\}) $\rightarrow$ \{1,\{2,3\}\} & 1.1    &    0.023&  6.3\\      % t= 1204.0 %carrv60 (mr2)
     &  42294 & (1,2),3 $\rightarrow$ (1,2,3) $\rightarrow$ (1,\{2,3\}) $\rightarrow$ \{1,\{2,3\}\} & 0.71    &    0.020&  5.3\\   % t= 1153.0 %carrv59:/runs/triples/bin/in_246_mr3/run246_hr_mr3
     &  84642 & (1,2),3 $\rightarrow$ (1,2,3) $\rightarrow$ (\{1,2\},3) $\rightarrow$ \{1,\{2,3\}\} & 3.4    &    0.096&  7.9      \\% t= 7268.0 %carrv60
  \hline
  249&  pm & (1,3),2 $\rightarrow$ (1,2,3) $\rightarrow$ (1,3),2 & 0.06, 30\\
     &  ss    & (1,3),2 $\rightarrow$ (\{1,2\},3)& 0\\
     &  10374 & (1,3),2 $\rightarrow$ (\{1,2\},3) &  $ < 0.1$& $ < 0.001$&  $< 0.1$\\  % t=  603.0 %carrv59:/runs/triples/bin/in_249/run249_redge3
     &  82812 & (1,3),2 $\rightarrow$ (\{1,2\},3) &  $ < 0.1$&  $ < 0.001$ & $< 0.1$\\  % t=  605.0 %carrv59:/runs/triples/bin/in_249_hr/run249_hr
  \hline
  250&  pm    & (1,2),3 $\rightarrow$ (1,2,3) $\rightarrow$ (1,2),3 & 2.3, 324\\
     &  ss    & (1,2),3 $\rightarrow$ (1,2,3) $\rightarrow$ (\{1,3\},2)& 0\\
     &  10254 & (1,2),3 $\rightarrow$ (1,2,3) $\rightarrow$ (\{1,3\},2)  &  0.4    &  $ < 0.001$ &  0.54\\      % t=  356.0 % carrv59 - checked by eye
  \hline
  253&  pm    & (1,2),3 & 2.0, 212\\
     &  ss    & (1,2),3 $\rightarrow$ (\{1,3\},2) & 0\\
     &  10374 & (1,2),3 $\rightarrow$ \{1,2\},3  &   1.0, 176 &   0.032&  5.3\\      % t=14680.0%carrv59:/runs3/triples/bin/in_253/run253
  \hline
  256&  pm & (1,3),2 $\rightarrow$ (1,2,3) $\rightarrow$ (1,3),2 & 39, 234\\
     &  ss    & (1,3),2 $\rightarrow$ (\{1,2\},3) $\rightarrow$ \{\{1,2\},3\}& 0\\
     &  10200 & (1,3),2 $\rightarrow$ (1,2,3) $\rightarrow$ (\{1,2\},3) $\rightarrow$ \{\{1,2\},3\} &  7    &    0.15&  11.\\      % t= 8197.0%carrv59:/runs3/triples/bin/in_256/run256
  \hline
  257&  pm & (1,2,3) $\rightarrow$ (1,2),3 & 11, 257 \\
     &  ss & (1,2,3) $\rightarrow$ (\{1,3\},2) $\rightarrow$ \{\{1,3\},2\} & 0\\
     &  10236 & (1,2,3) $\rightarrow$ (\{1,3\},2) $\rightarrow$ \{\{1,3\},2\} &  2.9    &    0.087&  36.\\      % t= 1749.0 %carrv58 - checked by eye
  \hline
  258&  pm    & (1,3),2 & 0.8, 79\\
     &  ss    & (1,3),2 $\rightarrow$ (\{1,2\},3) $\rightarrow$ \{\{1,2\},3\}& 0\\
     &  10104 & (1,3),2 $\rightarrow$ (1,2,3) $\rightarrow$ (\{1,2\},3) $\rightarrow$ \{\{1,2\},3\} &  0.6    &  $<0.001$&   2.0\\      % t=  784.0%carrv59:/runs3/triples/bin/in_258/run258
  \hline
  259&  pm & (1,2),3 & 52, 353\\
     &  ss    & (1,2),3 $\rightarrow$ (\{1,3\},2) $\rightarrow$ \{\{1,3\},2\}& 0\\
     &  10272 & (1,2),3 $\rightarrow$ (1,2,3) $\rightarrow$ (\{1,3\},2) $\rightarrow$ \{\{1,3\},2\}  &  5    &    0.085&  30.\\      % t=  749.0%carrv59:/runs3/triples/bin/in_259/run259
  \hline
  260 & pm    & (1,2),3 $\rightarrow$ (1,2,3)  $\rightarrow$ (1,2),3 & 31.0, 340\\
      & ss    & (1,2),3 $\rightarrow$ (1,\{2,3\}) & 0\\
      & 22518 & (1,2),3 $\rightarrow$ \{1,2\},3 & 44, 456 & 0.024 & 6.6  \\
  \hline
  261&  pm & (1,2,3) $\rightarrow$ (1,2),3 & 39, 198\\
    &  ss & (1,2,3) $\rightarrow$ (\{1,3\},2)& 0\\
    &  10092 & (1,2,3) $\rightarrow$ (\{1,3\},2)&  7.1    &    0.020&  22.\\      % t=  283.0 %carrv58
  \hline
  262&  pm & (1,2,3) $\rightarrow$ (1,2),3 & 69, 226\\
     &  ss & (1,2,3) $\rightarrow$ (1,\{2,3\})& 0\\
    &  10314 & (1,2,3) $\rightarrow$ (1,2),3 $\rightarrow$ \{2,1\},3&  68, 223&    0.011&  2.1\\      % t= 1009.0  %carrv58
  \hline
  267 & pm    & (1,3),2 $\rightarrow$ (1,2),3 & 403, 905 \\
      & ss    & (1,3),2 $\rightarrow$ (\{1,3\},2) $\rightarrow$ \{2,\{1,3\}\} & 0 \\
      & 14934 & (1,3),2 $\rightarrow$ (\{1,2\},3) $\rightarrow$ \{\{1,2\},3\} & $9.6$ & $0.063$ & $67.$ \\
  \hline
  298&  pm & (1,3),2 $\rightarrow$ (1,2,3) $\rightarrow$ (1,2),3 & 60, 200\\
     &  ss & (1,3),2 $\rightarrow$ (\{1,2\},3) $\rightarrow$ \{\{1,2\},3\}& 0\\
     &  13818 & (1,3),2 $\rightarrow$ (1,2,3) $\rightarrow$ (\{1,3\},2) $\rightarrow$ \{\{1,3\},2\}  &  7    &   0.21&  52.\\      % t= 1609.0  %carrv58
  \hline
  299 & pm &     (1,3),2 $\rightarrow$ (1,2,3) $\rightarrow$ (2,3), 1 & 128, 797 \\
      & ss &     (1,3),2 $\rightarrow$ (\{1,2\},3) $\rightarrow$ \{\{1,2\},3\} & 0\\ 
      & 10194 &  (1,3),2 $\rightarrow$ (\{1,2\}, 3) $\rightarrow$ \{\{1,2\},3\} & 6.9 & 0.15 & 310 \\
      & 102540 & (1,3),2 $\rightarrow$ (\{1,2\}, 3) $\rightarrow$ \{\{1,2\},3\} & 2.7 & 0.16 & 330 \\ 
  \hline
%---------------------------
\hline
\\* % 1 newline with no page break allowed
%  \end{tabular}
%  \end{minipage}
\caption{Summary of the 40 simulations for three-star interactions.
  The first column gives the case identification number. The second
  column either gives the number $N$ of SPH particles used to simulate
  this case, or names the treatment as ``pm'' (point mass) or ``ss''
  (sticky spheres).  The third column summarizes the interaction that
  resulted by listing all changes in the state of the system.  The fourth column lists the projected speed(s) at
  infinity of the resulting object(s), in units of km s$^{-1}$.  The
  fifth column gives the fractional mass loss 200 time units after the
  final change of state, while the sixth column lists the total energy
  ejected, in units of $10^{48}$ erg, at that same time.}
\label{tab:results}
\end{longtable}
%\end{landscape}
\twocolumn

\subsection{Selected cases}

In Table \ref{tab:results} we summarise the outcomes of all collisions
from Tables \ref{tab:initial_conditions2} and
\ref{tbl_eg_jcl_spz08:3s_runs}. Binaries are represented by $\rm
(1,2)$ while resonances are represented by $\rm (1,2,3)$, with the
masses satisfying $M_1>M_2>M_3$. The merger product between stars $\rm
1$ and $\rm 2$, due either to a binary coalescence or a direct
collision, is represented using braces, $\rm \{1,2\}$, where
the more massive component {\it at the time of the merger} is listed
first.
In addition, the notation can be embedded.  Consider, for example,
case 242 with the following interaction sequence: (1,2),3
$\rightarrow$ (1,2,3) $\rightarrow$ (\{1,3\},2) $\rightarrow$
\{2,\{1,3\}\}. The initial state (1,2),3 represents a primary 1 and a
secondary 2 in a binary being intruded upon by the least massive star
3.  The (1,2,3) indicates that there is neither an immediate retreat
of the intruder nor an immediate merger, but instead the three stars
move in a resonance interaction.  The state (\{1,3\},2) means that the
intruder has merged with the primary, leaving the merger product in a
binary with the secondary star.  Finally, \{2,\{1,3\}\} indicates that
these two remaining objects coalesce. Note that in this final state,
the secondary star indicated to the left of \{1,3\} within the outer
braces, because the former was more massive at the time of the merger
due to dynamical mass transfer during the final stages of binary
inspiral.

\begin{figure}
  \begin{center}
    \includegraphics[width=\columnwidth]{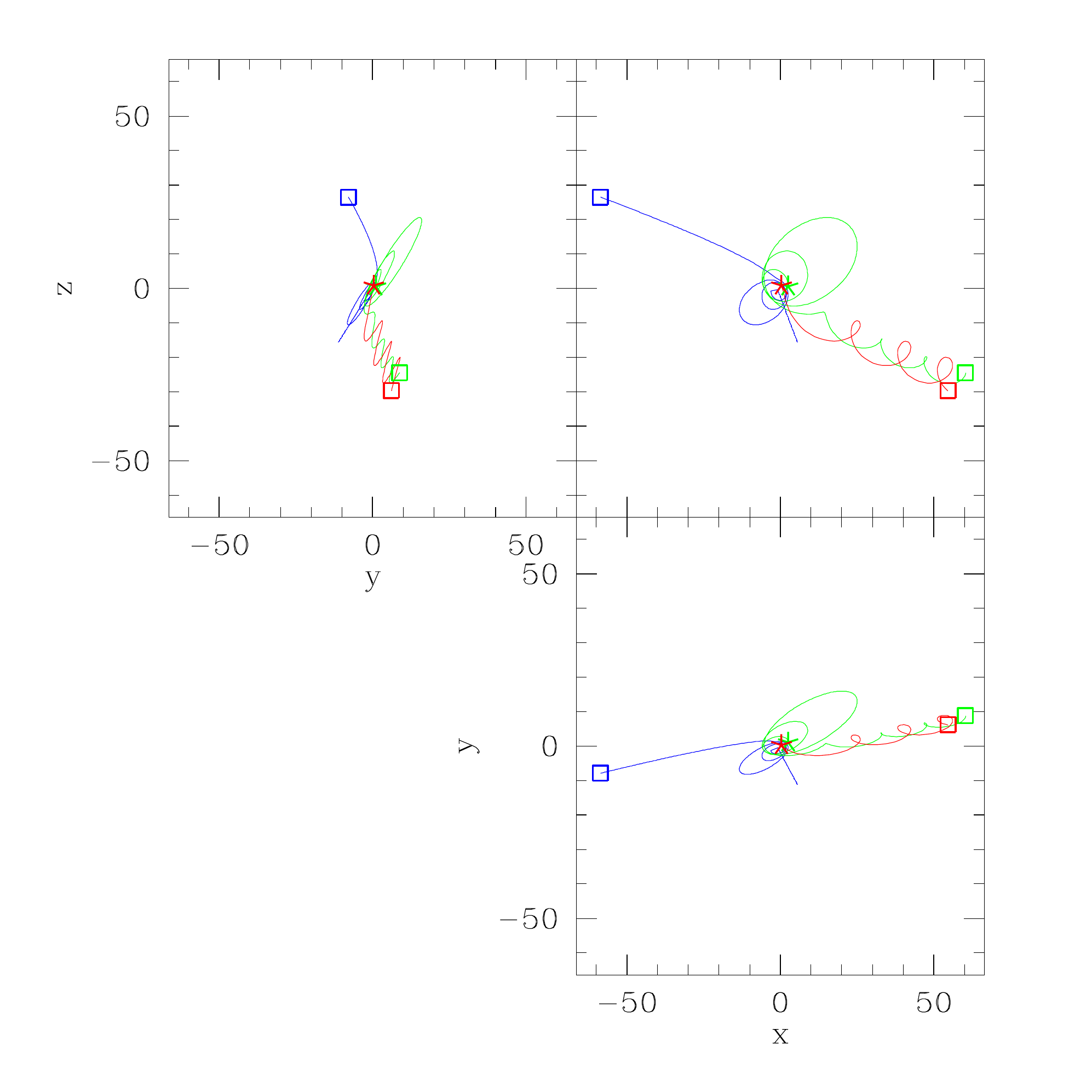}
  \end{center}
  \caption{Trajectories in the $xy$ (lower right), $xz$ (upper right),
    and $yz$ (upper left) planes for case 232, as given by our
    hydrodynamics calculation.  The initial conditions are marked by
    squares, while the final position of an object before merger is
    marked by a 5-point asterisk.}
  \label{fig:trajectories232}
\end{figure}

\begin{figure*}
  \begin{center}
    \includegraphics[scale=0.5]{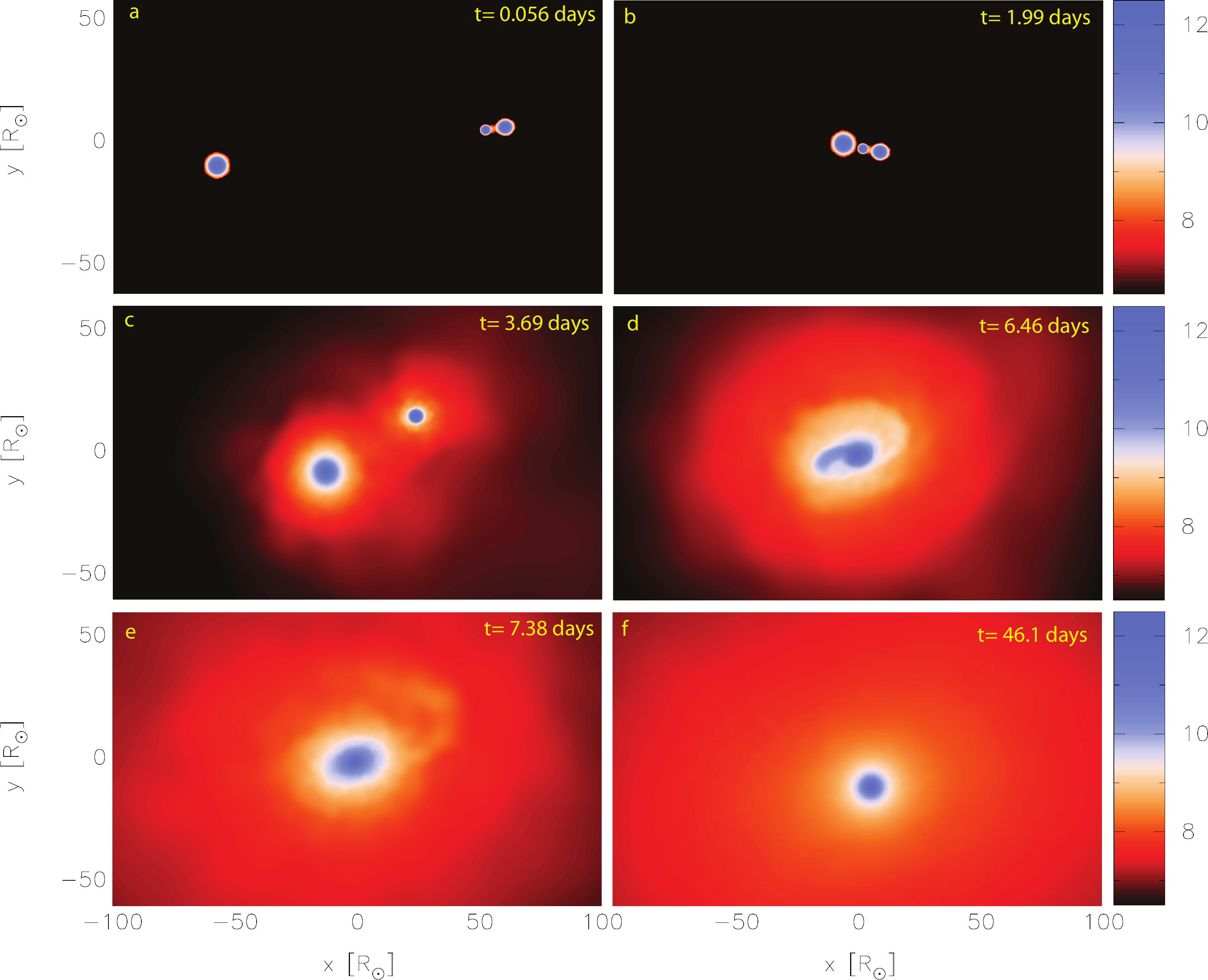}
  \end{center}
  \caption{Column density along lines of sight perpendicular to the
    $xy$ plane at various times for the same hydrodynamic calculation
    of case 232 presented in Figure \ref{fig:trajectories232}.
}
  \label{fig:case232}
\end{figure*}

\begin{figure}
  \begin{center}
    \includegraphics[width=\columnwidth]{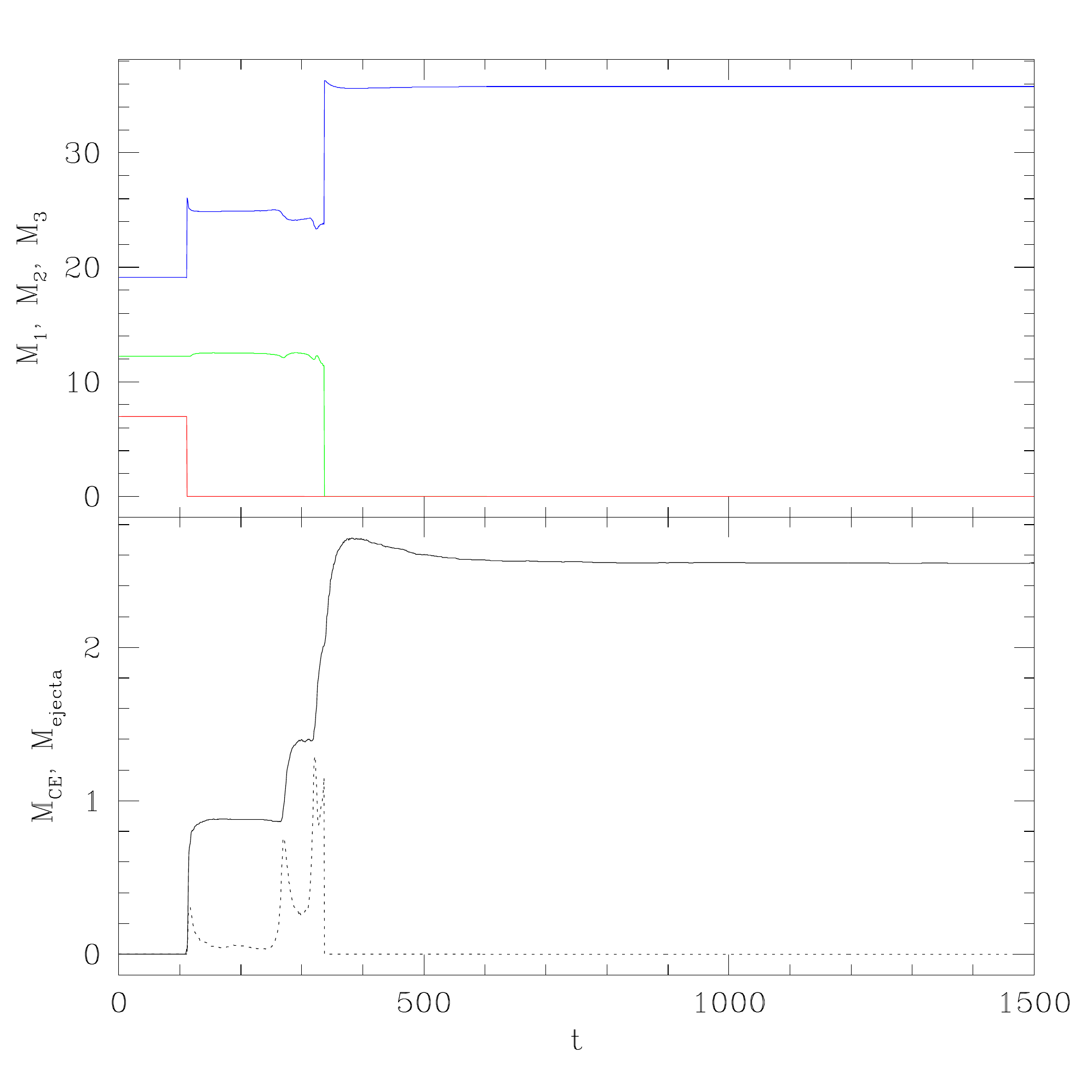}
  \end{center}
  \caption{Masses versus time for case 232.  The top frame shows
    the time evolution of the mass of the primary, secondary, and
    intruder.  The bottom frame shows the amount of mass ejected
    (solid curve) and bound in a circumbinary envelope (dotted
    curve).}
  \label{fig:case232_mvtt}
\end{figure}

Here, we present several cases in greater details. First, in Figures
\ref{fig:trajectories232} and \ref{fig:case232} we show trajectories
and column density plots respectively from calculations of case 232,
in which a $12.2+6.99 M_\odot$ binary collides with a $19.1 M_\odot$
intruder.  The set up of the initial conditions for these three
particular stars is described in \S\ref{sect:methods}
(Figs.\ \ref{fig:plnb073_m19}, \ref{fig:contact_binary} and
\ref{fig:contact_0_440_99999}).  Figure \ref{fig:case232}a shows the
three bodies shortly after the start of the calculation.  Figure
\ref{fig:case232}b shows the three bodies just prior to the impact and
merger of the intruder and the secondary from the binary. The first
apocentre passage in the resulting binary star is shown in Figure
\ref{fig:case232}c, while Figure \ref{fig:case232}d shows the binary
in the process of merger. In Figure \ref{fig:case232}e we show the
snapshot shortly after the fluid from the three stars has merged into
a single object, and finally, Figure \ref{fig:case232}f shows a
snapshot from near the end of our calculation: the merger product has
drifted away from the origin due to asymmetric mass loss. In this
case, the merger product has little angular momentum, and the
calculated mass loss quickly asymptotes to a constant value of
approximately 2.6$M_\odot$ (see Fig.\ \ref{fig:case232_mvtt}).

\begin{figure*}
  \begin{center}
    \includegraphics[scale=0.4]{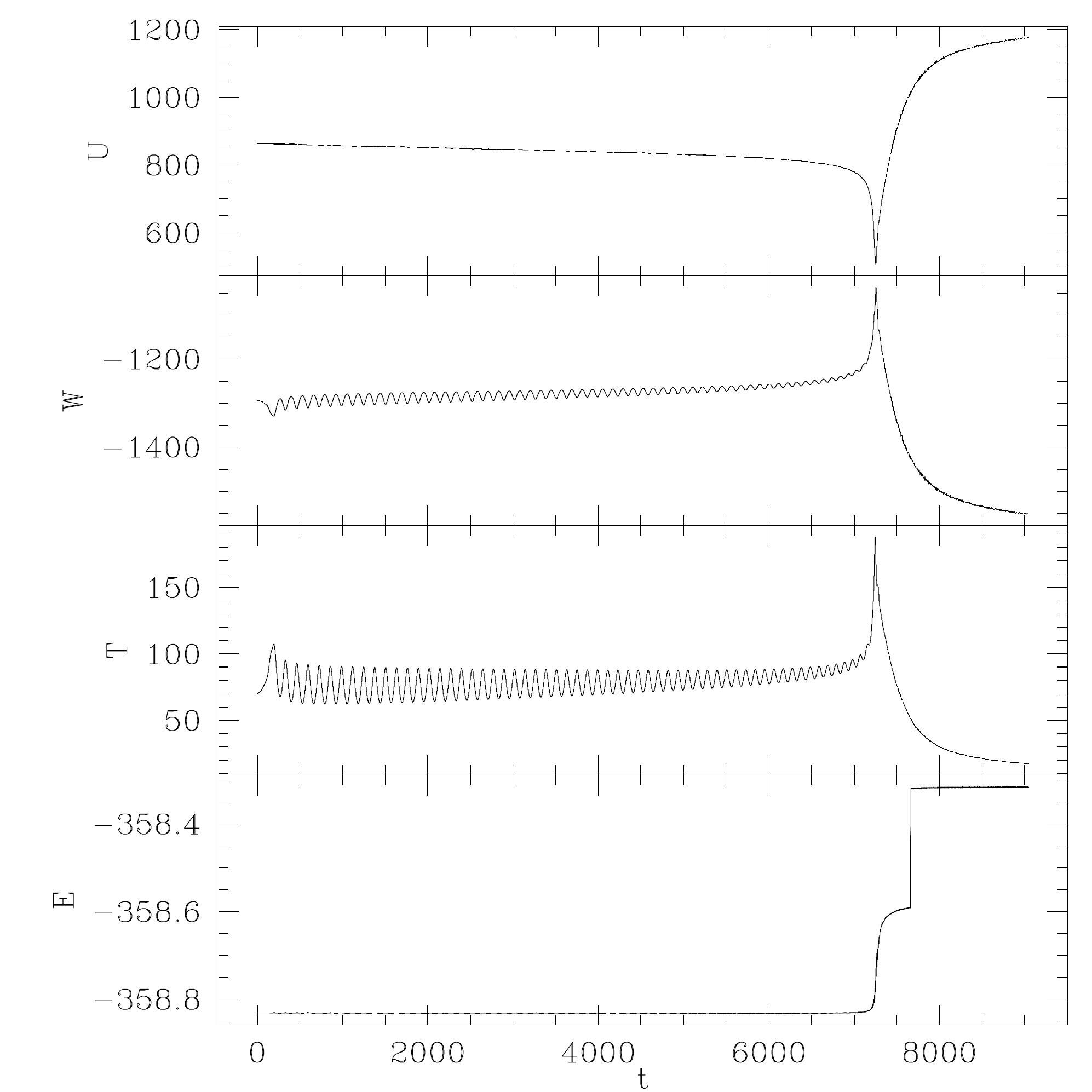}$\qquad$
    \includegraphics[scale=0.4]{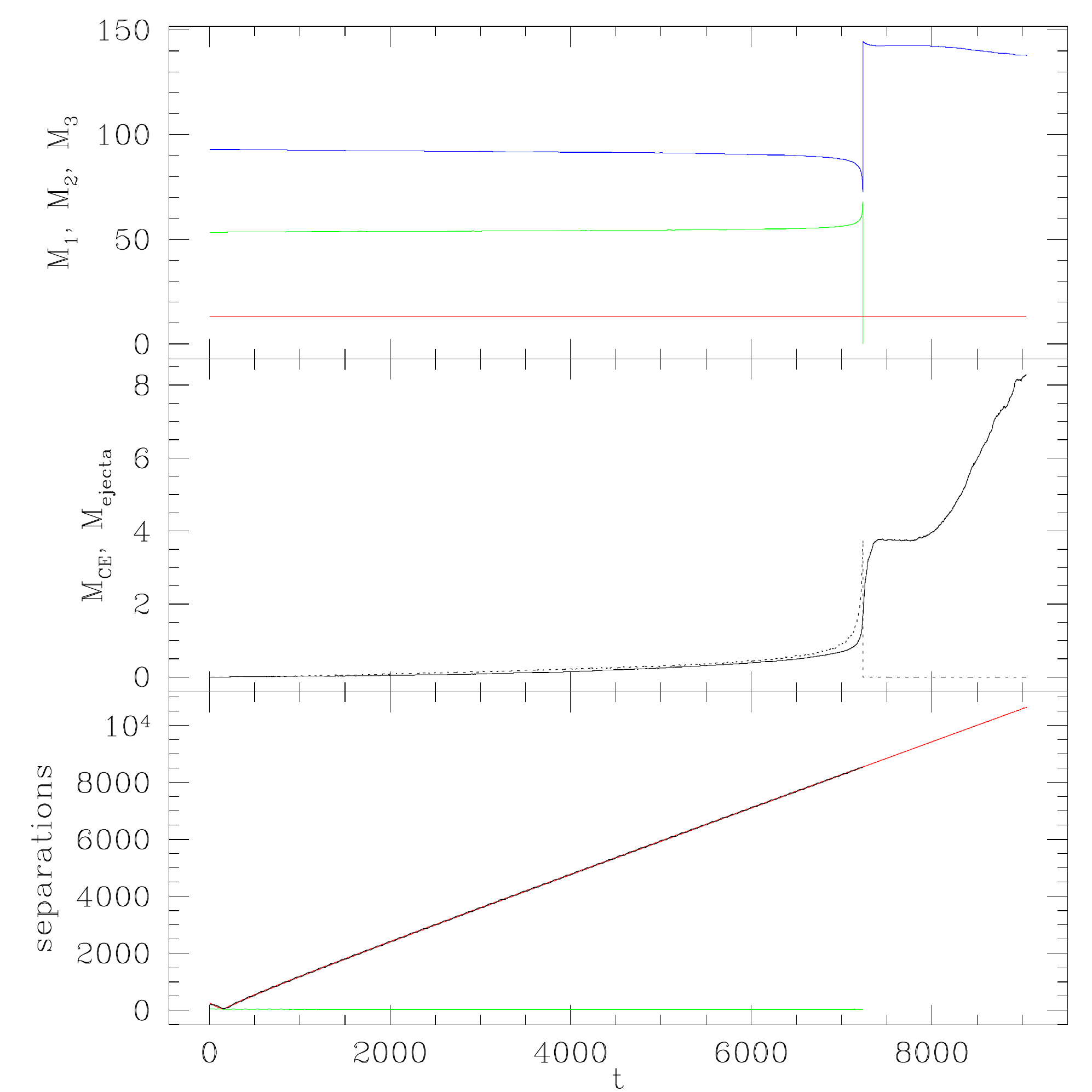}
  \end{center}
  \caption{In the left panel, we show the time evolution of internal
    energy $U$, gravitational potential energy $W$, kinetic energy $T$
    and total energy $E$ for case 260, while on the right panel we
    display the evolution of stellar masses and separations. The top
    frame shows the masses of components 1, 2, and 3, represented by
    blue, green, and red curves, respectively.  The middle frame plots
    the amount of mass ejected (solid curve) and bound in a
    circumbinary envelope (dotted curve).  The bottom curves show the
    separations between components 1 and 2 (green), between 1 and 3
    (red), and between 2 and 3 (black).}
  \label{fig:run260}
\end{figure*}

\begin{figure*}
  \begin{center}
    \includegraphics[scale=0.5]{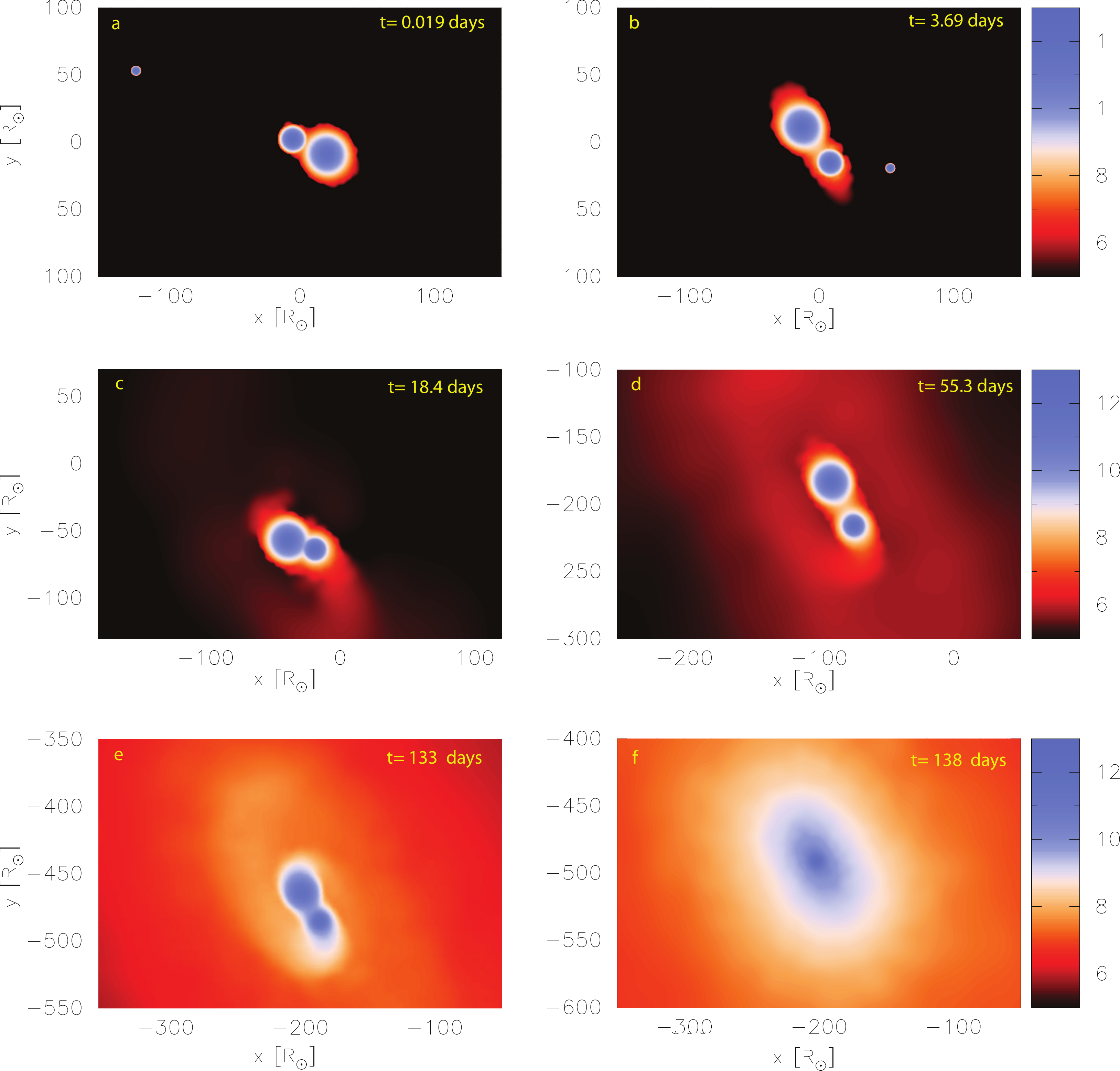}
  \end{center}
  \caption{Column density along lines of sight perpendicular to the
    $xy$ plane at various times for the same hydrodynamic calculation
    of case 260.}
  \label{fig:case260}
\end{figure*}

As another example, we consider case 260 in which a massive binary
(92.9 and $53.3 M_\odot$) is perturbed by a less massive intruder
($13.3 M_\odot$). Here, the intruder is a catalyst which triggers
binary merger. In Figure \ref{fig:run260}, we show time evolution
of energies (left panel) and global quantities (right panel), such as
the masses of individual stars and ejected fluid. In Figure
\ref{fig:case260}, we present time snapshots for this run. Figure
\ref{fig:case260}a shows a snapshot at the beginning of the
simulations, and Figure \ref{fig:case260}b at the moment of closest
approach between the intruder star and the binary. The binary merger
process is shown in Figures \ref{fig:case260}c, \ref{fig:case260}d and
\ref{fig:case260}e. It can be seen that fluid is gradually lost from
the L2 Lagrangian point. Finally, the merged binary is shown in Figure
\ref{fig:case260}f. In contrast to case 232, binary orbital
angular momentum is converted into spin of the product, explaining the
elongated shape of the collisions product in Figure
\ref{fig:case260}f. One may also notice that the collision product is
quickly drifting away from the centre with a velocity of 14 km/s. Most
of this kick velocity comes form the escaping intruder star rather
than from the asymmetric mass ejection.

\begin{figure*}
  \begin{center}
    \includegraphics[scale=0.35]{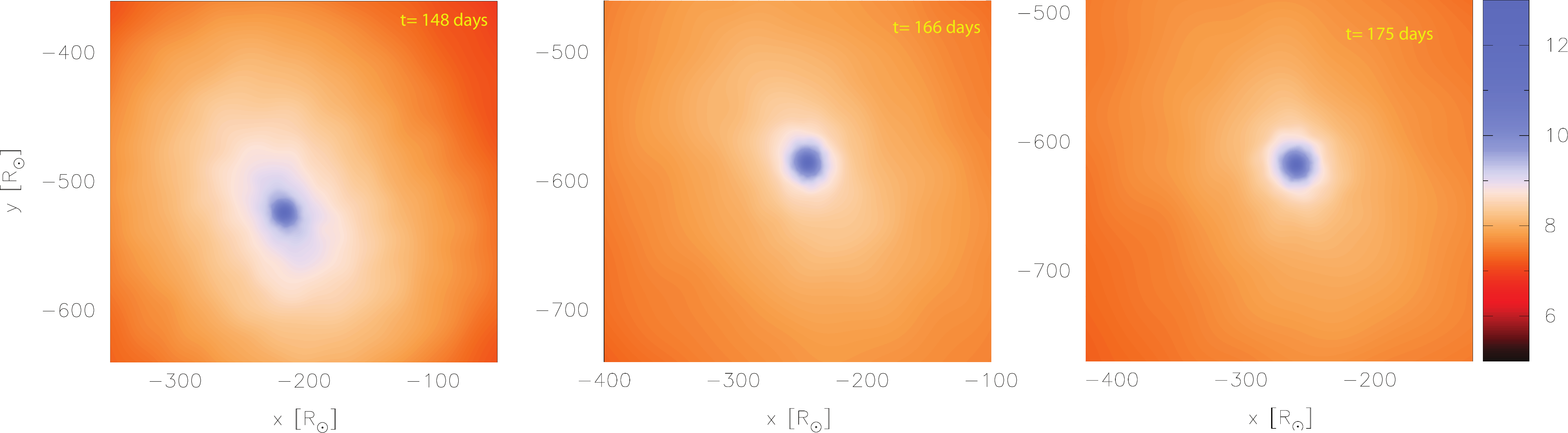}\\
  \end{center}
  \caption{Column density along lines of sight perpendicular to the
    $xy$ plane at various times for the same hydrodynamic calculation
    of case 260. While the inner regions of the
    product become more compact and spherically symmetric, the
    outer regions increase in size and maintain an elongated shape.}
  \label{fig:case260late}
\end{figure*}

The second episode of mass ejection, which occurs after the binary
merger as can be seen in the right panel of Figure \ref{fig:run260},
is an artifact of the artificial viscosity used in SPH. Initially, the
collision product is in the state of both differential rotation and
hydrostatic equilibrium.  Artificial viscosity tends to transfer
angular momentum from the rapidly rotating shells to slower ones
\citep{1999JCoPh.152..687L}, and this forces the product to be a solid
rotator. Since the inner regions of the collision product are spinning
much faster than the outer ones, the angular momentum is transferred
outwards. The net effect is that the inner regions of the collision
product contract, because of loss of the rotational support, but the
outer regions expand because of the continuously increasing supply of
angular momentum, and this case can be seen in
Fig. \ref{fig:case260late}. Eventually, these outer regions become
unbound and escape, and this results in the second episode of the mass
loss.  The mass loss we report in Table \ref{tab:initial_conditions2}
is before this second episode but after the first, corresponding to
the plateau $M_{ejecta}\approx 4 M_\odot$ near $t=7500$ in Figure \ref
{fig:run260}.

From the data of Table \ref{tab:results} it is evident that when only
two stars merge the mass loss remains below a few percent, and often
considerably smaller. It is known that mass loss in a parabolic
collision between two main-sequence stars is small
\citep{2005MNRAS.358.1133F, 2008MNRAS.383L...5G}. The mass loss
percentage is typically larger in cases where all three stars
ultimately merge, exceeding 10\% in the hydrodynamic simulations of
cases 212, 214, 223, 233, 236, 256, and 298. The hydrodynamic
evolution in these more extreme cases is qualitatively similar: the
first merger event is between the most massive star and one of the
other two, and typically occurs after a short resonant
interaction. The resulting merger product is enhanced in size by shock
heating and rotation, leaving its outermost layers loosely bound. The
third star, often after flung out to a large distance, can experience
several periastron passages through the envelope of the first merger
product before ultimately donating its fluid to the mix. In the
process, substantial amounts of gas are ejected from the diffuse
envelope at every periastron passage.

\begin{figure}
 \begin{center}
   \includegraphics[width=\columnwidth]{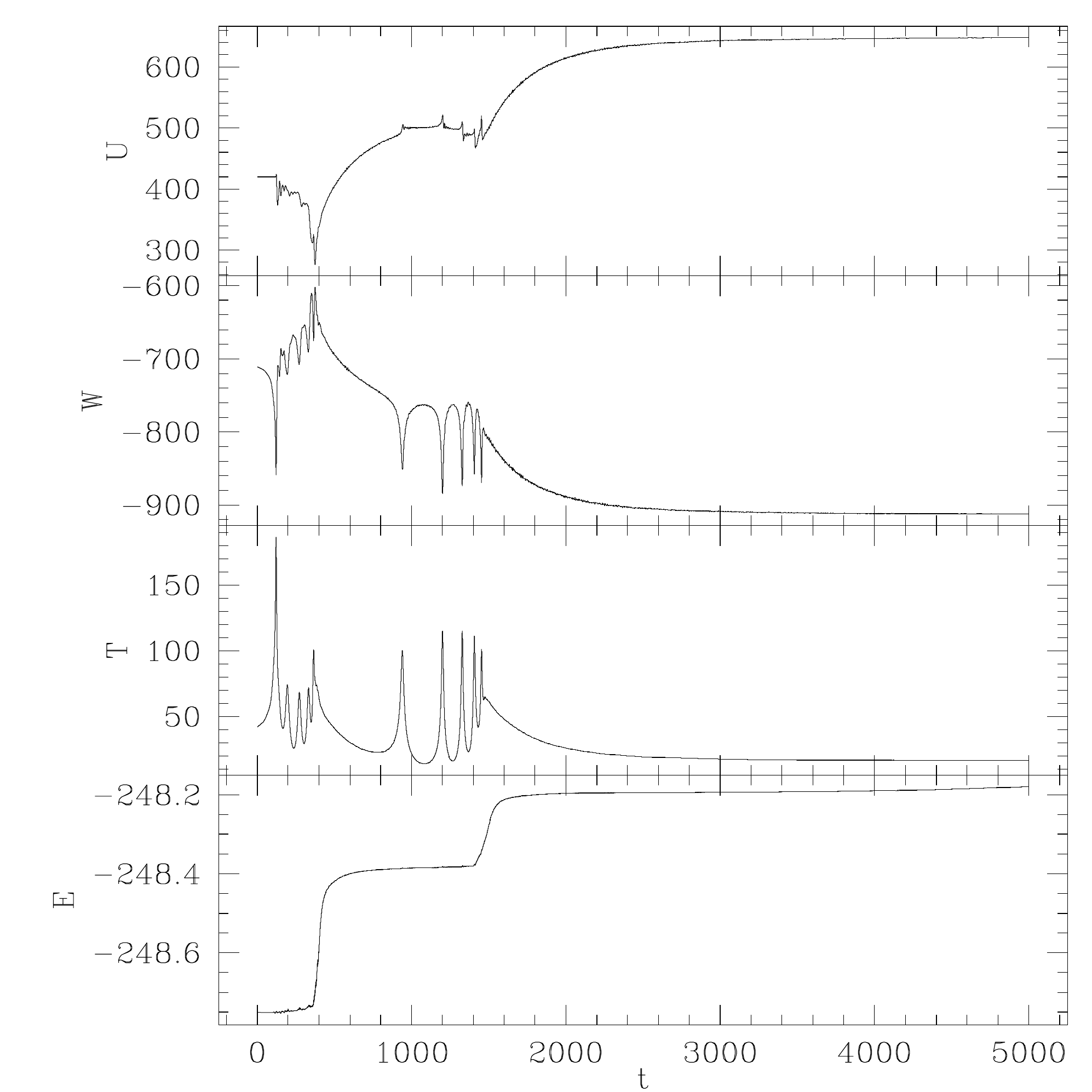}
 \end{center}
 \caption{Internal energy $U$, gravitational energy $W$, kinetic
   energy $T$, and total energy $E$ versus time for case 298.  Peaks
   in $T$ and associated dips in $W$ correspond to close passes or
   mergers between the stars.  Note that the total energy is conserved
   to about 0.2\% over the interval shown.}
 \label{fig:vnvt_0_5000_case298}
\end{figure}

\begin{figure}
  \begin{center}
    \includegraphics[width=\columnwidth]{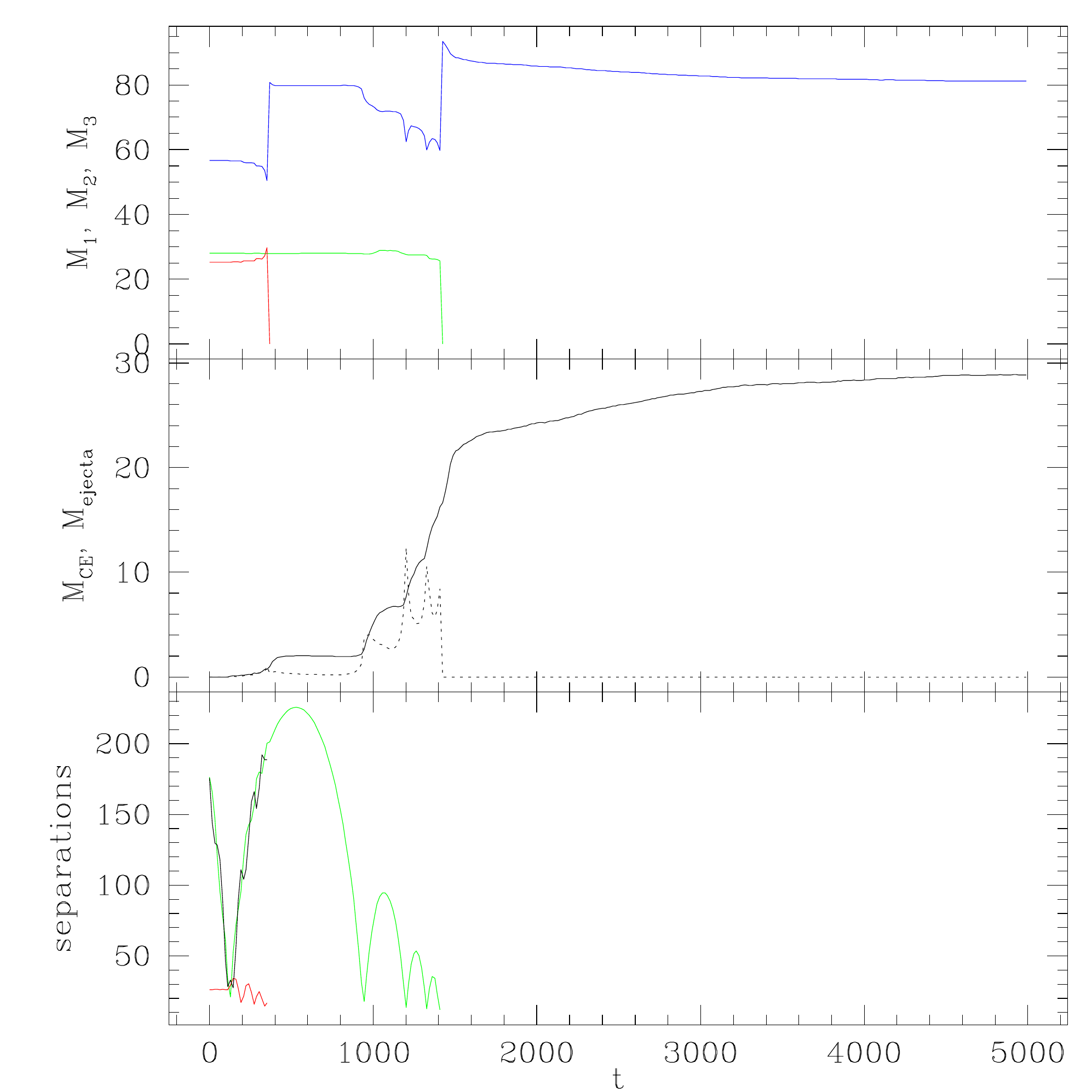}
  \end{center}
  \caption{Masses and separations versus time for the calculation
    displayed in Fig.\ \ref{fig:vnvt_0_5000_case298}, case
    298. The top frame shows the masses of components 1, 2, and 3,
    represented by blue, green, and red curves, respectively.  The
    middle frame plots the amount of mass ejected (solid curve) and
    bound in a circumbinary envelope (dotted curve).  The bottom
    curves show the separations between components 1 and 2 (green),
    between 1 and 3 (red), and between 2 and 3 (black).}
  \label{fig:lightmeatst_5000_case298}
\end{figure}

An example of this type of interaction is summarised in Figures
\ref{fig:vnvt_0_5000_case298} and \ref{fig:lightmeatst_5000_case298}
for case 298, which involves a $56.7 M_\odot$ + $25.3 M_\odot$ binary
is intruded upon by a 28.1 $M_\odot$ star.
The features of these curves can be associated with events during the
encounter.  In this situation the intruding star initiates a short
lived resonance that ends with the induced merger of the binary
components near $t=400$.  As can be seen in the middle frame of Figure
\ref{fig:lightmeatst_5000_case298}, approximately $2 M_\odot$ of fluid
is ejected in the process.  The intruder retreats on an eccentric
orbit, reaching an apastron separation of more than $200 R_\odot$ and
returning for its next pericentre passage shortly before $t=1000$.  As
the intruder moves through the outer layers of the first merger
product, its orbit decays and more mass is ejected.  By $t=1400$, the
three-body merger product is formed and more than $20 M_\odot$ has
been ejected in total.

\begin{figure}
  \begin{center}
    \includegraphics[width=\columnwidth]{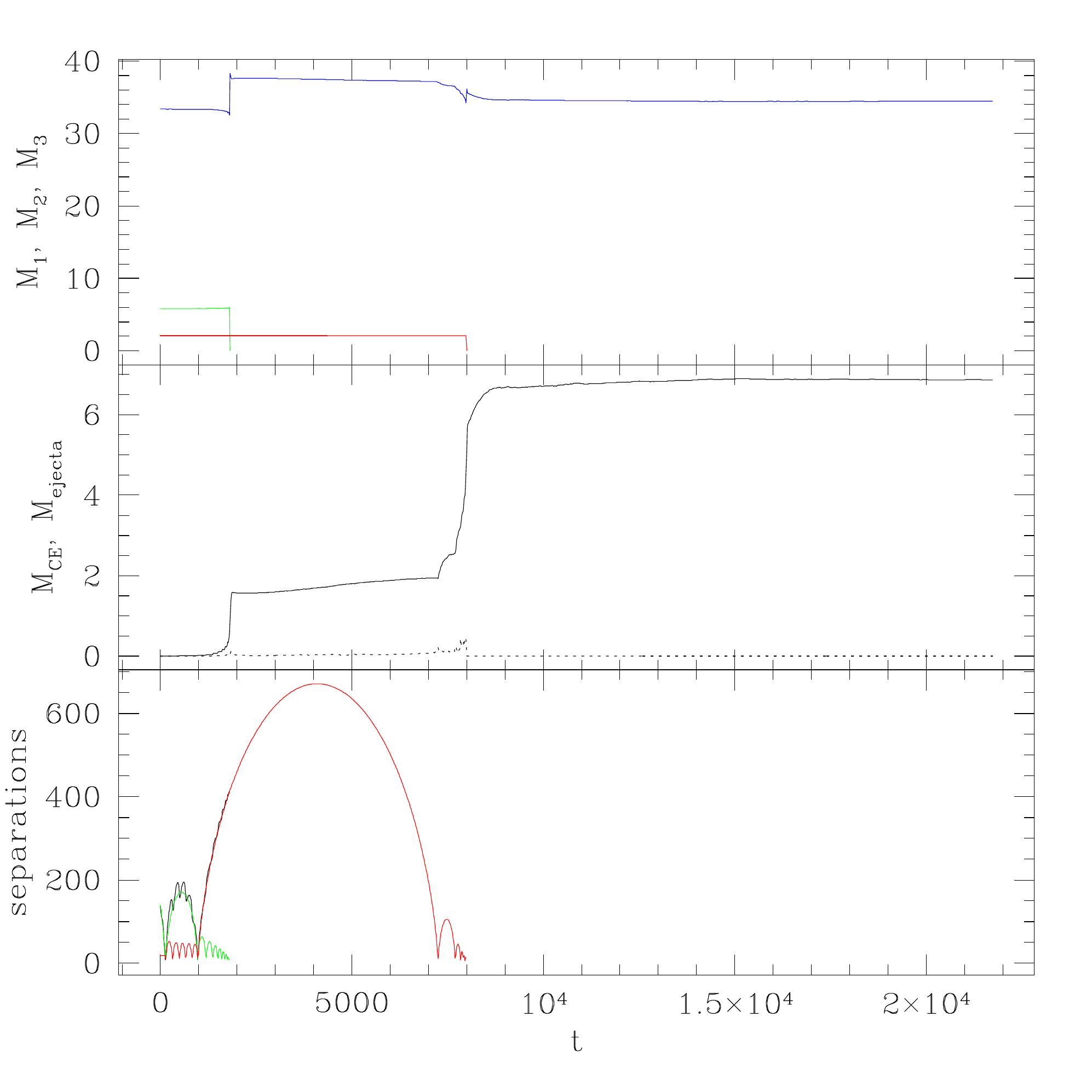}
  \end{center}
  \caption{Like Fig.\ \ref{fig:lightmeatst_5000_case298}, but for case
    256.}
  \label{fig:lightmeats_case256}
\end{figure}

Another double merger resulting in significant mass loss is summarised
in Figure \ref{fig:lightmeats_case256}, which shows the masses and
separations relevant to the hydrodynamic calculation of case 256 (a
$33.4 + 2.11 M_\odot$ binary and a $5.84 M_\odot$ intruder).  Here the
initial merger occurs between the two most massive stars near
$t=1800$, with about $1.5 M_\odot$ of fluid is ejected in the process.
The third star is left on a highly eccentric orbit, reaching an
apastron separation of more than $600 R_\odot$ at $t=4100$, and
returning for its next pericentre passage at $t=7200$.  With each
passage through the envelope of the first merger product, the orbit of
the third star decays and more mass is ejected until ultimately, at
$t=8000$, the three-body merger product is formed.

\begin{figure}
  \begin{center}
    \includegraphics[width=\columnwidth]{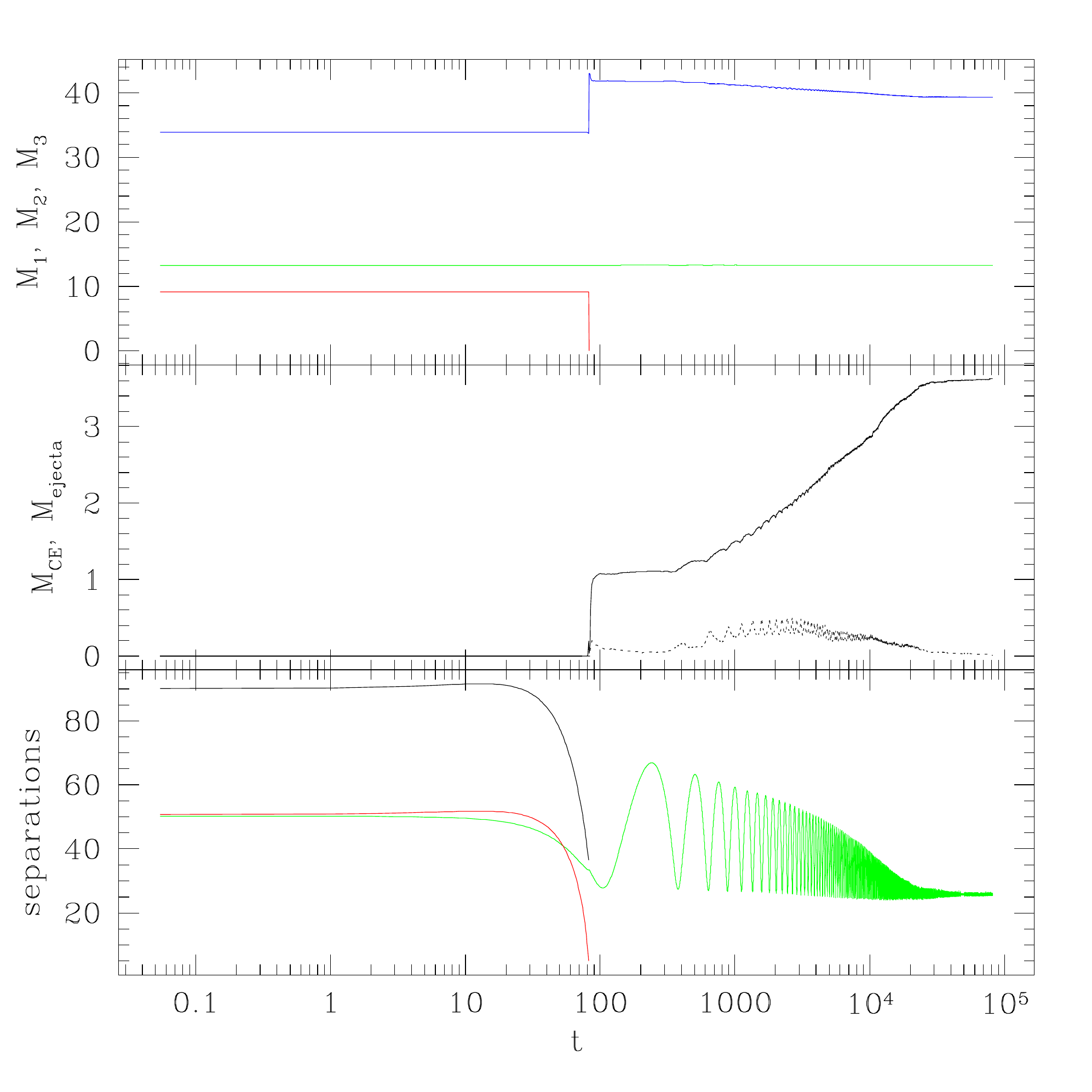}
  \end{center}
  \caption{Like Fig.\ \ref{fig:lightmeatst_5000_case298} and
    Fig.\ \ref{fig:lightmeats_case256}, but for case 261, and with
    time plotted on a logarithmic scale so that the long term
    evolution and circularization of the resulting binary can be more
    easily observed.}
  \label{fig:loglightmeats2000}
\end{figure}

In cases 250 and 261 the impact of the relatively low-mass intruder
into the primary causes the outer layers of the latter to expand and
overflow its Roche lobe, resulting ultimately in a stable
binary. Figure \ref{fig:loglightmeats2000} shows masses and
separations of stars for case 261, which begins with the three stars
in a resonant interaction. At $t=83$, the lowest mass star is absorbed
into the largest star. The collision immediately ejects $1 M_\odot$ of
material and leaves the two remaining stars in an eccentric binary
($e\approx 0.4$).  A fraction of a solar mass is also placed into a
circum-binary envelope: this fluid is not gravitationally bound to
either star individually but rather to the remaining binary as a
whole.  As the binary grinds through the envelope, the orbit gradually
circularises, as can been seen by examining the separation curve in
the bottom frame of Figure \ref{fig:loglightmeats2000}.  By $t\approx
6\times 10^4$, the envelope has been effectively removed and the
binary has essentially reached a steady state with an orbital period
of 113 time units (50 hours) and a separation of $26 R_\odot$.  The
calculation for this case lasted more than $3.3\times 10^6$ iterations
and covered a timespan of over 80000 time units (over 4 years simulation
time).  During this calculation, total energy and angular momentum
were conserved to better than 0.1\%.

% \subsection{Resolution study}

\subsection{The effect of numerical resolution}

\begin{figure*}
  \begin{center}
    \includegraphics[scale=0.4]{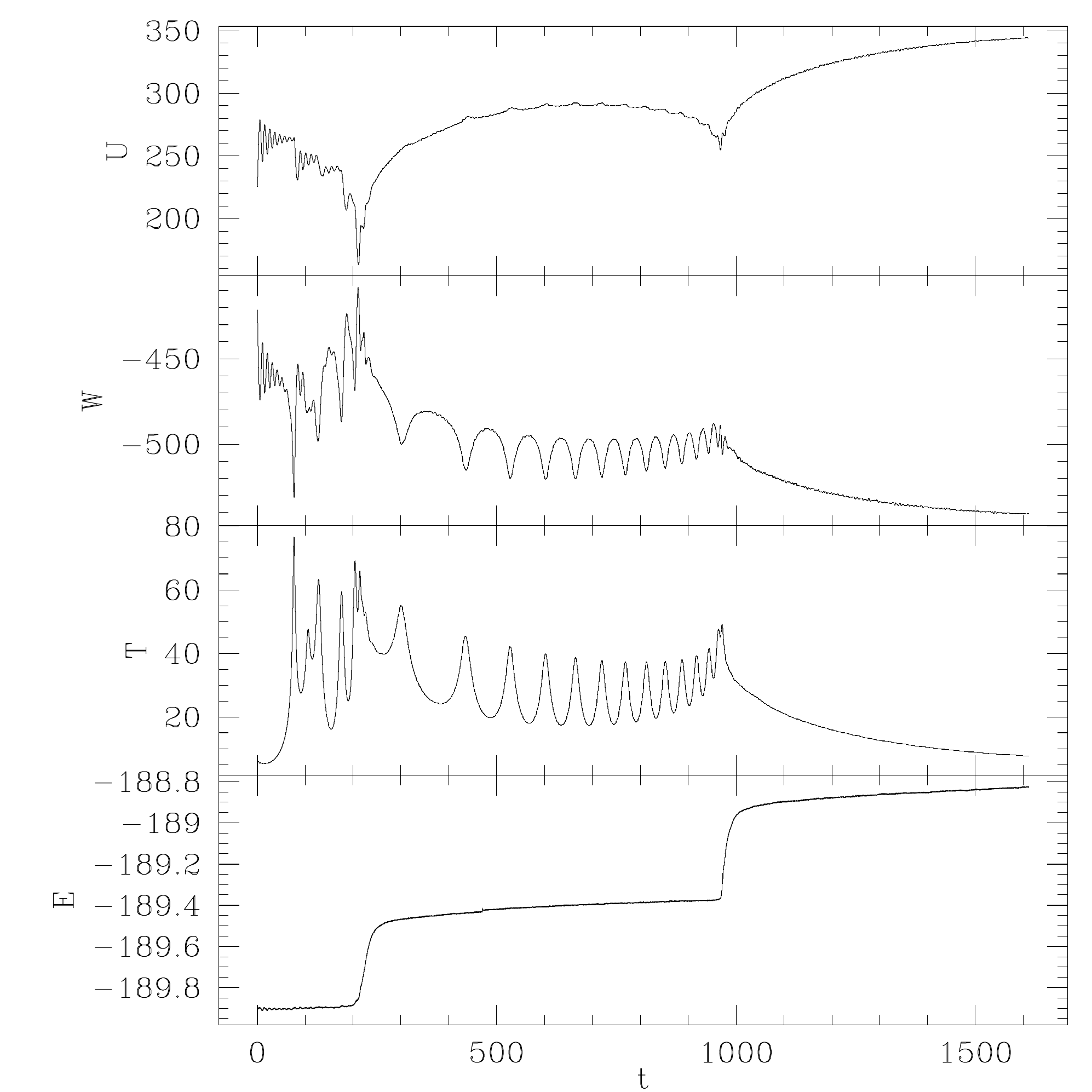}$\,$
    \includegraphics[scale=0.4]{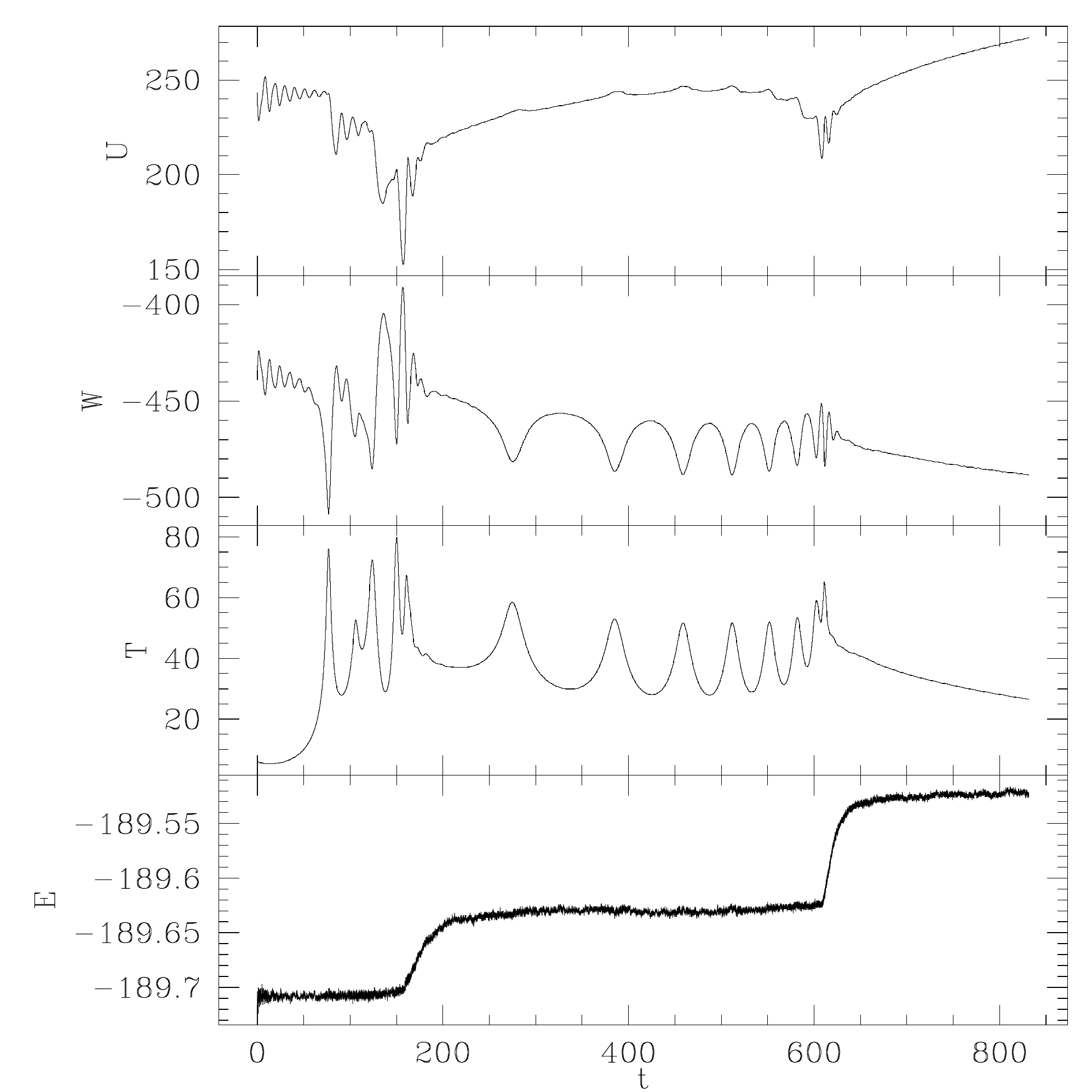}$\,$
  \end{center}
  %carrv59:/runs3/triples/three_singles/new_integrator
  \caption{Internal energy $U$, gravitational potential energy $W$,
    kinetic energy $T$, and total energy $E$ versus time $t$ for two simulations of case
    204 that differ in resolution: $N=11946$ (left panel) and 60024
    (right panel).}
  \label{fig:run204_res1}
\end{figure*}

\begin{figure*}
  \begin{center}
    \includegraphics[scale=0.4]{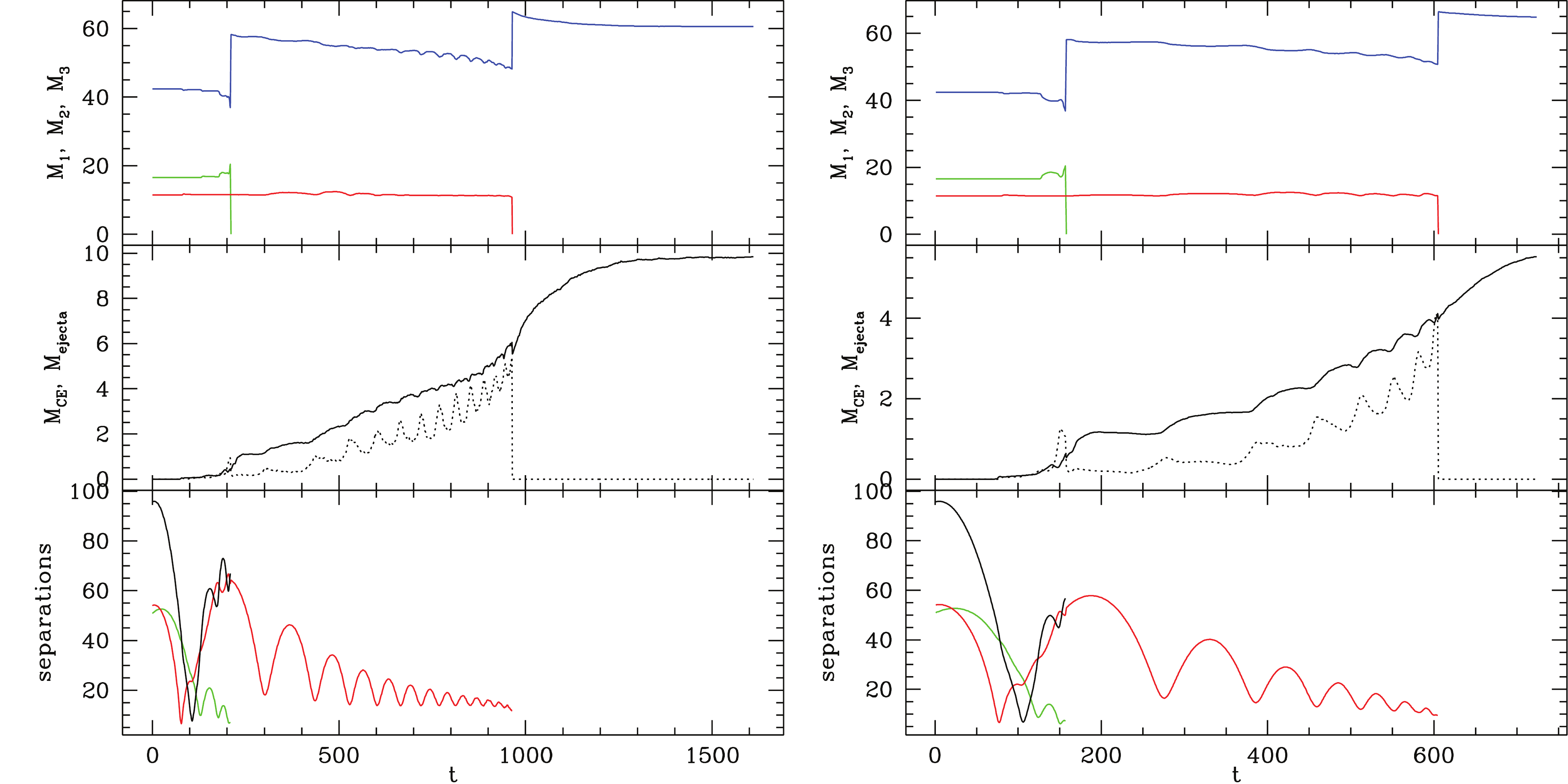}$\,$
  \end{center}
  %carrv59:/runs3/triples/three_singles/new_integrator
  \caption{Masses and separations versus time for two simulations of
    case 204: $N=11946$ (left panel) and 60024 (right panel). The
    top frame shows the masses of components 1, 2, and 3, represented
    by blue, green, and red curves, respectively.  The middle frame
    plots the amount of mass ejected (solid curve) and bound in a
    circumbinary envelope (dotted curve).  The bottom curves show the
    separations between components 1 and 2 (green), between 1 and 3
    (red), and between 2 and 3 (black).
}
  \label{fig:run204_res2}
\end{figure*}

Because of the longevity of three-body interactions, most of our
simulations are limited to $N \approx (1$--$2)\times 10^4$ particles.
Even with this relatively low number of particles, a single simulation
may take a few weeks to complete, as it typically needs to span at
least several thousand time units.  To test whether our results are
affected by numerical artifacts, we recalculated a few of the
simulations in high resolution. In most cases, the results are only
weakly dependent on the resolution. In particular, a case of interest
is case 204, which begins with three single stars in the middle of the
resonance interaction. In Figure \ref{fig:run204_res1} we present the
time evolution of the energy for two resolutions. One may see from the
kinetic energy plot that the first close interaction occurs at $t
\simeq 75$. The further behaviour of the three stars bear
characteristics of typical resonant interactions, with kinetic and
gravitational potential energy exhibiting aperiodic oscillations of
different magnitudes until $t \simeq 200$. At this time two of the
three stars merge (Table \ref{tab:results}) and binary continues to
decay. In the high resolution case (right panel in Figure
\ref{fig:run204_res1}), the merger occurs somewhat earlier than in the
low resolution case. Because this kind of interaction is chaotic, it is
well known that the details at the level of trajectories are
resolution sensitive \citep{1993A&A...272..430D,
  2005MNRAS.358.1133F}. However, the final outcome is consistent
between the two resolutions: all three stars eventually
merge. Moreover the mass and energy of the ejecta, as well as the kick
velocity of the merger product, change by at most a factor of two. In
Figure \ref{fig:run204_res2} we show the time evolution of masses of
three stars, the mass of ejected fluid and the separation between
stars.

\begin{figure*}
  \begin{center}
    \includegraphics[scale=0.4]{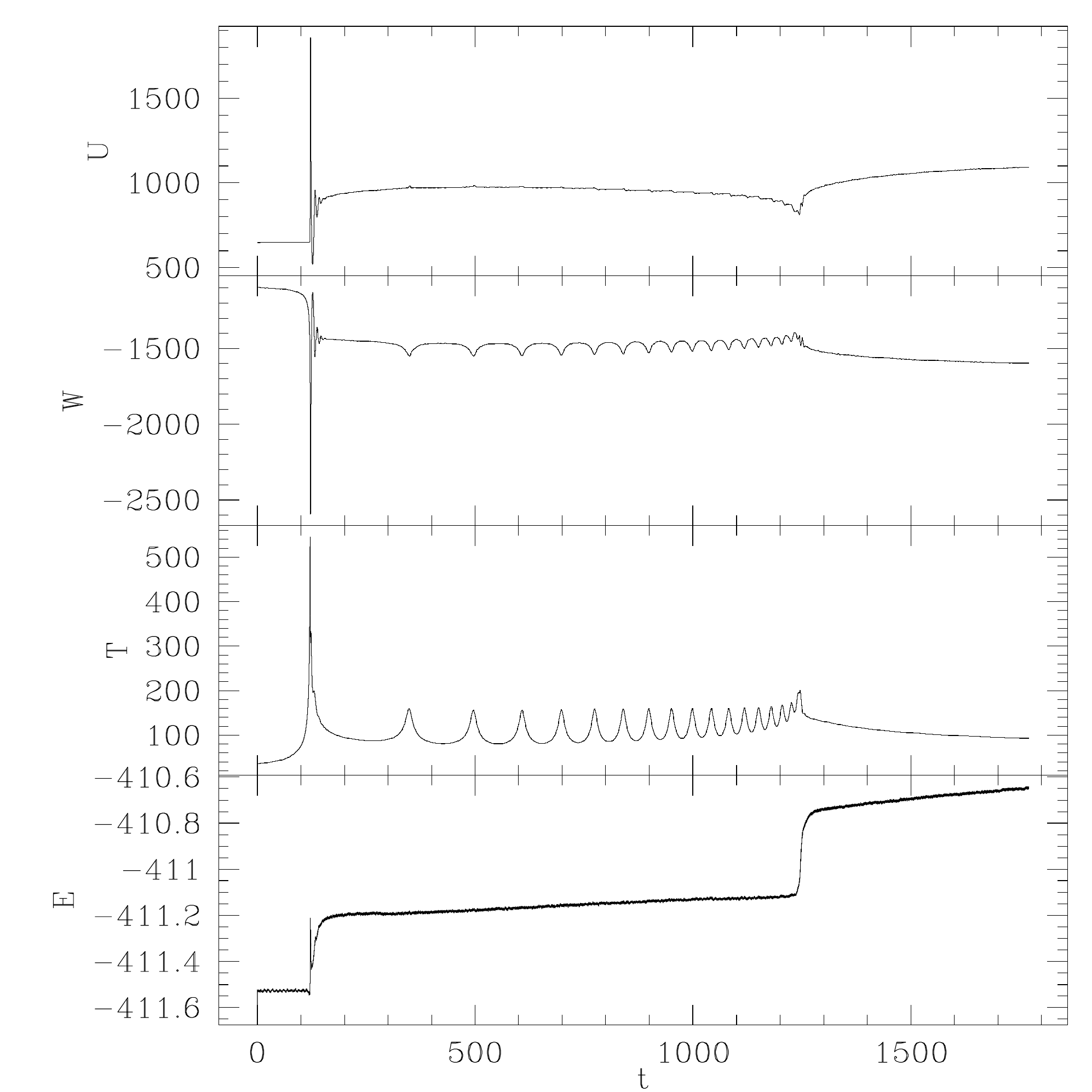}$\,$
    \includegraphics[scale=0.4]{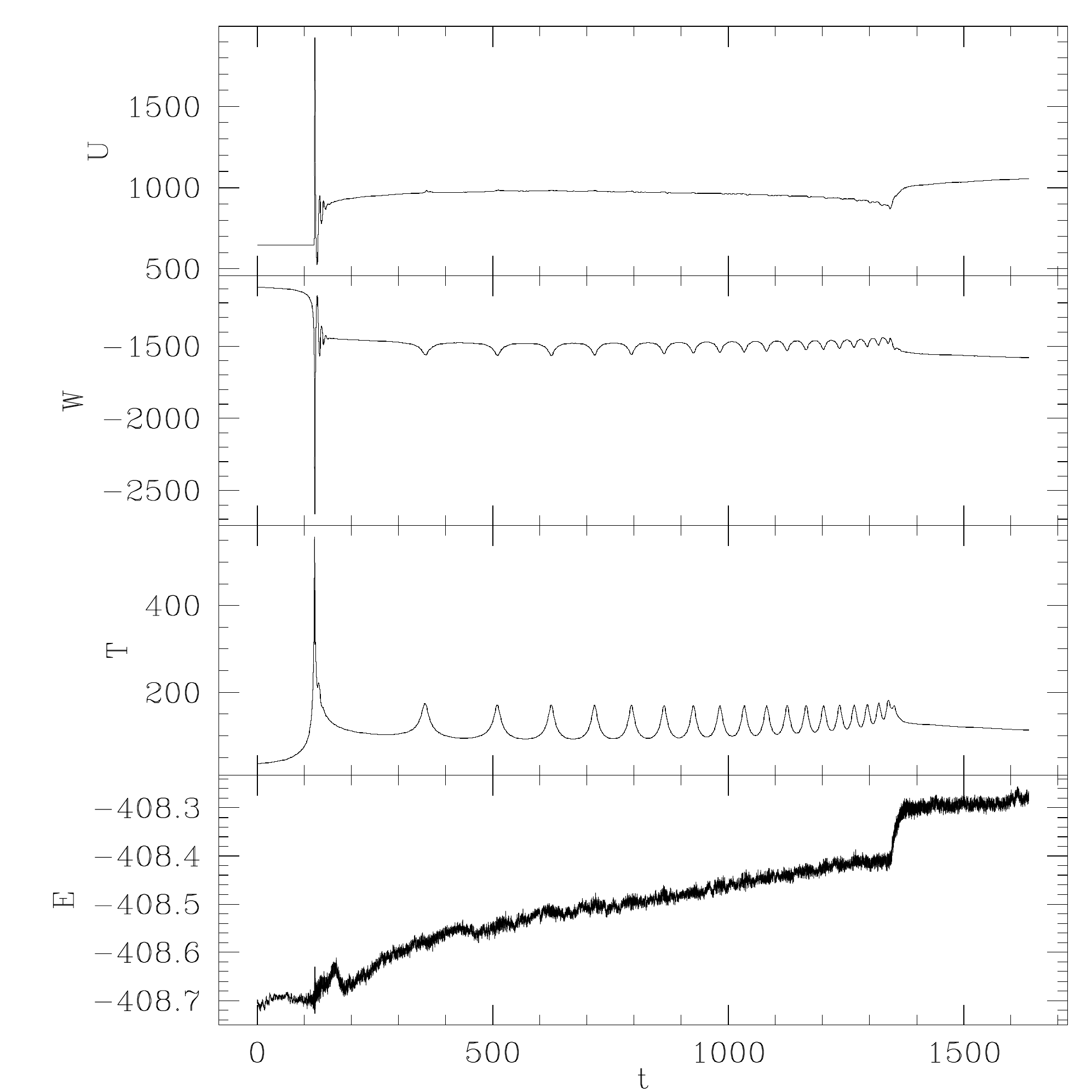}$\,$
  \end{center}
  %carrv59:/runs3/triples/three_singles/new_integrator
  \caption{Internal energy $U$, gravitational potential energy $W$,
    kinetic energy $T$, and total energy $E$ versus time $t$ for two simulations of case
    299 that differ in resolution: $N=10194$ (left panel) and 102540
    (right panel).}
  \label{fig:run299_res1}
\end{figure*}

\begin{figure*}
  \begin{center}
    \includegraphics[scale=0.4]{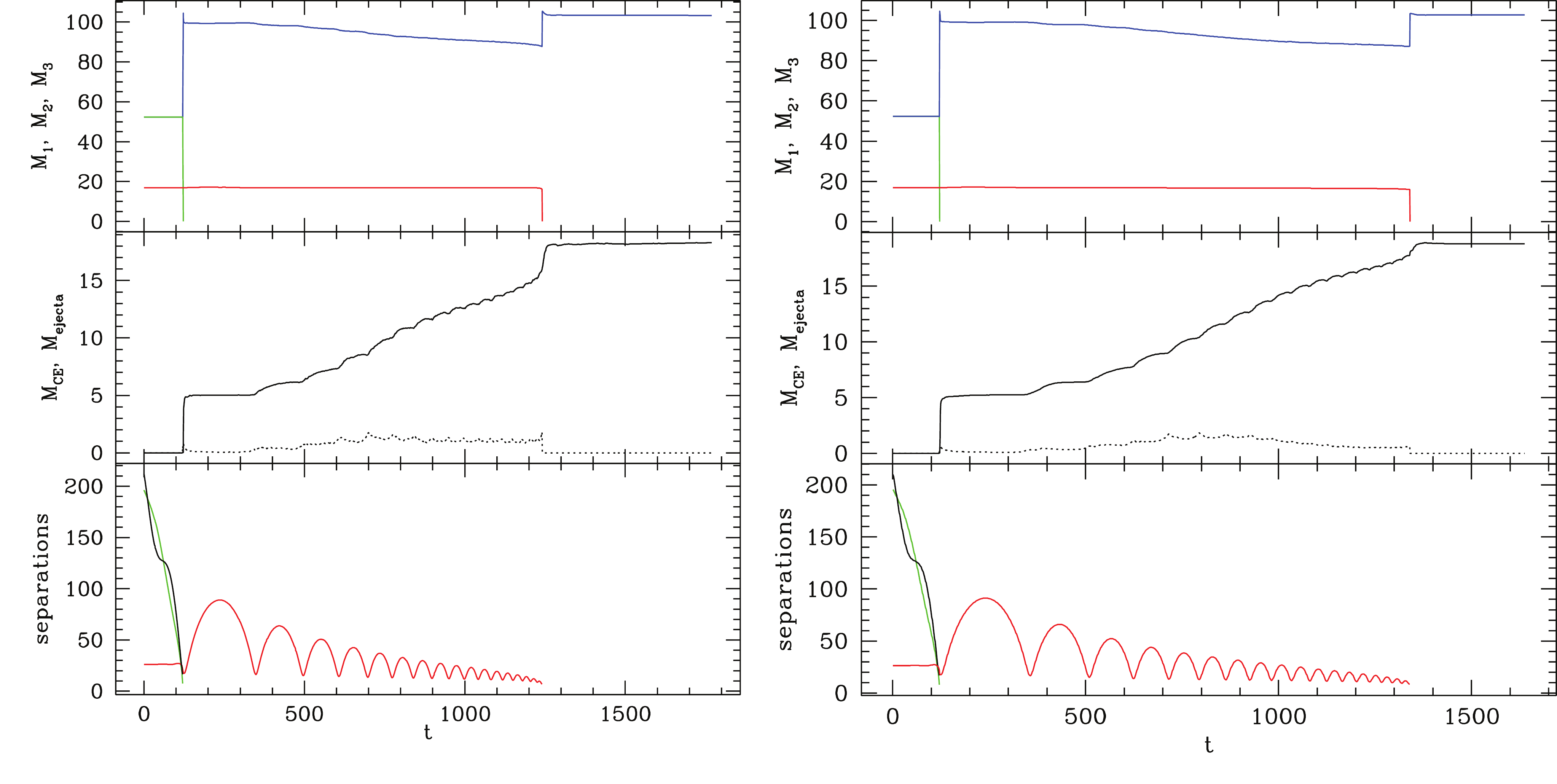}$\,$
  \end{center}
  %carrv59:/runs3/triples/three_singles/new_integrator
  \caption{Masses and separations versus time for two simulations of
    case 299: $N=10194$ (left panel) and 102540 (right panel).  Line
    types are as in Fig.\ \ref{fig:run204_res2}.}
  \label{fig:run299_res2}
\end{figure*}

Another interesting case is 299, in which a massive binary (52.3 +16.9
$M_\odot$) is intruded upon by a massive star ($52.3 M_\odot$). In
Figures \ref{fig:run299_res1} and \ref{fig:run299_res2} we show the
time evolution of energies and global quantities, such as the masses
of stars, the ejecta mass, and the stellar separations.
Even though there are some differences, the general agreement
between these two simulations is excellent. The merger between two of the three stars (the intruder
and the primary of the binary) occurs at $t \simeq 110$, and further
binary decay lasts for more than 1200 units. Mass loss and energy of
ejected fluid are consistent between these two runs of different
resolution.

\begin{figure}
  \begin{center}
    \includegraphics[width=\columnwidth]{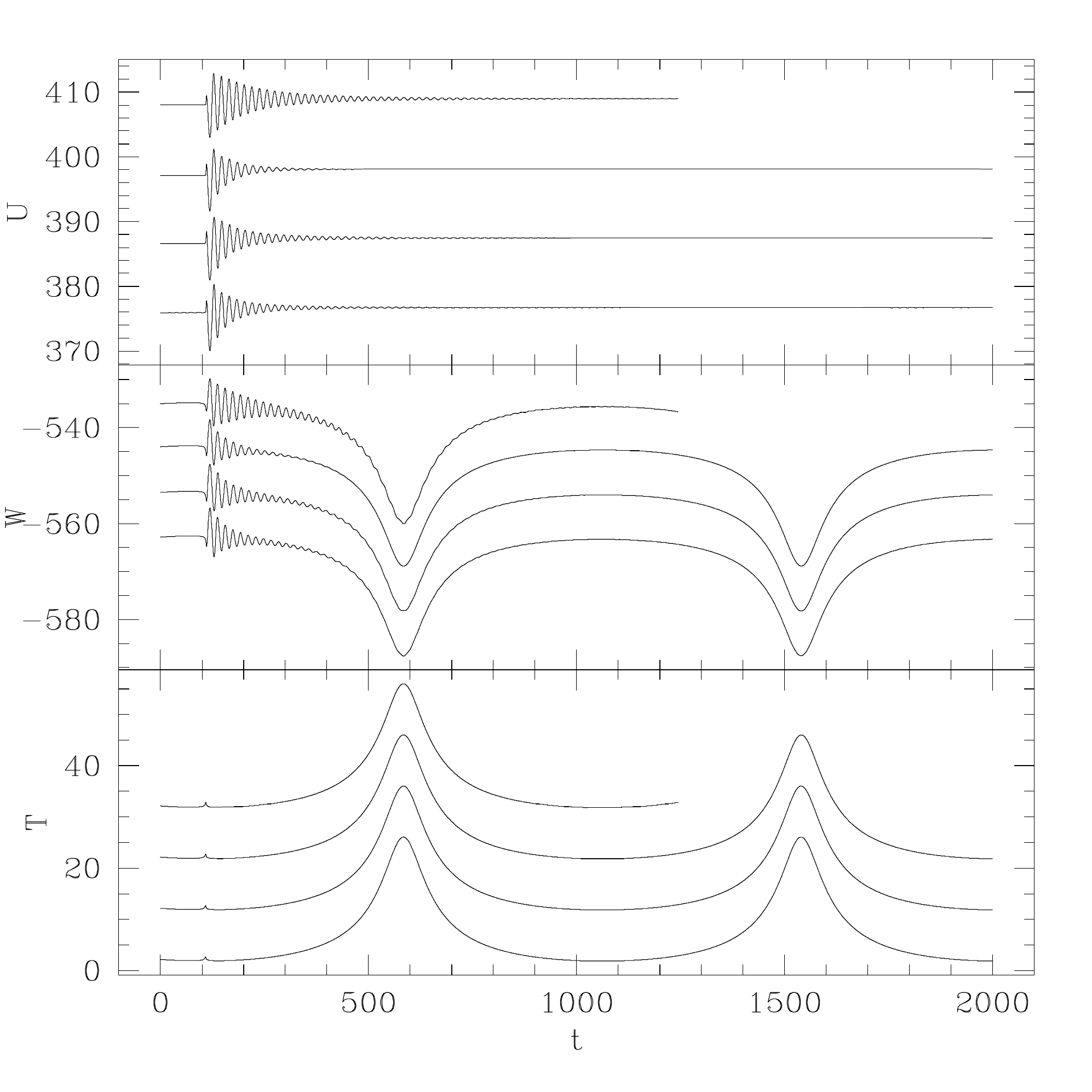}
  \end{center}
  %carrv59:/runs3/triples/three_singles/new_integrator
  \caption{Internal energy $U$, gravitational potential energy $W$ and
    kinetic energy $T$ versus time $t$ for four simulations of case
    202 that differ in resolution: $N=6138$ (bottom curve), 11466
    (second from bottom), 22380 (third from bottom), 91956 (top).  The
  energy scale on the left axis corresponds to the low resolution
  $N=6138$ case: the other energy curves have been offset by 10, 20,
  and 30 energy units to facilitate the comparison.}
  \label{fig:res_2000_10}
\end{figure}

In Figure \ref{fig:res_2000_10}, we examine the effects of resolution
for four separate simulations of case 202 with the number of particles
varying by a factor of 15 from the lowest resolution treatment to the
highest resolution. The agreement is excellent, with even the lowest
resolution simulation capturing all important aspects of the orbital
dynamics.  The small bump in the kinetic energy $T$ shortly after the
time $t=100$ corresponds to the absorption of the $0.120M_\odot$ star
into the $57.9M_\odot$ star, which excites oscillations in the merger
product that are visible in the internal energy $U$ and gravitational
potential energy $W$ curves.  The merger product is left orbiting the
$29.9M_\odot$ star in a stable binary with eccentricity $e=0.583$ and
semimajor axis $a=127R_\odot$: the peaks in $T$ and simultaneous dips
in $W$ correspond to the periastron passages.

%%%%%%%%%%%%%%%%%%%%%%%%%%%%%%%%%%%%%%%%%%%

\begin{figure*}
  \begin{center}
    \includegraphics[width=\columnwidth]{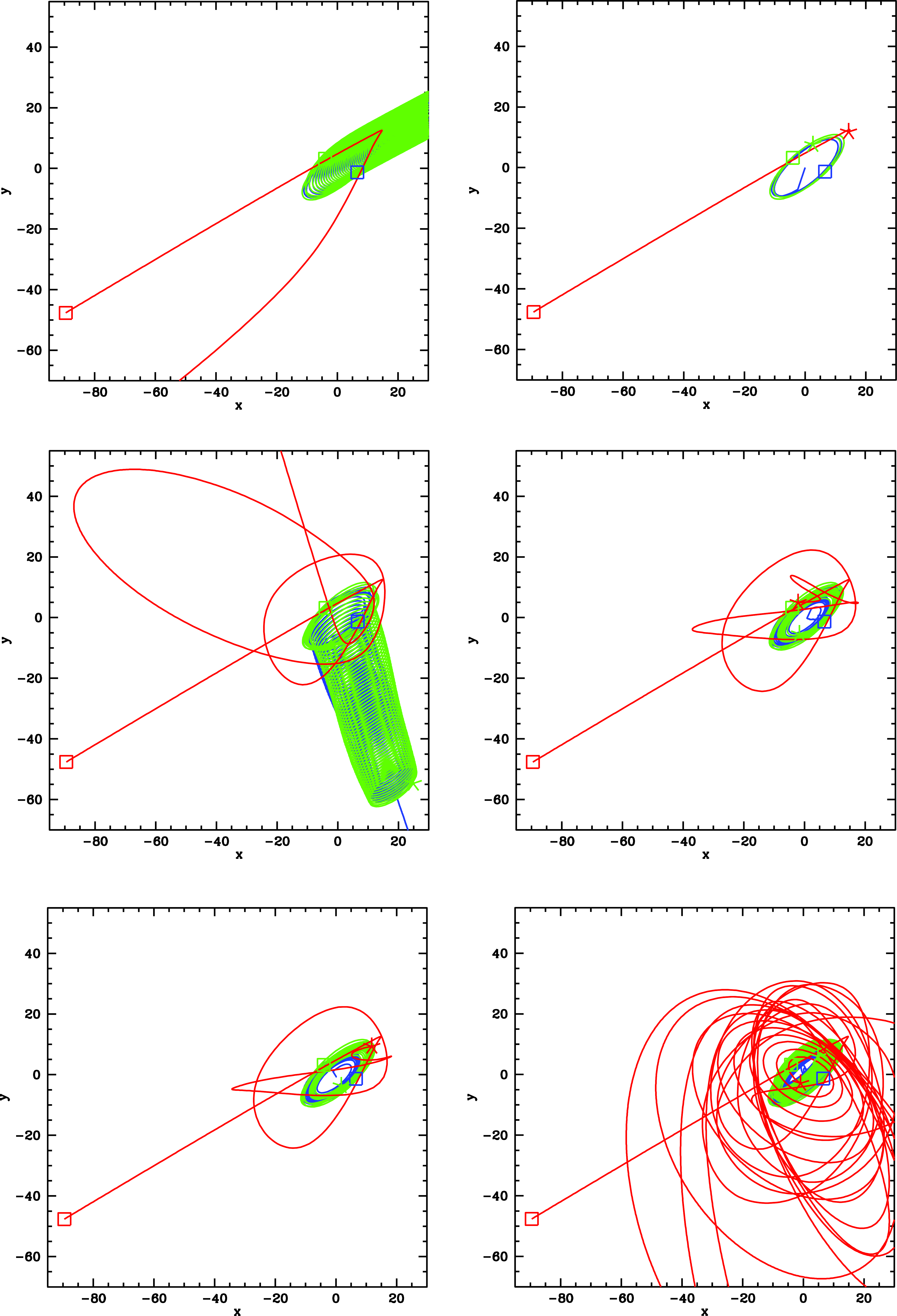}
  \end{center}
  \caption{Trajectories projected onto the $xy$ plane for case 246 as
    calculated in a pure point mass approximation (top left), in a
    sticky sphere approximation (top right), by our hydrodynamics code
    with $N=10554$ (middle left), with $N=21204$ (middle right), with
    $N=42294$ (bottom left), and with $N=84642$ (bottom right).  We
    adopt the convention that the trajectory of the most massive star
    is represented by the blue curve, the intermediate mass star by
    the green curve, and the lowest mass star by the red curve.  The
    initial conditions are marked by squares, while the final position
    of an object before it merges with another one is marked by a
    5-point asterisk.}
\label{fig:trajectories}
\end{figure*}

In Figure \ref{fig:trajectories} we show the projected trajectories of
the three stars in case 246 of masses 42.2, 38.3 and
$1.37M_\odot$, as calculated with a point-mass integrator (top left
frame), by  using sticky spheres (top right frame) and with the
hydrodynamics code (bottom four frames) with different resolution.  In
all cases, the $1.37M_\odot$ intruder approaches the circular binary
on a hyperbolic trajectory with eccentricity $e=1.09$.  In the point
mass approximation, the intruder reaches a minimum separation of
$4.90R_\odot$ from the secondary and then slingshots back outward on a
trajectory with eccentricy $e=1.05$.  The interaction increases the
semimajor axis of the binary slightly to $28.5R_\odot$, while also
perturbing its eccentricity to $e=0.0424$. In the sticky sphere
approximation, a merger between the intruder and the secondary of the
binary occurs during the initial pericenter passage, followed shortly
thereafter by a second merger with the primary.

The case plays out qualitatively differently when the
hydrodynamics is followed.  The intruder again passes to a minimum
separation of about $5R_\odot$ from the core of the secondary, well
within its $11R_\odot$ stellar radius, and then begins to retreat.
The impact, however, transfers energy into oscillations of the
secondary and the intruder is not moving fast enough to escape further
than about $40R_\odot$ from the secondary.  The hydrodynamic
calculations indicate that the intruder makes a second pericenter passage
through the secondary, but these calculations deviate depending
on the resolution: the resulting trajectories do not converge as the
number of particles is increased up to $N=84642$ due to the chaotic
nature of the orbits.

\begin{figure}
  \begin{center}
    \includegraphics[width=\columnwidth]{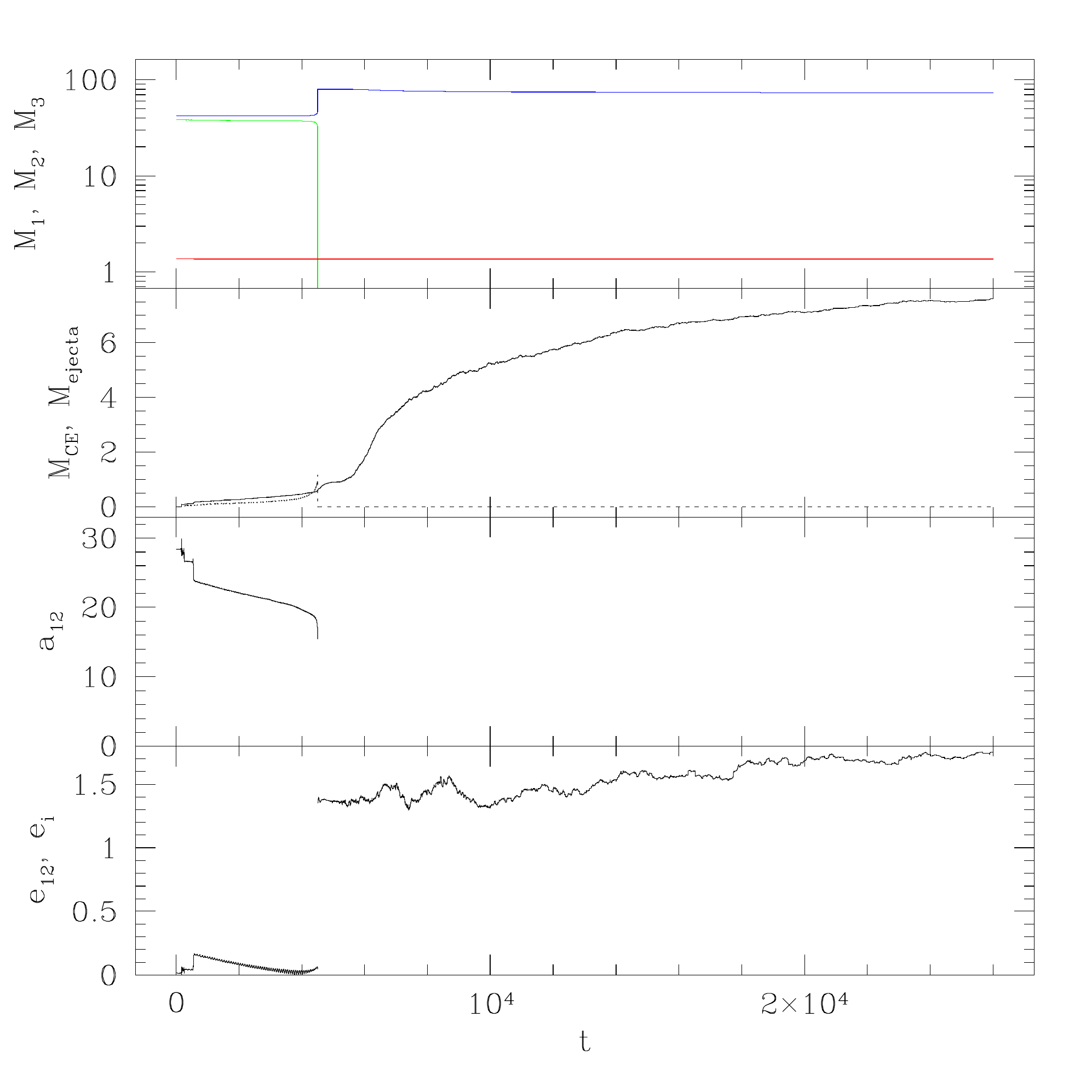}
  \end{center}
  \caption{Evolution versus time of, from the top of the figure to the
    bottom, the stellar masses, mass in common envelope (dotted curve)
    as well as in ejecta (solid curve), semimajor axis $a_{12}$ of the
    binary, and eccentricity $e_{12}$ of the binary ($t<4500$) as well
    as eccentricity $e_i$ of the third star as it departs from the
    merger product ($t>4500$) for the $N=10554$ SPH calculation of
    case 246.}
\label{fig:meavt}
\end{figure}

In the case of our relatively low-resolution $N=10554$ calculation of
case 246, the intruder is shot out to a distance of over
$100R_\odot$.  Finally, the intruder makes one final pass through the
secondary, and is ejected out of the system on a trajectory with
$e=1.3$.  The removal of orbital energy from the binary initiates a
mass transfer instability.  The primary canibalizes the secondary and,
as the binary merges, $0.06M_\odot$ of material is ejected.
At this time, the blue and green curves in Figure
\ref{fig:trajectories} merge into a single blue curve (see the lower
right hand corner of the middle left frame).  Masses and orbital parameters
for this calculation are shown in Figure \ref{fig:meavt}.

The $N=21204$ and $42294$ calculations of case 246 yield
qualitatively similar results.  After the third pericenter passage of
the intruder through the secondary, the two stars merge.  The
resulting binary, surrounded by an envelope of gas removed from the
secondary by the impacts, ultimately merges.  In our highest
resolution calculation of this case ($N=84642$), the intruder does
not immediately merge with either star in the binary, but rather the
three stars move around one another in a long-lived resonance
interaction before ultimately all three stars merge.

\section{Discussion and Conclusions}\label{sect:discussion}

We present a set of hydrodynamical simulations of 40 close encounters
between three stars. The initial conditions are taken from the
high-precision direct $N$-body simulations of
\citet{2008MNRAS.384..376G}, who studied the onset of collision
runaway in young star clusters. Most of the collisions (31) involve a
massive binary star intruded upon by, generally, a lower mass
star. The rest of the collisions (9) occur between three single stars
which are in the middle of the resonant interaction. All the
simulations were carried out with both the SPH method and in the
sticky sphere approximation.

If only initial and final states are of interest, the sticky sphere
method provides the appropriate outcome of the encounter in about 3
out of every 4 cases.  In the cases where sticky spheres result in a
merger between three stars, our hydrodynamic simulations tend to give
a similar result. However, if one is interested in mass loss, close
inspection reveals that in a considerable amount of mass can be
ejected in double mergers.  In addition, the collision product
acquires a kick velocity, which is usually a result of the asymmetric
mass ejection. The kick velocity can be sifficietly high to eject the
merger product to the cluster halo and even to escape. In cases
%% (scenarios \simon{Which other scenario did you have in mind here:}
%% XXX, 211 and 262),
where only two stars merge and the third escapes, the kick velocity is
large enough that the collision product could be ejected out of the
star cluster completely. Therefore, it is not completely unreasonable
to expect collision products to be observed in the outer regions of
young star cluster, and the Pistol star in the Quintuplet cluster
\citep{1998ApJ...506..384F} may well be a merger product resulting from
an encounter between a single and a binary star.

The sticky sphere approximation, however, fails in several cases. On
occasion, this approximation predicts the formation of a binary with a
merger product as one of the components (cases 214, 253, 260 and 262),
an interesting outcome from either an observational or theoretical
point of view.  Detailed hydrodynamic modelling of the same cases,
however, show that a complete merger is a more likely outcome, if the
interaction is mild; otherwise, the outcome is two unbound stars. In
another case, the sticky sphere method predicts either one (case
207) or two collisions (case 222) in a system, but the
hydrodynamic simulations predict a fly-by. These are the cases for
grazing encounters which result in the ejection of the intruder star.
If the semi-major axis of the binary is sufficiently large, binaries
tend to avoid mergers and become eccentric instead.

For those situations in which the sticky sphere algorithm predicts a
single merger event, the result is incorrect in almost half of the
situations. It is important to keep in mind that the condition for a
merger in the sticky sphere approximation is energy independent, and
therefore if two stars with large enough velocities have a grazing
collision, this method will incorrectly predict a complete merger.

Thus in an environment with high velocity dispersion, such as galactic
nuclei in which the velocity dispersion is typically at least an order
of magnitude larger than in the cores of young massive star clusters,
the sticky sphere approximation may fail more often.  In such
environments, the merger cross-section is reduced, as grazing
interactions between stars may not necessarily lead to mergers
\citep{2005MNRAS.358.1133F}. While this could be improved by a more
sophisticated effective radius of the merger product (we use simply
$R_1+R_2$), it is unlikely that simple recipes can correctly reproduce
the richness of the hydrodynamic results, especially if one is
interested in the close interaction between three or more stars.

All of our collision products posses some amount of angular
momentum. In some cases, the angular momentum is large enough that the
shape of the collision product substantially deviates from spherical
symmetry. Evolving such an object is a challenge for stellar evolution
codes, given that even the evolution of non-rotating massive
collisions product is a formidable task \citep{2008PHD_GLEBBEEK}. In
addition, there still exist problems on even hydrodynamical grounds,
as some of our rotating collision products are gradually losing mass
even at the termination of our hydrodynamic calculations. The reason
for this mass loss is due to spurious transport of angular momentum
outward caused by artificial viscosity \citep{1999JCoPh.152..687L}, as
described in \S \ref{sect:results}. The precise timescale of this
effect depends on numerical parameters and the treatment of artificial
viscosity.  For example, in case 220, the progression of the stellar
collisions is essentially the same in the N=20178 and N=46296
calculations.  In the higher resolution simulation, however, the
angular momentum transport and resulting mass loss in the final
collision product progresses more slowly. It is worth noting, however,
that {\it physical} angular momentum transport mechanisms, such as
stellar winds and magnetic braking, would have a similar qualitative
effect but on a longer timescale \citep{2005MNRAS.358..716S}.

Stellar collisions in a young dense star cluster are expected to occur
in the first few million years of the cluster lifetime
\citep{1999A&A...348..117P}. At this age, the star cluster may still
be embedded in a natal gas \citep{2003ARA&A..41...57L}, and therefore
if the ejecta is energetic enough the state of the gas may be
considerably disturbed, and such mechanism has recently been proposed
within the context of globular clusters
\citep{2008ApJ...680L.113U}. In the case of young star clusters, our
results suggest that ejecta emanating from stellar collisions is
energetic enough to significantly disturb and even eject the remaining
gas. Indeed, a young massive star cluster with a star formation
efficiency of about 50\% has about $10^{49}-10^{50}$ ergs in binding
energy of the remaining gas. Our results show that the energy of the
ejected fluid in stellar collisions exceeds $10^{49}$ ergs, and in two
cases (cases 220 and 299) even $10^{50}$ ergs. Since collisions are
expected to occur in the core of a star cluster, it would be just a
matter of a few collisions to significantly perturb or largely expel
the natal gas from the central region. In the case of a runaway merger
\citep{2004Natur.428..724P, 2006MNRAS.368..141F}, we therefore expect
that the gas will be expelled form the central regions before the end
of runaway.

\section*{Acknowledgements}

JCL thanks Alex Brown for useful conversations.
We acknowledge the use of the 3D virtual collaboration environment
{\tt Qwaq Forums}, which greatly facilitated the remote collaboration
leading to this paper.  This work was supported by the Netherlands
Research School for Astronomy (NOVA), the National Science Foundation
under Grant No.\, 0703545 No.\, PHY05-51164 and by NWO (grants
\#635.000.303 and \#643.200.503).  Column density plots in this paper
were produced using SPLASH, a publicly available visualisation tool
for SPH \citep{2007PASA...24..159P}.  This work used the MUSE
framework, which is available at http://muse.li.

\appendix
\section{Derivation of SPH equations of motion}\label{appendix:derivation}

The use of non-equal mass particles in the simulations allows us to
resolve both the core and the envelope of parent stars. However,
during the merger process, particles of significantly different mass
from two or more
parent stars mix, and the standard constraint between density and the
smoothing length, $h_i = f(\rho_i, C_i)$, becomes inappropriate. Such a
constraint naturally involves a constant with dimensionality of mass,
$C_i$.  This constant is usually determined during the set up of
the initial conditions and therefore reflects the initial mass
resolution of particle $i$, that is the initial total mass of the neighbours of that particle.
However, as the particle migrates from one region to
another, the mass resolution of the particle should adapt to its
new environment. If this does not happen, the particle may have too
few or too many neighbours, depending on whether it migrates into a
region with, respectively, an average particle mass significantly larger or smaller
than in its initial environment.
To mend this, we present a new approach
that keeps the number of neighbors roughly constant.
Here, we can draw an analogy with
finite-difference hydrodynamics, either on fixed or moving meshes: the
number of neighbouring cells that a given grid cell interacts with is
also roughly constant (exactly constant on a fixed mesh) and is, to some
degree, independent of the local fluid conditions.

\begin{figure}
  \begin{center}
    \includegraphics[width=\columnwidth]{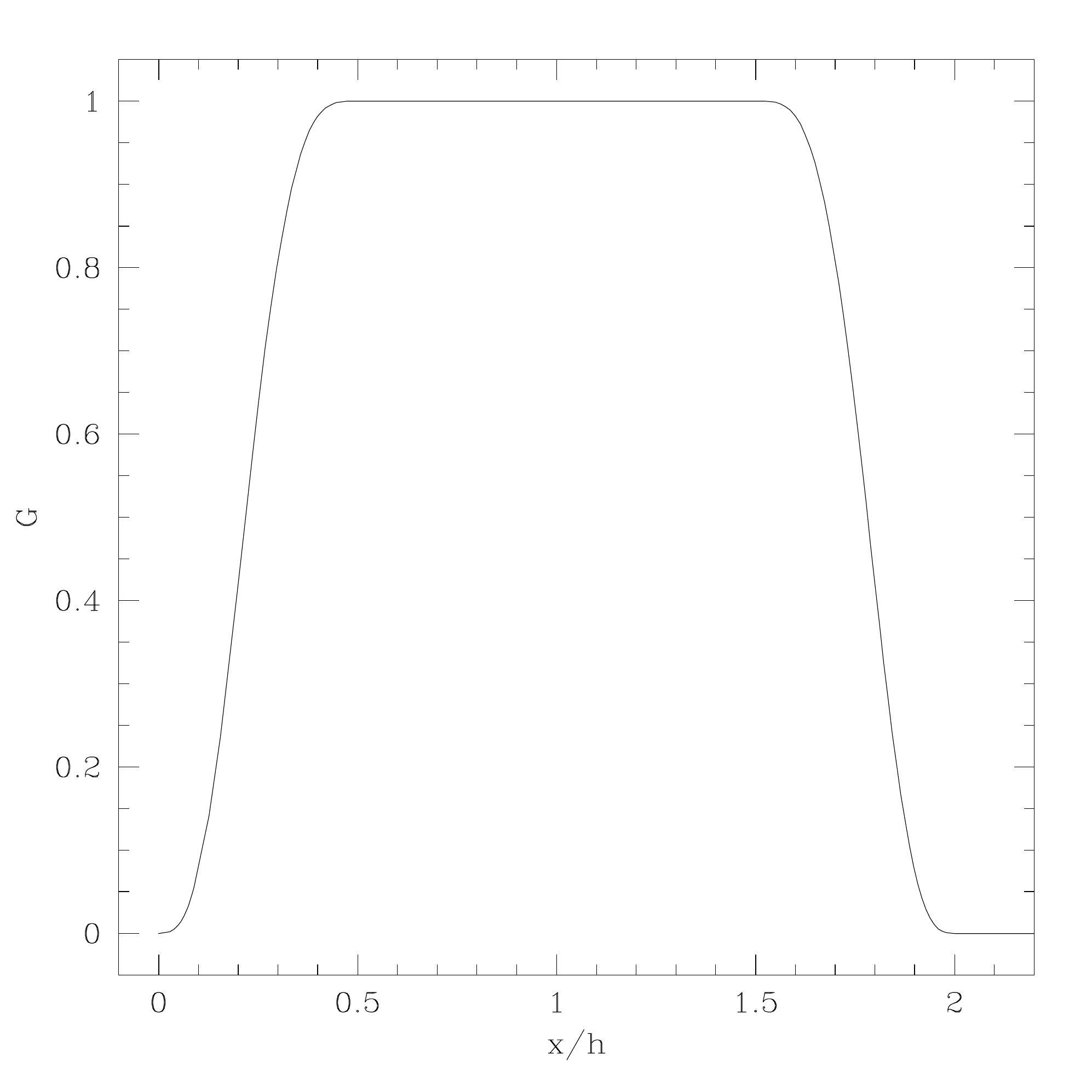}
  \end{center}
  \caption{Plotted here versus normalized separation $x/h$, the
  weighting function $G$ used to help keep neighbor numbers roughly
  constant even when non-equal mass particles are used.}
  \label{fig:gvsr}
\end{figure}

We propose a continuous constraint between an estimate of the
number of neighbours and the smoothing length.
Relaxing the
condition that the neighbour number estimate be an integer, we weight each
neighbour with a function that depends on its distance from the
particle, $G(r_{ij}/h_i)$, where $r_{ij}$ is separation between the
particle $i$ and the neighbour $j$.
Using such a weight function, we estimate the number of neighbours of a
given particle $i$ as
\begin{equation}
  N_i = \sum_j G(|{\bf r}_i - {\bf r}_j|, h_i) \equiv \sum_j
  G_{ij}(h_i).
  \label{eq:constraintN}
\end{equation}
We find empirically that the following
function provides satisfactory results:
\begin{equation}
   G(x, h) \equiv V(4h - 4|x - h|, h),
\end{equation}
where $0 \le x < 2h$, otherwise it is equal to zero, and
\begin{equation}
  V(x, h) \equiv 4\pi\int_0^x x^2 W(x, h)\,dx.
\end{equation}
Here, $W(x,h)$ is an SPH smoothing kernel with a compact support of
$2h$.  We use the kernel of
\citet{1985A&A...149..135M}, for which the weighting function $G$
takes on the form shown in Figure \ref{fig:gvsr}.
Setting $N_i$ to be constant, equation (\ref{eq:constraintN}) provides
the particle with a necessary constraint between $h_i$ and its
estimated instantaneous number of neighbours in a continuous way.  In
the calculations presented in this paper, we choose $N_i=22$, which
typically results in about 35 to 40 actual neighbors enclosed by the
smoothing kernel.  (It is not surprising that the actual number of
neighbors is consistently larger than the chosen $N_i$, as can be seen
by noting from Fig.\ \ref{fig:gvsr} that $G\le 1$.)

Because our constraint allows particles smoothing lengths to be a
function of particle coordinates, the variational formulation of SPH
can be used to derive equations of motion \citep{2002MNRAS.333..649S,
2002MNRAS.335..843M,
2007MNRAS.374.1347P}. In particular, we consider the SPH Lagrangian
\begin{equation}
  {\cal L} = \frac{1}{2}\sum_j m_j v_j^2 - \sum_j m_j u_j - \frac{1}{2}\sum_jm_j\phi_j.
\end{equation}
Here, $m_j$ is the mass of SPH particle $j$, $v_j$ and $u_j$ its
velocity and specific internal energy respectively, and $\phi_j$ is its
gravitational potential, which is defined as
\begin{equation}
  \phi_j = \sum_k m_k g(|{\bf r}_j - {\bf r}_k|, h_j)  \equiv \sum_k m_k g_{jk}(h_j),
\end{equation}
where $g(x,h)$ is the gravitational potential between two SPH
particles of unit mass. The Euler-Lagrange
equations resulting from this Lagrangian are
\begin{equation}
  m_i\dot{\bf v}_i = -\sum_jm_j\left(\frac{\partial
    u}{\partial\rho}\right)_{s,j}\frac{d\rho_j}{d{\bf r}_i} -
  \frac{1}{2}\sum_j m_j\frac{d\phi_j}{d{\bf r}_i}.
\end{equation}
Here, the first term is the hydrodynamic force, $m_i{\bf a}_{{\rm
    h},i}$, the second term is the gravitational force, $m_i {\bf
  a}_{{\rm g},i}$, and the partial derivative, $(\partial
u/\partial\rho)_s$, is evaluated at constant entropy $s$. Using the SPH
definition of density,
\begin{equation}
  \rho_j = \sum_k m_k W(|{\bf r}_j - {\bf r}_k|, h_j) \equiv \sum_k
  m_k W_{jk}(h_j),
\end{equation}
we derive its gradient
\begin{eqnarray}
  \frac{d\rho_j}{d{\bf r}_i} & = & \sum_k m_k\nabla_i
  W_{ik}(h_i)\delta_{ij} + m_i\nabla_i W_{ij}(h_j) \nonumber \\ & &
  +\sum_k m_k\frac{\partial W_{jk}(h_j)}{\partial h_j}\frac{dh_j}{d{\bf
      r}_i}.
\end{eqnarray}
Differentiating Eq. \ref{eq:constraintN} with respect to ${\bf r}_i$
we find
\begin{equation}
  \chi_j\frac{dh_j}{d{\bf r}_i} =
  -\sum_k\nabla_iG_{jk}(h_i)\delta_{ij} - \nabla_iG_{ij}(h_j),
\end{equation}
where
\begin{equation}
  \chi_j \equiv \sum_k \frac{\partial G_{jk}(h_j)}{\partial h_j}.
\end{equation}
With these equations, it is straightforward to derive accelerations
due to pressure 
\begin{eqnarray}
  {\bf a}_{{\rm h},i} = &
  -\sum_j m_j {P_i\over\rho_i^2}\left[\nabla_iW_{ij}(h_i) - {\omega_i\over \chi_im_j}\nabla_iG_{ij}(h_i)\right] \\
  & -\sum_j m_j {P_j\over\rho_j^2}\left[\nabla_iW_{ij}(h_j) - {\omega_j\over \chi_jm_i}\nabla_iG_{ij}(h_j)\right],
  \label{eq:a_hydro}
\end{eqnarray}
and due to gravity
\begin{eqnarray}
  {\bf a}_{{\rm g},i} &=& -{1\over 2}\sum_j m_j\left[\nabla_i g_{ij}(h_i) + \nabla_i g_{ij}(h_j)\right]\\
                   & & +{1\over 2}\sum_j m_j{\Psi_i\over\chi_im_j}\nabla_iG_{ij}(h_i)\\
                   & & +{1\over 2}\sum_j m_j{\Psi_j\over\chi_jm_i}\nabla_iG_{ij}(h_j).
  \label{eq:a_grav}
\end{eqnarray}
Here, we define two more quantities:
\begin{equation}
  \omega_j \equiv \sum_k m_k {\partial W_{jk}(h_j)\over\partial h_j}
\end{equation}
and
\begin{equation}
  \Psi_i \equiv \sum_k m_k {\partial g_{ik}(h_i)\over\partial h_i}.
\end{equation}

Following the approach of \citet{2002MNRAS.335..843M} (see their \S2.3), we find the rate of change of the specific internal energy to be
\begin{equation}
  {du_i\over dt} =  {P_i\over\rho_i^2}
  \sum_j m_j ({\bf v_i}-{\bf v_j})\cdot  
\left[\nabla_iW_{ij}(h_i) - {\omega_i\over \chi_im_j}\nabla_iG_{ij}(h_i)\right],
  \label{eq:dudt}
\end{equation}
which guarantees conservation of total energy and entropy in the
absence of shocks.  In order to handle shock waves while maintaining
energy conservation, we augment these equations with artificial
viscosity \citep{1997JCoPh.136..298M}.  For the calculations of this
paper, we implement a variation on the artificial viscosity term
proposed by
\citet{1995JCoPh.121..357B}:
\begin{equation}
\Pi_{ij}=
\left({p_i\over\rho_i^2}+
      {p_j\over\rho_j^2}
\right)
\left(-\alpha\mu_{ij}+
      \beta\mu_{ij}^2\right)\,,
\label{piDB}
\end{equation}
with $\alpha=1$ and $\beta=2$. In our treatment,
\begin{equation}
\mu_{ij}=
  {({\bf v}_i-{\bf v}_j)\cdot
   ({\bf r}_i-{\bf r}_j)\over
   |{\bf r}_i -{\bf r}_j|}\,
  {f_i+f_j\over c_i+c_j}
\label{muDB}
\end{equation}
if $({\bf v}_i-{\bf v}_j)\cdot({\bf r}_i-{\bf r}_j)<0$; otherwise $\mu_{ij}=0$.
Here $c_i$ is the sound speed at particle $i$.
See \citet{2006ApJ...640..441L} for the definition of the form factor $f_i$ and
for additional details on how the artificial viscosity is encorporated.

The evolution equations are integrated using a symplectic integrator
with shared symmetrised timesteps, as in \cite{2005MNRAS.364.1105S}.
Our shared timestep is determined as
\begin{equation}
  \Delta t={\rm Min}_i \left[\left(\Delta t_{1,i}^{-1}+ \Delta
    t_{2,i}^{-1} \right)^{-1}\right]\,,
  \label{good.dt}
\end{equation}
where for each SPH particle $i$, we use
\begin{equation}
  \Delta t_{1,i}=C_{N,1}
         {h_i\over
           {\rm Max}\left[{\rm Max}_j\left(\kappa_{ij}\right),
             {\rm Max}_j\left(\kappa_{ji}\right)\right]}
         \label{dt1}
\end{equation}
with
\begin{equation}
  \kappa_{ij}\equiv\left[\left( {p_i\over \rho_i^2}+
    \frac12\Pi_{ij}\right)\rho_i\right]^{1/2}\,,
\end{equation}
and
\begin{equation}
  \Delta t_{2,i}=C_{N,2}{u_i\over \left|du_i/dt\right|}\,.
  \label{dt2}
\end{equation}
For the simulations presented in this paper, $C_{N,1}=0.2$~to~0.3 and
$C_{N,2}=0.05$.  The Max$_j$ function in equation~(\ref{dt1}) refers
to the maximum of the value of its expression for all SPH particles
$j$ that are neighbors with $i$.  The denominator of
equation~(\ref{dt1}) is an approximate upper limit to the signal
propagation speed near particle $i$.  The incorporation of $\Delta
t_2$ enables us to treat shocks without drastically
decreasing the timestep during intervals in which the flow is
subsonic.

\section{Initial Conditions}\label{sect:appendix}

In Table \ref{tab:initial_conditions_all}, we summarize the raw initial conditions
of our calculations in order to facilitate comparisons with any future works.

\begin{table*}
 \begin{minipage}{190mm}
 \caption{ The first column gives the case identification number. The
   second, third, and fourth columns give the masses $M_1$, $M_2$, and
   $M_3$ of the colliding stars.  Columns 5 through 7 and columns 8
   through 10 give the position and velocity, respectively, of star 1
   in Cartesian coordinates.  Likewise, Columns 11 through 13 and
   columns 14 through 16 give the position and velocity of star 2.
   The position and velocity of star 3 can be determined from the
   constraints that the center of mass be at the origin and that the
   net momentum is zero.  All quantities are in solar units.  }
%\tiny
 \begin{tabular}{llllllllllllllll}
\hline
id &$M_1$     & $M_2$    & $M_3$    &  $x_1$   & $y_1$    & $z_1$    & $v_{x,1}$& $v_{y,1}$& $v_{z,1}$& $x_2$    & $y_2$    & $z_2$    & $v_{x,2}$& $v_{y,2}$& $v_{z,2}$ \\ \hline
201 &  84.1    &  27.1    & 0.250    &  81.6    &  33.9    & -7.56    &-0.00102  & 0.0690   & 0.0777   & -254.    & -108.    &  22.0    & 0.00207  &-0.206    &-0.237    \\
202 &  57.9    &  29.9    & 0.120    & -52.5    &  13.8    & -37.1    & 0.0150   & 0.148    &-0.0508   &  102.    & -26.4    &  71.9    &-0.0302   &-0.289    & 0.0977   \\
203 &  47.1    &  36.3    &  1.09    & -7.32    &  8.64    & -9.06    & 0.474    &-0.998E-01&-0.469    &  10.5    & -10.5    &  12.3    &-0.620    & 0.121    & 0.640    \\ % -33.1    & -23.4    & -18.1    & 0.215    & 0.277    & -1.07
206 &  24.6    &  21.9    &  20.6    &  17.0    &  46.8    &  20.8    &-0.850    & 0.404E-03&-0.386    &  23.4    &  52.8    &  11.0    & 0.791    &-0.688    & 0.275    \\ % -45.1    & -112.    & -36.5    & 0.176    & 0.728    & 0.169    
207 &  42.2    &  18.2    & 0.651    &  9.29    &  1.86    & -1.35    &-0.434E-02&-0.969E-02&-0.412    & -19.6    & -8.11    &-0.774    & 0.293E-02& 0.524E-01& 0.993    \\ % -52.9    &  106.    &  109.    & 0.199    &-0.838    & -1.07    
208 &  86.7    & 0.513    & 0.161    & -1.86    & -1.27    & 0.139    & 0.382E-02& 0.489E-02& 0.505E-03&  298.    &  220.    & -21.2    &-0.524    &-0.454    &-0.303E-01\\ %  49.3    & -16.9    & -7.36    &-0.387    & -1.19    &-0.176
211 &  61.7    &  18.4    &  8.89    & -28.1    &  1.04    &  13.5    & 0.384E-01&-0.676E-01& 0.288E-01&  117.    & -5.11    & -46.2    &-0.553    & 0.356    & 0.620    \\ % -48.2    &  3.38    &  2.07    & 0.880    &-0.269    & -1.49    
212 &  87.6    &  27.1    &  22.7    &  29.8    &  18.3    & -7.61    & 0.204    & 0.155E-01& 0.212    &  15.4    &  6.21    &  21.9    & -1.21    &-0.598    &-0.737    \\ % -133.    & -77.9    &  3.29    & 0.657    & 0.653    & 0.609E-01
213  &  76.8    &  13.6    & 0.227    & -2.97    & -3.52    & -2.06    &-0.137    & 0.199    &-0.810E-01&  17.9    &  16.8    &  11.8    & 0.769    & -1.10    & 0.458    \\ % -71.4    &  185.    & -11.0    & 0.330    & -1.06    &-0.243E-01
214 &  17.8    &  16.8    &  9.49    &  2.63    &  13.0    &  8.76    &-0.343    & 0.322    &-0.150    & -6.44    & -3.78    & -2.14    & 0.264    &-0.0197   &-0.350    \\
217 &  86.4    &  28.9    & 0.110    & -11.9    &  4.92    &-0.182    &-0.0434   &-0.0713   & 0.367    &  34.8    & -15.7    &  1.25    & 0.131    & 0.216    & -1.10    \\
219 &  36.6    &  10.7    &  9.10    &  15.9    & -7.02    & -6.37    & 0.109    & 0.155    & 0.0155   & -44.7    &  24.4    &  27.7    &-0.170    & 0.111    &-0.0360   \\
220 &  84.3    &  68.3    &  32.7    & -12.1    & -7.97    & -41.0    & 0.569    & 0.631    & 0.369    &  6.17    & -43.2    & -13.6    &-0.605    &-0.636    &-0.102    \\
222 &  22.8    &  11.1    &  5.28    & -14.1    & -3.42    &  2.74    & 0.205    &-0.245    &-0.146    & -41.8    & 0.696    & -17.9    &-0.209E-01& 0.593    & 0.325    \\ %  149.    &  13.3    &  25.8    &-0.837    &-0.190    &-0.542E-01
223 &  28.6    &  19.4    &  4.57    & -31.3    & -10.8    &  37.9    & 0.118    & 0.198    &-0.453    &  55.2    &  16.7    & -66.3    &-0.208    &-0.0781   & 0.500    \\
224 &  48.1    &  22.0    & 0.200    & -5.22    & -4.58    &  2.42    & 0.328    &-0.136    & 0.427    &  10.8    &  9.03    & -5.32    &-0.712    & 0.309    &-0.937    \\ %  71.7    &  110.    &  2.48    &-0.654    & -1.10    & 0.190    
227 &  16.0    &  5.62    & 0.174    & -17.9    & -19.4    &  12.1    & 0.0888   & 0.116    &-0.0455   &  52.3    &  55.7    & -35.0    &-0.268    &-0.353    & 0.121    \\
231 &  26.1    &  25.8    & 0.411    &  63.5    &  5.88    & -11.6    &-0.415    & 0.0894   & 0.137    & -63.5    & -6.03    &  11.8    & 0.418    &-0.103    &-0.141    \\
232 &  19.1    &  12.2    &  6.99    & -58.6    & -7.88    &  26.5    & 0.359    & 0.0774   &-0.120    &  60.4    &  8.80    & -24.4    & 0.0430   &-0.215    &-0.245    \\
233 &  47.6    &  28.9    &  2.94    &  7.42    & -88.9    &  24.3    &-0.100    & 0.302    &-0.999E-01& -9.56    &  134.    & -38.0    & 0.179    &-0.379    & 0.213    \\ % -26.2    &  124.    & -19.5    &-0.135    & -1.16    &-0.480
236 &  40.5    &  31.4    &  29.3    & -36.0    & -41.0    &  10.9    &-0.114    & 0.849    &-0.345    & -52.5    & -54.1    &  1.16    & 0.500    &-0.563    & 0.538    \\
241 &  41.7    &  28.1    &  11.3    &  26.2    &  46.6    &  46.4    &-0.228    &-0.205    &-0.405    & -27.0    & -53.1    & -49.8    & 0.760E-01& 0.104    & 0.878    \\ % -29.5    & -40.1    & -47.4    & 0.653    & 0.499    &-0.686
242 &  41.1    &  23.5    & 0.490    & -3.06    & -2.81    &  5.16    &-0.0758   &-0.516    &-0.365    &  7.38    &  7.01    & -10.6    & 0.124    & 0.891    & 0.646    \\
245 &  79.1    &  43.5    &  16.0    & -48.5    & -66.2    & -40.1    & 0.191    & 0.255    & 0.369    &  65.5    &  81.8    &  59.6    &-0.590    &-0.277    &-0.375    \\ %  61.8    &  105.    &  36.4    & 0.657    &-0.507    &-0.806
246 &  42.2    &  38.3    &  1.37    &  6.56    & -1.36    &  15.6    & 0.554    & 0.546    &-0.159    & -4.02    &  3.20    & -10.3    &-0.631    &-0.613    & 0.147    \\
249 &  74.7    & 0.154    & 0.110    &-0.718    & 0.256    & 0.0579   & 0.00154  &-0.00229  &-0.000616 &  328.    & -120.    & -66.3    &-0.506    & 0.217    & 0.183    \\
250 &  44.0    &  31.9    & 0.550    &  1.70    & 0.0166   &  15.0    & 0.0346   &-0.617    & 0.0335   & -2.35    & -1.72    & -20.4    &-0.0446   & 0.850    &-0.0481   \\
253 &  53.4    &  8.55    & 0.583    &-0.664    & -1.74    &  1.73    & 0.176    &-0.0823   &-0.104    & -2.64    &  14.5    & -14.7    & -1.11    & 0.501    & 0.666    \\
256 &  33.4    &  5.84    &  2.11    &  16.3    & -5.81    &  9.70    &-0.0106   & 0.0614   &-0.0617   & -101.    &  37.5    & -52.8    & 0.489    &-0.109    & 0.396    \\
257 &  97.3    &  24.9    &  5.18    & -6.21    &  13.7    & -3.31    & 0.166    & 0.119    & 0.0784   &  30.5    & -40.7    &  13.0    &-0.710    &-0.439    &-0.401    \\
258 &  90.4    & 0.929    & 0.546    & -1.35    & -1.00    &-0.482    & 0.00898  & 0.0159   & 0.000210 &  116.    &  103.    &  47.8    &-0.603    &-0.669    &-0.527    \\
259 &  55.9    &  21.7    &  11.4    &  13.0    &  21.9    &  7.96    & 0.208    &-0.140    &-0.465    & -4.92    &  5.50    & -2.09    &-0.651    &-0.117    &  1.03    \\
260 &  92.9    &  53.3    &  13.3    &  20.3    & -8.98    &  28.8    &-0.200    & 0.648    & 0.229    & -4.97    &  2.34    & -5.17    & 0.192    &-0.939    &-0.588    \\ % -122.    &  53.2    & -180.    & 0.628    &-0.762    & 0.753
261 &  33.9    &  13.2    &  9.17    & -4.19    &  3.93    &  7.70    &-0.122    & 0.00813  &-0.131    &  11.1    & -42.0    & -5.62    & 0.180    &-0.0511   & 0.562    \\
262 &  31.5    &  29.3    &  18.4    & -37.9    & -23.2    & -65.6    &-0.249    & 0.0274   & 0.575    & -6.01    & -8.40    & -8.67    & 0.343    & 0.228    &-0.274    \\
267 &  28.6    &  19.1    &  14.2    & -54.5    &  16.3    &-0.486    & 0.354    & 0.313    & 0.450E-01&  107.    & -25.5    & -2.16    &-0.261    & 0.788E-01&-0.149E-01\\ % -33.3    &  1.34    &  3.87    &-0.363    &-0.736    &-0.706E-01
298 &  56.7    &  28.1    &  25.3    &  28.0    &  35.2    & -9.25    & 0.220    & 0.0918   &-0.184    & -93.3    & -91.4    &  9.93    & 0.736    & 0.356    & 0.0203   \\
299 &  52.3    &  52.3    &  16.9    &  82.3    & -4.64    & -2.27    &-0.165    &-0.155    &-0.195    & -114.    & 0.720E-04& 0.175E-03& 0.628    &-0.208E-06&-0.128E-06\\ %  97.7    &  14.3    &  7.03    & -1.43    & 0.480    & 0.603
\hline
  \end{tabular}
  \end{minipage}
  \label{tab:initial_conditions_all}
\end{table*}

\bibliographystyle{mn2e}

\begin{thebibliography}{}

\bibitem[\protect\citeauthoryear{{Adams}, {Davies} \& {Sills}}{{Adams}
  et~al.}{2004}]{2004MNRAS.348..469A}
{Adams} T.,  {Davies} M.~B.,    {Sills} A.,  2004, \mnras, 348, 469

\bibitem[\protect\citeauthoryear{{Balsara}}{{Balsara}}{1995}]{1995JCoPh.121..3%
57B}
{Balsara} D.~S.,  1995, Journal of Computational Physics, 121, 357

\bibitem[\protect\citeauthoryear{{Baumgardt} \& {Kroupa}}{{Baumgardt} \&
  {Kroupa}}{2007}]{2007MNRAS.380.1589B}
{Baumgardt} H.,  {Kroupa} P.,  2007, \mnras, 380, 1589

\bibitem[\protect\citeauthoryear{{Belleman}, {B{\'e}dorf} \& {Portegies
  Zwart}}{{Belleman} et~al.}{2008}]{2008NewA...13..103B}
{Belleman} R.~G.,  {B{\'e}dorf} J.,    {Portegies Zwart} S.~F.,  2008, New
  Astronomy, 13, 103

\bibitem[\protect\citeauthoryear{{Benz} \& {Hills}}{{Benz} \&
  {Hills}}{1987}]{1987ApJ...323..614B}
{Benz} W.,  {Hills} J.~G.,  1987, \apj, 323, 614

\bibitem[\protect\citeauthoryear{{Cleary} \& {Monaghan}}{{Cleary} \&
  {Monaghan}}{1990}]{1990ApJ...349..150C}
{Cleary} P.~W.,  {Monaghan} J.~J.,  1990, \apj, 349, 150

\bibitem[\protect\citeauthoryear{{Cohn}, {Hut} \& {Wise}}{{Cohn}
  et~al.}{1989}]{1989ApJ...342..814C}
{Cohn} H.,  {Hut} P.,    {Wise} M.,  1989, \apj, 342, 814

\bibitem[\protect\citeauthoryear{{Davies}, {Benz} \& {Hills}}{{Davies}
  et~al.}{1993}]{1993ApJ...411..285D}
{Davies} M.~B.,  {Benz} W.,    {Hills} J.~G.,  1993, \apj, 411, 285

\bibitem[\protect\citeauthoryear{{Davies}, {Benz} \& {Hills}}{{Davies}
  et~al.}{1994}]{1994ApJ...424..870D}
{Davies} M.~B.,  {Benz} W.,    {Hills} J.~G.,  1994, \apj, 424, 870

\bibitem[\protect\citeauthoryear{{Davies}, {Blackwell}, {Bailey} \&
  {Sigurdsson}}{{Davies} et~al.}{1998}]{1998MNRAS.301..745D}
{Davies} M.~B.,  {Blackwell} R.,  {Bailey} V.~C.,    {Sigurdsson} S.,  1998,
  \mnras, 301, 745

\bibitem[\protect\citeauthoryear{{Davies}, {Ruffert}, {Benz} \&
  {Muller}}{{Davies} et~al.}{1993}]{1993A&A...272..430D}
{Davies} M.~B.,  {Ruffert} M.,  {Benz} W.,    {Muller} E.,  1993, \aap, 272,
  430

\bibitem[\protect\citeauthoryear{{Eggleton}}{{Eggleton}}{1971}]{1971MNRAS.151.%
.351E}
{Eggleton} P.~P.,  1971, \mnras, 151, 351

\bibitem[\protect\citeauthoryear{{Figer}, {Najarro}, {Morris}, {McLean},
  {Geballe}, {Ghez} \& {Langer}}{{Figer} et~al.}{1998}]{1998ApJ...506..384F}
{Figer} D.~F.,  {Najarro} F.,  {Morris} M.,  {McLean} I.~S.,  {Geballe} T.~R.,
  {Ghez} A.~M.,    {Langer} N.,  1998, \apj, 506, 384

\bibitem[\protect\citeauthoryear{{Fregeau}, {Cheung}, {Portegies Zwart} \&
  {Rasio}}{{Fregeau} et~al.}{2004}]{2004MNRAS.352....1F}
{Fregeau} J.~M.,  {Cheung} P.,  {Portegies Zwart} S.~F.,    {Rasio} F.~A.,
  2004, \mnras, 352, 1

\bibitem[\protect\citeauthoryear{{Fregeau}, {G{\"u}rkan}, {Joshi} \&
  {Rasio}}{{Fregeau} et~al.}{2003}]{2003ApJ...593..772F}
{Fregeau} J.~M.,  {G{\"u}rkan} M.~A.,  {Joshi} K.~J.,    {Rasio} F.~A.,  2003,
  \apj, 593, 772

\bibitem[\protect\citeauthoryear{{Freitag} \& {Benz}}{{Freitag} \&
  {Benz}}{2005}]{2005MNRAS.358.1133F}
{Freitag} M.,  {Benz} W.,  2005, \mnras, 358, 1133

\bibitem[\protect\citeauthoryear{{Freitag}, {G{\"u}rkan} \& {Rasio}}{{Freitag}
  et~al.}{2006}]{2006MNRAS.368..141F}
{Freitag} M.,  {G{\"u}rkan} M.~A.,    {Rasio} F.~A.,  2006, \mnras, 368, 141

\bibitem[\protect\citeauthoryear{{Fukushige}, {Taiji}, {Makino}, {Ebisuzaki} \&
  {Sugimoto}}{{Fukushige} et~al.}{1996}]{1996ApJ...468...51F}
{Fukushige} T.,  {Taiji} M.,  {Makino} J.,  {Ebisuzaki} T.,    {Sugimoto} D.,
  1996, \apj, 468, 51

\bibitem[\protect\citeauthoryear{{Gaburov}, {Gualandris} \& {Portegies
  Zwart}}{{Gaburov} et~al.}{2008}]{2008MNRAS.384..376G}
{Gaburov} E.,  {Gualandris} A.,    {Portegies Zwart} S.,  2008, \mnras, 384,
  376

\bibitem[\protect\citeauthoryear{{Gaburov}, {Harfst} \& {Portegies
  Zwart}}{{Gaburov} et~al.}{2009}]{2009arXiv0902.4463G}
{Gaburov} E.,  {Harfst} S.,    {Portegies Zwart} S.,  2009, ArXiv e-prints

\bibitem[\protect\citeauthoryear{{Gaburov}, {Lombardi} \& {Portegies
  Zwart}}{{Gaburov} et~al.}{2008}]{2008MNRAS.383L...5G}
{Gaburov} E.,  {Lombardi} J.~C.,    {Portegies Zwart} S.,  2008, \mnras, 383,
  L5

\bibitem[\protect\citeauthoryear{{Glebbeek}}{{Glebbeek}}{2008}]{2008PHD_GLEBBE%
EK}
{Glebbeek} E.,  2008, PhD thesis, Utrecht University

\bibitem[\protect\citeauthoryear{{Glebbeek}, {Gaburov}, {de Mink}, {Pols} \&
  {Portegies Zwart}}{{Glebbeek} et~al.}{2009}]{2009arXiv0902.1753G}
{Glebbeek} E.,  {Gaburov} E.,  {de Mink} S.~E.,  {Pols} O.~R.,    {Portegies
  Zwart} S.~F.,  2009, ArXiv e-prints

\bibitem[\protect\citeauthoryear{{Glebbeek} \& {Pols}}{{Glebbeek} \&
  {Pols}}{2008}]{2008A&A...488.1017G}
{Glebbeek} E.,  {Pols} O.~R.,  2008, \aap, 488, 1017

\bibitem[\protect\citeauthoryear{{Glebbeek}, {Pols} \& {Hurley}}{{Glebbeek}
  et~al.}{2008}]{2008A&A...488.1007G}
{Glebbeek} E.,  {Pols} O.~R.,    {Hurley} J.~R.,  2008, \aap, 488, 1007

\bibitem[\protect\citeauthoryear{{G{\"u}rkan}, {Fregeau} \&
  {Rasio}}{{G{\"u}rkan} et~al.}{2006}]{2006ApJ...640L..39G}
{G{\"u}rkan} M.~A.,  {Fregeau} J.~M.,    {Rasio} F.~A.,  2006, \apjl, 640, L39

\bibitem[\protect\citeauthoryear{{Hamada} \& {Iitaka}}{{Hamada} \&
  {Iitaka}}{2007}]{2007astro.ph..3100H}
{Hamada} T.,  {Iitaka} T.,  2007, ArXiv Astrophysics e-prints

\bibitem[\protect\citeauthoryear{{Harrington}}{{Harrington}}{1970}]{1970AJ....%
.75.1140H}
{Harrington} R.~S.,  1970, \aj, 75, 1140

\bibitem[\protect\citeauthoryear{{Heggie}}{{Heggie}}{1975}]{1975MNRAS.173..729%
H}
{Heggie} D.~C.,  1975, \mnras, 173, 729

\bibitem[\protect\citeauthoryear{{Heggie} \& {Hut}}{{Heggie} \&
  {Hut}}{1993}]{1993ApJS...85..347H}
{Heggie} D.~C.,  {Hut} P.,  1993, \apjs, 85, 347

\bibitem[\protect\citeauthoryear{{Heggie}, {Trenti} \& {Hut}}{{Heggie}
  et~al.}{2006}]{2006MNRAS.368..677H}
{Heggie} D.~C.,  {Trenti} M.,    {Hut} P.,  2006, \mnras, 368, 677

\bibitem[\protect\citeauthoryear{{Hills}}{{Hills}}{1992}]{1992AJ....103.1955H}
{Hills} J.~G.,  1992, \aj, 103, 1955

\bibitem[\protect\citeauthoryear{{Hut}}{{Hut}}{1983}]{1983ApJ...268..342H}
{Hut} P.,  1983, \apj, 268, 342

\bibitem[\protect\citeauthoryear{{Hut} \& {Bahcall}}{{Hut} \&
  {Bahcall}}{1983}]{1983ApJ...268..319H}
{Hut} P.,  {Bahcall} J.~N.,  1983, \apj, 268, 319

\bibitem[\protect\citeauthoryear{{Lada} \& {Lada}}{{Lada} \&
  {Lada}}{2003}]{2003ARA&A..41...57L}
{Lada} C.~J.,  {Lada} E.~A.,  2003, \araa, 41, 57

\bibitem[\protect\citeauthoryear{{Lombardi}, {Sills}, {Rasio} \&
  {Shapiro}}{{Lombardi} et~al.}{1999}]{1999JCoPh.152..687L}
{Lombardi} J.~C.,  {Sills} A.,  {Rasio} F.~A.,    {Shapiro} S.~L.,  1999,
  Journal of Computational Physics, 152, 687

\bibitem[\protect\citeauthoryear{{Lombardi}, {Thrall}, {Deneva}, {Fleming} \&
  {Grabowski}}{{Lombardi} et~al.}{2003}]{2003MNRAS.345..762L}
{Lombardi} J.~C.,  {Thrall} A.~P.,  {Deneva} J.~S.,  {Fleming} S.~W.,
  {Grabowski} P.~E.,  2003, \mnras, 345, 762

\bibitem[\protect\citeauthoryear{{Lombardi} Jr., {Proulx}, {Dooley},
  {Theriault}, {Ivanova} \& {Rasio}}{{Lombardi}
  et~al.}{2006}]{2006ApJ...640..441L}
{Lombardi} Jr. J.~C.,  {Proulx} Z.~F.,  {Dooley} K.~L.,  {Theriault} E.~M.,
  {Ivanova} N.,    {Rasio} F.~A.,  2006, \apj, 640, 441

\bibitem[\protect\citeauthoryear{{Lombardi} Jr., {Rasio} \&
  {Shapiro}}{{Lombardi} et~al.}{1995}]{1995ApJ...445L.117L}
{Lombardi} Jr. J.~C.,  {Rasio} F.~A.,    {Shapiro} S.~L.,  1995, \apjl, 445,
  L117

\bibitem[\protect\citeauthoryear{{Lombardi} Jr., {Warren}, {Rasio}, {Sills} \&
  {Warren}}{{Lombardi} et~al.}{2002}]{2002ApJ...568..939L}
{Lombardi} Jr. J.~C.,  {Warren} J.~S.,  {Rasio} F.~A.,  {Sills} A.,    {Warren}
  A.~R.,  2002, \apj, 568, 939

\bibitem[\protect\citeauthoryear{{McMillan}}{{McMillan}}{1986}]{1986ApJ...306.%
.552M}
{McMillan} S.~L.~W.,  1986, \apj, 306, 552

\bibitem[\protect\citeauthoryear{{Monaghan}}{{Monaghan}}{1992}]{1992ARA&A..30.%
.543M}
{Monaghan} J.~J.,  1992, \araa, 30, 543

\bibitem[\protect\citeauthoryear{{Monaghan}}{{Monaghan}}{1997}]{1997JCoPh.136.%
.298M}
{Monaghan} J.~J.,  1997, Journal of Computational Physics, 136, 298

\bibitem[\protect\citeauthoryear{{Monaghan}}{{Monaghan}}{2002}]{2002MNRAS.335.%
.843M}
{Monaghan} J.~J.,  2002, \mnras, 335, 843

\bibitem[\protect\citeauthoryear{{Monaghan} \& {Lattanzio}}{{Monaghan} \&
  {Lattanzio}}{1985}]{1985A&A...149..135M}
{Monaghan} J.~J.,  {Lattanzio} J.~C.,  1985, \aap, 149, 135

\bibitem[\protect\citeauthoryear{{Portegies Zwart}, {Gaburov}, {Chen} \&
  {G{\"u}rkan}}{{Portegies Zwart} et~al.}{2007}]{2007MNRAS.378L..29P}
{Portegies Zwart} S.,  {Gaburov} E.,  {Chen} H.-C.,    {G{\"u}rkan} M.~A.,
  2007, \mnras, 378, L29

\bibitem[\protect\citeauthoryear{{Portegies Zwart}, {McMillan} \& {Harfst}
  S.}{{Portegies Zwart} et~al.}{2009}]{2009NewA...14..369P}
{Portegies Zwart} S.,  {McMillan} S.,    {Harfst} S. e.~a.,  2009, New
  Astronomy, 14, 369

\bibitem[\protect\citeauthoryear{{Portegies Zwart}, {Baumgardt}, {Hut},
  {Makino} \& {McMillan}}{{Portegies Zwart} et~al.}{2004}]{2004Natur.428..724P}
{Portegies Zwart} S.~F.,  {Baumgardt} H.,  {Hut} P.,  {Makino} J.,
  {McMillan} S.~L.~W.,  2004, \nat, 428, 724

\bibitem[\protect\citeauthoryear{{Portegies Zwart}, {Belleman} \&
  {Geldof}}{{Portegies Zwart} et~al.}{2007}]{2007NewA...12..641P}
{Portegies Zwart} S.~F.,  {Belleman} R.~G.,    {Geldof} P.~M.,  2007, New
  Astronomy, 12, 641

\bibitem[\protect\citeauthoryear{{Portegies Zwart}, {Makino}, {McMillan} \&
  {Hut}}{{Portegies Zwart} et~al.}{1999}]{1999A&A...348..117P}
{Portegies Zwart} S.~F.,  {Makino} J.,  {McMillan} S.~L.~W.,    {Hut} P.,
  1999, \aap, 348, 117

\bibitem[\protect\citeauthoryear{{Portegies Zwart} \& {McMillan}}{{Portegies
  Zwart} \& {McMillan}}{2002}]{2002ApJ...576..899P}
{Portegies Zwart} S.~F.,  {McMillan} S.~L.~W.,  2002, \apj, 576, 899

\bibitem[\protect\citeauthoryear{{Price}}{{Price}}{2007}]{2007PASA...24..159P}
{Price} D.~J.,  2007, Publications of the Astronomical Society of Australia,
  24, 159

\bibitem[\protect\citeauthoryear{{Price} \& {Monaghan}}{{Price} \&
  {Monaghan}}{2007}]{2007MNRAS.374.1347P}
{Price} D.~J.,  {Monaghan} J.~J.,  2007, \mnras, 374, 1347

\bibitem[\protect\citeauthoryear{{Rasio} \& {Lombardi} Jr.}{{Rasio} \&
  {Lombardi}}{1999}]{1999JCoAM.109..213R}
{Rasio} F.~A.,  {Lombardi} Jr. J.~C.,  1999, Journal of Computational and
  Applied Mathematics, 109, 213

\bibitem[\protect\citeauthoryear{{Rasio} \& {Shapiro}}{{Rasio} \&
  {Shapiro}}{1994}]{1994ApJ...432..242R}
{Rasio} F.~A.,  {Shapiro} S.~L.,  1994, \apj, 432, 242

\bibitem[\protect\citeauthoryear{{Sills}, {Adams} \& {Davies}}{{Sills}
  et~al.}{2005}]{2005MNRAS.358..716S}
{Sills} A.,  {Adams} T.,    {Davies} M.~B.,  2005, \mnras, 358, 716

\bibitem[\protect\citeauthoryear{{Sills}, {Faber}, {Lombardi} Jr., {Rasio} \&
  {Warren}}{{Sills} et~al.}{2001}]{2001ApJ...548..323S}
{Sills} A.,  {Faber} J.~A.,  {Lombardi} Jr. J.~C.,  {Rasio} F.~A.,    {Warren}
  A.~R.,  2001, \apj, 548, 323

\bibitem[\protect\citeauthoryear{{Sills}, {Karakas} \& {Lattanzio}}{{Sills}
  et~al.}{2008}]{2008arXiv0811.2974S}
{Sills} A.,  {Karakas} A.,    {Lattanzio} J.,  2008, ArXiv e-prints

\bibitem[\protect\citeauthoryear{{Springel}}{{Springel}}{2005}]{2005MNRAS.364.%
1105S}
{Springel} V.,  2005, \mnras, 364, 1105

\bibitem[\protect\citeauthoryear{{Springel} \& {Hernquist}}{{Springel} \&
  {Hernquist}}{2002}]{2002MNRAS.333..649S}
{Springel} V.,  {Hernquist} L.,  2002, \mnras, 333, 649

\bibitem[\protect\citeauthoryear{{Sugimoto} \& {Bettwieser}}{{Sugimoto} \&
  {Bettwieser}}{1983}]{1983MNRAS.204P..19S}
{Sugimoto} D.,  {Bettwieser} E.,  1983, \mnras, 204, 19P

\bibitem[\protect\citeauthoryear{{Suzuki}, {Nakasato}, {Baumgardt},
  {Ibukiyama}, {Makino} \& {Ebisuzaki}}{{Suzuki}
  et~al.}{2007}]{2007ApJ...668..435S}
{Suzuki} T.~K.,  {Nakasato} N.,  {Baumgardt} H.,  {Ibukiyama} A.,  {Makino} J.,
     {Ebisuzaki} T.,  2007, \apj, 668, 435

\bibitem[\protect\citeauthoryear{{Umbreit}, {Chatterjee} \& {Rasio}}{{Umbreit}
  et~al.}{2008}]{2008ApJ...680L.113U}
{Umbreit} S.,  {Chatterjee} S.,    {Rasio} F.~A.,  2008, \apjl, 680, L113

\end{thebibliography}

\end{document}